\documentclass[a4paper,11pt]{article}
\pdfoutput=1 % if your are submitting a pdflatex (i.e. if you have
\usepackage{jheppub} % for details on the use of the package, please
\usepackage{csquotes}
\usepackage[T1]{fontenc} % if needed
\usepackage[english]{babel}
\usepackage{graphicx}			    	%inserire immagini
\usepackage{slashed}
\usepackage{caption}
\usepackage{amsmath}
\usepackage{amsthm}
\usepackage{amsfonts}
\usepackage{mathtools}
\usepackage{subcaption}
\usepackage{tikz} 
\usepackage{tikz-feynman}
\usepackage{tikz-3dplot}
\tikzfeynmanset{compat=1.1.0}
\usepackage{placeins}
\usepackage{scalerel}
\usepackage{verbatim}
\usepackage{mathtools}

\usetikzlibrary{shapes.misc}
\tikzset{cross/.style={cross out, draw=black, minimum size=2*(#1-\pgflinewidth), inner sep=0pt, outer sep=0pt},
	cross/.default={5pt}}

\DeclareMathOperator{\Det}{Det}
\DeclareMathOperator{\Tr}{Tr}

\theoremstyle{plain}
\theoremstyle{plain}

\theoremstyle{definition}
\makeatletter
\def\@fpheader{\relax}
\makeatother
\tikzset{
	skel/.style={line width=0.9pt, black},
	pinch/.style={line width=0.8pt, black, dash pattern=on 2.2pt off 2.4pt},
}
\newcommand{\ophairpin}[3]{%
	\begin{scope}[shift={(#1:#2)}, rotate=#1-90]
		\draw[skel] (-0.32,#3) .. controls (-0.28,0.40) and (-0.12,0.03) .. (0,0)
		.. controls (0.12,0.03) and (0.28,0.40) .. (0.32,#3);
		\fill (0,0) circle (1.6pt);
\end{scope}}

\title{Large-$N$ expansion in SU($N$) Yang-Mills-like theories}

\author[a]{Marco Bochicchio,}
\author[b,a]{Mauro Papinutto,}
\author[c]{Francesco Scardino}

\affiliation[a]{Physics Department, INFN Roma1, 
	Piazzale A. Moro 2, Roma, I-00185, Italy}
\affiliation[b]{Physics Department, Sapienza University, Piazzale A. Moro 2, Roma, I-00185, Italy}
\affiliation[c]{Department of Physics, School of Mathematics and Natural Sciences, Bergische Universit\"at Wuppertal, Gau\ss stra\ss e 20, 42119 Wuppertal, Germany}

\emailAdd{marco.bochicchio@roma1.infn.it}
\emailAdd{mauro.papinutto@roma1.infn.it}
\emailAdd{scardino@uni-wuppertal.de}

\abstract{It has been known for fifty years that bare U($N$) pure Yang-Mills (YM) theory admits the large-$N$ 't Hooft expansion \cite{tHooft:1973alw} for the $n$-point connected correlators of gauge-invariant single-trace operators, where the corresponding Feynman diagrams are topologically classified by a sum over the genus $g$ of $n$-punctured smooth orientable closed Riemann surfaces. In fact, the 't Hooft expansion extends \emph{verbatim} to U($N$) YM-like theories whose Lagrangians only involve fields valued in the U($N$) Lie algebra -- including supersymmetric (SUSY) extensions of pure YM theory such as $\mathcal{N}=1,4$ SUSY YM. The aim of the present paper is to work out systematically the topology of the large-$N$ expansion in SU($N$) -- as opposed to U($N$) -- YM-like theories, where a number of subtleties inevitably arise. Indeed, since the U($1$) factor inside U($N$) YM-like theories decouples, the large-$N$ U($N$) and SU($N$) expansions should either essentially coincide or intrinsically differ -- in the latter case necessarily by extra, possibly singular, topologies, as the 't Hooft expansion in U($N$) YM-like theories already exhausts all smooth topologies. In fact, we demonstrate that 't Hooft smooth topologies are generically incomplete in the large-$N$ expansion of both SU($N$) and renormalized U($N$) YM-like theories, because renormalization inevitably separates the interacting SU($N$) sector from the free U($1$) sector. Specifically, we interpret the extra topologies arising in SU($N$) theories for correlators of higher-twist quasi-partonic operators as singular pinched closed surfaces, thereby extending our previous investigation \cite{BPSL} of correlators of twist-$2$ operators, for which extra pinched tori arise.}

\begin{document} 
	
	\definecolor{c969696}{RGB}{150,150,150}
	\maketitle	% make the title area
	\flushbottom

	\section{Introduction and outlook}
	
	Fifty years ago, 't Hooft classified Feynman diagrams in the large-$N$ expansion of pure U($N$) Yang-Mills (YM) theory by oriented closed Riemann surfaces, obtained by a suitable gluing of propagators and vertices in a double-line representation \cite{tHooft:1973alw}. \par
	Most remarkably, the large-$N$ weight of Feynman diagrams contributing to $n$-point connected correlators of gauge-invariant single-trace operators turns out to be $N^\chi$, where $\chi$ is the Euler characteristic $\chi=2-2g-n$ of the corresponding Riemann surface with $g$ handles and $n$ punctures.
	The matching of topologies and weights has suggested \cite{tHooft:1973alw,Veneziano:1976wm,Aharony:1999ti} the existence of a string-theory interpretation \cite{Veneziano:1974dr,Aharony:1999ti}, with closed-string coupling $g_s=\frac{1}{N}$, of the 't Hooft expansion, on which we will briefly comment at the end of this section. \par
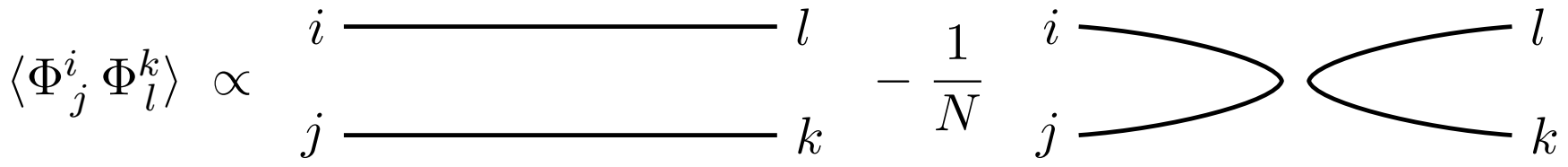
\begin{figure}[h!]
		\centering
		$\langle \Phi^i_{\,j}\,\Phi^k_{\,l}\rangle\;\propto\;$
		\begin{tikzpicture}[baseline={(0,-0.1)},
			skel/.style={line width=0.9pt, black}]
			% leading contraction: delta^i_l delta^k_j
			\draw[skel] (-1.6,0.4) -- (1.6,0.4);
			\draw[skel] (-1.6,-0.4) -- (1.6,-0.4);
			\node[left]  at (-1.6, 0.4) {$i$};
			\node[left]  at (-1.6,-0.4) {$j$};
			\node[right] at ( 1.6, 0.4) {$l$};
			\node[right] at ( 1.6,-0.4) {$k$};
		\end{tikzpicture}
		$\;-\;\dfrac{1}{N}\;$
		\begin{tikzpicture}[baseline={(0,-0.1)},
			skel/.style={line width=0.9pt, black}]
			% u(1) projector: delta^i_j delta^k_l
			\draw[skel] (-1.6,-0.4) .. controls (-0.90,-0.35) and (-0.14,-0.15) .. (-0.10,0)
			.. controls (-0.14,0.15) and (-0.90,0.35) .. (-1.6,0.4);
			\draw[skel] ( 1.6,-0.4) .. controls ( 0.90,-0.35) and ( 0.14,-0.15) .. ( 0.10,0)
			.. controls ( 0.14,0.15) and ( 0.90,0.35) .. ( 1.6,0.4);
			\node[left]  at (-1.6, 0.4) {$i$};
			\node[left]  at (-1.6,-0.4) {$j$};
			\node[right] at ( 1.6, 0.4) {$l$};
			\node[right] at ( 1.6,-0.4) {$k$};
		\end{tikzpicture}
		\caption{Double-line representation of the propagator in eq. \eqref{eq:11}.}
		\label{fig:propdoubleline}
	\end{figure}

	Since the early days of the 't Hooft expansion, it was clear that the 't Hooft argument may need to be modified in the SU($N$) as opposed to the U($N$) case, due to the fact that in the former case the propagator has a $\frac{1}{N}$ correction -- reported in eq. \eqref{eq:11} and depicted in fig. \ref{fig:propdoubleline} -- compared to the latter.
	Indeed, in \cite{Aharony:1999ti} it was remarked that \enquote{in an SU($N$) theory there is
	a small mixing term
	\begin{equation}\label{eq:11}
		\langle \Phi^i_j \Phi^k_l\rangle\propto\left(\delta^i_l \delta^k_j-\frac{1}{N}\delta^i_j\delta^k_l\right)\,,
	\end{equation}
	which makes the expansion slightly more complicated, but this involves only subleading
	terms in the large-$N$ limit so we will neglect this difference.}\par
	One of the aims of the present paper is precisely not to neglect the above difference.
	We point out that the large-$N$ expansion of the SU($N$) theory either essentially coincides with the U($N$) one or intrinsically differs, since the 't Hooft classification already exhausts all the smooth topologies of Riemann surfaces.\par
	In fact, we observe that all the vacuum diagrams in the SU($N$) theory coincide with the ones of the U($N$) theory because, if the subleading part of the propagator is connected to an action vertex, the corresponding diagram vanishes (fig. \ref{fig:proof}),  
	as the subleading propagator involves a U($1$) projection, while the vertices arise from the commutators of the SU($N$) theory.
	Yet, it is much less clear whether the decoupling of the subleading propagator extends to correlators of single-trace local operators. \par
	Indeed, the effect of the subleading propagator on the topology of Feynman diagrams contributing to the large-$N$ expansion of correlators of local operators has not been explored in the literature until now, with the notable exception of \cite{BPSL}.\par
	In \cite{BPSL} it was discovered that only for very special local operators do the topologies of the large-$N$ SU($N$) and U($N$) expansions coincide, thanks to a mechanism similar to the aforementioned one for vacuum diagrams.\par 
	Yet, in general, the topologies of the large-$N$ expansion of the SU($N$) and U($N$) theories do not coincide \cite{BPSL}. Specifically, for correlators of $2$-gluon twist-$2$ operators with maximal spin components along the $p_+$ direction, certain new -- geometrically planar -- diagrams occur at subleading $\frac{1}{N}$ order, contrary to the U($N$) theory, where geometrically planar diagrams only occur at leading order. \par 
	In \cite{BPSL} it was argued that, after gluing oppositely oriented lines of such subleading diagrams, singular surfaces that have the topology of pinched tori arise, this being the most conservative interpretation of the topology of the new Feynman diagrams in the SU($N$) theory, as we will discuss in more detail in the present paper. \par
	In fact, the careful reader may wonder how it is possible that pinched tori -- being the nodal degeneration of smooth tori, which are nonplanar diagrams -- may arise from geometrically planar diagrams. The apparent paradox is solved by observing that the normalization\footnote{The normalization of a surface with nodes is obtained by cutting the nodes to obtain possibly disconnected pieces with marked points corresponding to the nodes \cite{Chan2021ModuliSO,Witten:2012ga}.} of pinched tori consists of punctured spheres, which are actually planar diagrams. \par
	The aim of the present paper is to extend the above analysis to a large class of multi-gluon local operators in order to establish that the 't Hooft expansion, consisting only of smooth surfaces, is actually incomplete in the SU($N$) theory. For multi-gluon higher-twist operators, nonplanar pinched diagrams occur in perturbation theory as well, which give rise by gluing to higher-genus pinched Riemann surfaces, whose normalization includes higher-genus smooth punctured surfaces, contrary to the twist-$2$ case, which only includes punctured spheres. \par
	While the main focus of our analysis is the perturbative expansion of SU($N$) YM-like theories on $R^4$, it is interesting to compare it with the large-$N$ expansion of the free energy in two-dimensional YM theory on a \emph{compact} Riemann surface, for which an exact nonperturbative solution is available. In the latter case, it is known that, in the large-$N$ expansion, in both the U($N$) and SU($N$) theories, smooth surfaces that are ramified coverings of the compact space-time occur. However, in the SU($N$) case, extra singular topologies -- including \enquote{zero-size tubes} essentially equivalent to our pinches -- arise \cite{M93,G93,G932,A23,A25}. \par
	In any case, our findings have considerable consequences -- perturbative and nonperturbative -- both in field theory and for the existence of a conjectured string solution of the large-$N$ SU($N$) YM theory and its $\mathcal{N}=1$ SUSY extension, for the correlators and the S-matrix in four dimensions, which have already been partially discussed in \cite{BPSL,BPS42}. \par
	The implications for the string interpretation of the $\mathcal{N}=4$ SUSY SU($N$) YM theory will appear in a forthcoming paper.
	
	\section{Plan of the paper}
	In section \ref{sec:s1}, we review the 't Hooft expansion in the bare U($N$) YM theory.\par
	In section \ref{sec:s3}, we demonstrate that the large-$N$ expansion for vacuum diagrams and a special class of correlators of local operators in SU($N$) YM theory  coincides with the 't Hooft expansion in the U($N$) theory, in the sense that the contribution of all potentially extra topologies vanishes.\par
	In section \ref{sec:s5}, we recall the new topologies arising from correlators of twist-$2$ operators in SU($N$) and renormalized U($N$) YM theories.\par
	In section \ref{sec:s6}, we recall the renormalization properties of  quasi-partonic operators of arbitrary twist in SU($N$) and U($N$) YM theories.\par
	In section \ref{sec:s7}, we study the new topologies that arise for the above  $3$- and $4$-gluon quasi-partonic operators in SU($N$)  YM theory.\par
	In section \ref{sec:conc}, we summarize our conclusions.\par
	In the appendices, we provide definitions and ancillary computations to make the paper self-contained.

	\section{The 't Hooft expansion in bare U($N$) YM theory}\label{sec:s1}

We recall the fundamental properties of the 't Hooft expansion \cite{tHooft:1973alw,Veneziano:1976wm} in large-$N$ U($N$) YM theory. 
The gauge field is valued in a representation of the $u(N)$ Lie algebra,
\begin{equation}
	A_\mu=A^\alpha_\mu T^\alpha_R\,,
\end{equation}
with $\alpha=0,\ldots,N^2-1$, where $T^a_R$ ($a=1,\ldots,N^2-1$) are matrices in an irreducible representation of the $su(N)$ Lie algebra normalized as
\begin{align}
	\Tr T^a_R T^b_R = T(R)\delta^{ab} \,,
\end{align}
with Dynkin index
\begin{equation}
	T(R)=\frac{d(R)C_2(R)}{N^2-1}\,,
\end{equation}
where $d(R)$ is the dimension of the representation $R$ and $C_2(R)$ is its Casimir. Here, $T^0_R$ is the $u(1)$ generator, which we find convenient to normalize in the same way
\begin{align}
	\Tr T^0_R T^0_R = T(R)\,.
\end{align}
By defining 
\begin{align}
	F_{\mu\nu} = \partial_\mu A_\nu - \partial_\nu A_\mu +i\left[A_\mu,A_\nu\right]\,,
\end{align}
it follows that the Lagrangian density,
\begin{equation}
	\Tr_R F^2 = T(R)F^\alpha_{\mu\nu}F^\alpha_{\mu\nu}\,,
\end{equation}
is independent of the representation $R$ up to an overall normalization. Typically, in the literature, $R$ is chosen to be the fundamental representation\footnote{Sometimes, instead, the adjoint representation is used.}, so that $T^a_F=\lambda^a$ (appendix~\ref{app:group}) and the Euclidean action with Wilsonian normalization reads
\begin{align}\label{eq:action}
	S_{YM} =\frac{1}{2 g_{YM}^2}\int d^4x\Tr_F F^2= \frac{N}{2 g^2}\int d^4x\Tr F^2\,,
\end{align}
where $g^2 = Ng^2_{YM}$ is the 't Hooft coupling \cite{tHooft:1973alw}.
Using eq.~\eqref{eq:doubled}, the action becomes
\begin{equation}
	S_{YM} = \frac{N}{4 g^2}\int d^4x\,  \big(F_{\mu\nu}^aF_{\mu\nu}^a+ F_{\mu\nu}^0F_{\mu\nu}^0 \big)\,.
\end{equation}
Clearly, since the photon field $A^0\lambda^0$ commutes with all the gluon fields $A^a\lambda^a$ according to eq.~\eqref{eq:u(1)}, the $0$-component of the field strength is simply its Abelian part
\begin{align}
	F^0_{\mu\nu} = \partial_\mu A^0_\nu - \partial_\nu A^0_\mu \,.
\end{align}
Therefore, the trace $\Tr A$ of the gauge field decouples from the interacting part of the theory, and the action splits into an interacting SU($N$) part and a free U($1$) part
\begin{equation}
	S^{U(N)}_{YM}(A^a,A^0)  =S_{YM}^{SU(N)}(A^a)+ S_{YM}^{U(1)}(A^0)\,.
\end{equation}
The propagators and vertices of the U($N$) theory admit the 't Hooft double-line representation (figs.~\ref{fig:unprop} and \ref{fig:3vertex}) \cite{tHooft:1973alw,Veneziano:1976wm,marino2015instantons,maltoni2003color} and give rise to the 't Hooft topological classification of Feynman diagrams by rewriting eq.~\eqref{eq:action} as
\begin{align}
	S_{YM} = \frac{N}{2 g^2}\int d^4x\ (F_{\mu\nu})^i_{\,j}(F_{\mu\nu})^j_{\,i}\,,
\end{align}
where $A^i_{\, j}=A^\alpha (\lambda^\alpha)^i_{\, j}$ and $(F_{\mu\nu})^i_{\,j}=F_{\mu\nu}^\alpha (\lambda^\alpha)^i_{\, j}$ are $N \times N$ Hermitian matrices (appendix~\ref{app:group}).

\subsection{Propagators and vertices in the double-line representation}

Following 't Hooft, we work out the propagator and the $3$- and $4$-point vertices in the double-line representation in the U($N$) theory, writing the gluon field $(A_\mu)^i_{\, j}$ as an $N \times N$ Hermitian matrix as in section~\ref{sec:s1}.
Therefore, the U($N$) propagator reads
\begin{equation}
	\langle (A_{\mu})^{i}_{\, j}(x)(A_{\nu})^{k}_{\, l}(y)\rangle=\frac{1}{2}\delta^{i}_{\, l}\delta^{k}_{\, j}D_{\mu\nu}(x-y)\,,
\end{equation}
and the color structure of the $3$- and $4$-point vertices is given by the traceless tensors
\begin{align}
	\label{vertices}
	&(\mathcal{V}_3)^{i\,k\,m}_{\,j\,l\,n} = f^{abc} (\lambda^a)^i_{\, j} (\lambda^b)^k_{\, l} (\lambda^c)^m_{\, n}= -\frac{i}{4}\left(\delta^i_{\,l}\delta^k_{\,n}\delta^m_{\,j}-\delta^i_{\,n}\delta^k_{\,j}\delta^m_{\,l}\right)\,,\nonumber\\
	&(\mathcal{V}_4)^{i\,k\,m\,p}_{\,j\,l\,n\,q}=  f^{abe}f^{cde} (\lambda^a)^i_{\, j} (\lambda^b)^k_{\, l} (\lambda^c)^m_{\, n}(\lambda^d)^p_{\, q}\nonumber\\
	&\qquad\qquad=\frac{1}{8} \left(\delta^i_{\,l}\delta^m_{\,j}\delta^k_{\,q}\delta^p_{\,n}-\delta^i_{\,l}\delta^p_{\,j}\delta^k_{\,n}\delta^m_{\,q}+\delta^i_{\,n}\delta^k_{\,j}\delta^p_{\,l}\delta^m_{\,q}-\delta^i_{\,q}\delta^k_{\,j}\delta^m_{\,l}\delta^p_{\,n}\right)\,,
\end{align}
where we have used eq.~\eqref{eq:doubled} to express them in this representation, which consists of drawing a line for every color index rather than for every gluon.\par
\begin{figure}[h!]
	\centering
	\includegraphics[width=0.7\linewidth]{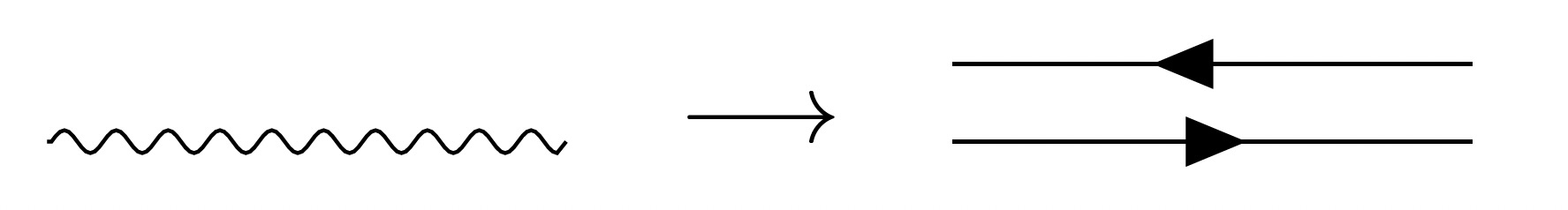}
	\caption{The gluon propagator is replaced by two lines with opposite orientations.}
	\label{fig:unprop}
\end{figure}
\begin{figure}[h!]
	\centering
	\includegraphics[width=1\linewidth]{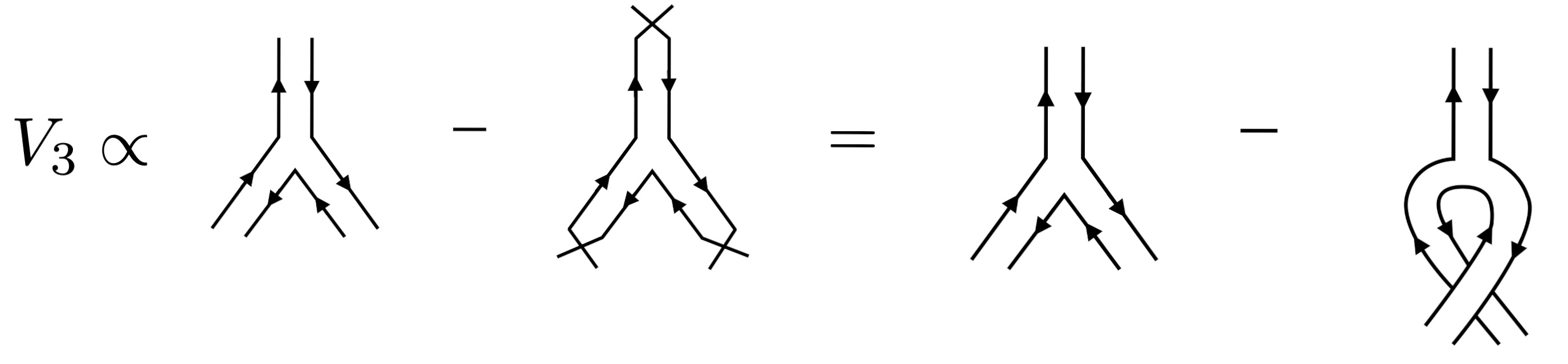}
	\caption{The $3$-gluon vertex in the double-line representation \cite{marino2015instantons}.}
	\label{fig:3vertex}
\end{figure}
\begin{figure}[h!]
	\centering
	\includegraphics[width=0.8\linewidth]{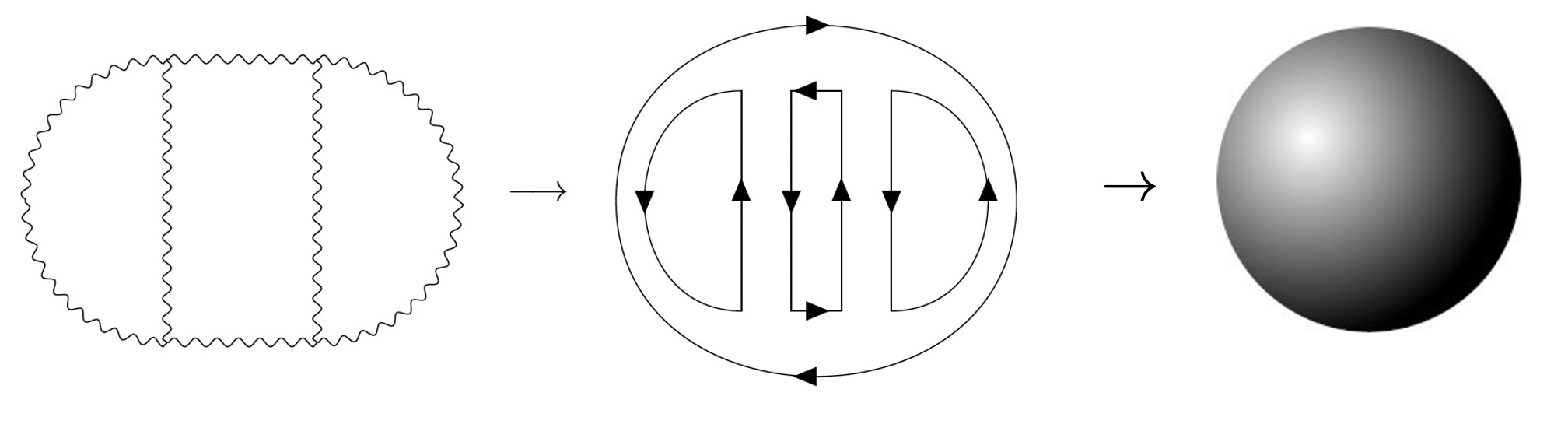}
	\caption{Double-line representation of a planar vacuum diagram. The three internal loops bound three topological disks, which are glued together to form a single inner disk. The inner and outer disks are then glued together to form a sphere.}
	\label{fig:thooftdouble}
\end{figure}

Thus, the gluon propagator is represented by a double line (fig.~\ref{fig:unprop}), while the $3$-gluon vertex is displayed in fig.~\ref{fig:3vertex}. The color structure of the $4$-point vertex is simply the square of that of the $3$-point vertex.
Hence, in the double-line representation, Feynman diagrams become oriented polygons (fig.~\ref{fig:thooftdouble}), whose line orientations provide a prescription for gluing them together.\par
\begin{figure}[h!]
	\centering
	\includegraphics[width=1\linewidth]{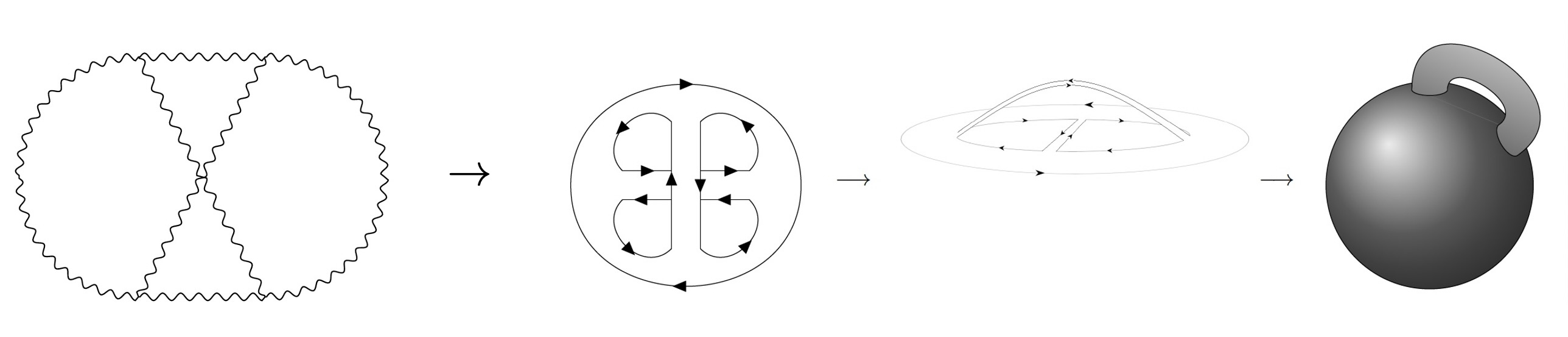}
	\caption{Double-line representation of a nonplanar vacuum diagram. The second drawing shows an outer disk and two smaller inner disks (projected onto the same plane) connected by a strip viewed from above. In the third drawing, the disks lie on the same plane while the strip arches into the third dimension. The two inner disks are then glued together to form a larger disk, which is subsequently glued to the outer disk to form a sphere. Finally, the two sides of the strip with opposite orientations are glued together to form a handle, yielding a torus as the final topology.}
	\label{fig:nonplanar}
\end{figure}

More precisely, the 't Hooft prescription consists of gluing an edge of one polygon to an edge of another if they have opposite orientations. In this way, 't Hooft constructs closed, oriented two-dimensional surfaces from the original Feynman diagrams (figs.~\ref{fig:thooftdouble} and \ref{fig:nonplanar}).\par 
Specifically, a sphere arises from planar diagrams by gluing inner and outer disks (fig.~\ref{fig:thooftdouble}), whereas handles arise from nonplanar diagrams by gluing together opposite edges of arching strips to form higher-genus Riemann surfaces (fig.~\ref{fig:nonplanar}).\par
The $N$-dependence of Feynman diagrams, $N^\chi$, follows from the topology of the glued polygons \cite{tHooft:1973alw}.
Perturbatively, each term in the 't Hooft expansion consists of summing all Feynman graphs that tile surfaces with the same topology, so that each order in $\frac{1}{N}$ is associated with an infinite sum of topologically equivalent Feynman diagrams. 
For example, the generating functional of connected vacuum diagrams admits the expansion
\begin{equation}
	\label{wconnect}
	W=\sum_{g=0}^{\infty} N^{2-2g}\ \mathcal{W}_{g}\,,
\end{equation}
where each $\mathcal{W}_g$ contains an infinite sum of topologically equivalent connected vacuum Feynman diagrams.

		\section{Cases where the topologies of the large-$N$ expansion of SU($N$) and U($N$) YM theories coincide}\label{sec:s3}

\subsection{Vacuum diagrams in large-$N$ SU($N$) YM theory}

In the large-$N$ expansion of the SU($N$) theory, gluons contribute to vacuum diagrams as follows. The gluon propagator reads
\begin{equation}
	\langle (A_{\mu})^{i}_{\, j}(x)(A_{\nu})^{k}_{\, l}(y)\rangle=\frac{1}{2}\left(\delta^{i}_{\, l}\delta^{k}_{\, j}-\frac{1}{N}\delta^{i}_{\, j}\delta^{k}_{\, l}\right)D_{\mu\nu}(x-y)
\end{equation}
and receives two contributions. The leading large-$N$ contribution is proportional to the identity in $u(N)$, $I^{i\,k}_{\,j\,l}=\delta^i_{\, l}\delta^k_{\, j}$, while the subleading contribution is proportional to the projector onto the $u(1)$ subalgebra inside $u(N)$,
$P^{i\,k}_{\, j\, l}=\frac{1}{N}\delta^i_{\, j}\delta^k_{\, l}$ (eq.~\eqref{eq:projectors}), so that
\begin{equation}
	\langle (A_{\mu})^{i}_{\, j}(x)(A_{\nu})^{k}_{\, l}(y)\rangle=\frac{1}{2}\left(I^{i\,k}_{\,j\,l}-P^{i\,k}_{\, j\, l}\right)D_{\mu\nu}(x-y)\,.
\end{equation}
When the subleading part of the gluon propagator is attached to the vertices of the action (appendix~\ref{app:ymlightcone}),
$V_3 \propto \frac{g}{\sqrt{N}}f^{abc}$ (fig.~\ref{fig:3vertex}) and 
$V_4 \propto \frac{g^2}{N}f^{abe}f^{ecd}$, its contribution vanishes \cite{marino2015instantons,maltoni2003color}.
\begin{figure}[h!]
	\centering
	\includegraphics[width=0.7\linewidth]{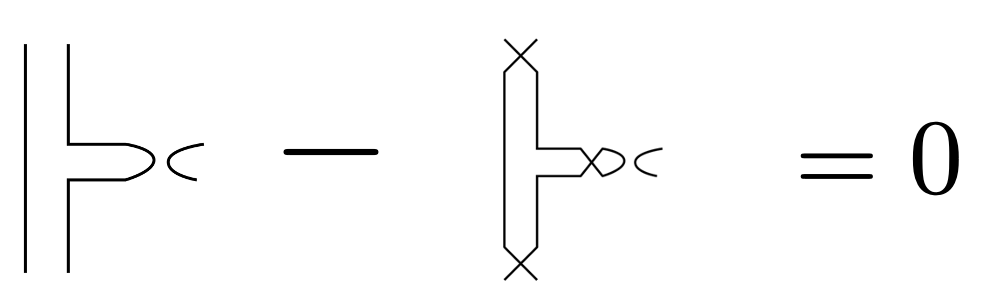}
	\caption{Decoupling of the subleading propagator from the $3$-gluon vertex.}
	\label{fig:proof}
\end{figure}
This is straightforward \cite{maltoni2003color}, since the subleading propagator is a $u(1)$ contribution, whereas the action vertices are purely nonabelian. Explicitly,
\begin{align}
	&(\mathcal{V}_3)^{i\,k\,m}_{\,j\,l\,n}P^{i\,k'}_{\, j\, l'} = -\frac{i}{4N}\left(\delta^i_{\,l}\delta^k_{\,n}\delta^m_{\,j}-\delta^i_{\,n}\delta^k_{\,j}\delta^m_{\,l}\right)\delta^i_{\, j}\delta^{k'}_{\, l'} =-\frac{i}{4N}\left(\delta^m_{\,l}\delta^k_{\,n}-\delta^k_{\,n}\delta^m_{\,l}\right)\delta^{k'}_{\, l'} =0\,,\nonumber\\
	&(\mathcal{V}_4)^{i\,k\,m\,p}_{\,j\,l\,n\,q}P^{i\,k'}_{\, j\, l'}  =\frac{1}{8N} \left(\delta^i_{\,l}\delta^m_{\,j}\delta^k_{\,q}\delta^p_{\,n}-\delta^i_{\,l}\delta^p_{\,j}\delta^k_{\,n}\delta^m_{\,q}+\delta^i_{\,n}\delta^k_{\,j}\delta^p_{\,l}\delta^m_{\,q}-\delta^i_{\,q}\delta^k_{\,j}\delta^m_{\,l}\delta^p_{\,n}\right)\delta^i_{\, j}\delta^{k'}_{\, l'}\nonumber\\
	&\qquad\qquad =\frac{1}{8N} \left(\delta^m_{\,l}\delta^k_{\,q}\delta^p_{\,n}-\delta^p_{\,l}\delta^k_{\,n}\delta^m_{\,q}+\delta^k_{\,n}\delta^p_{\,l}\delta^m_{\,q}-\delta^k_{\,q}\delta^m_{\,l}\delta^p_{\,n}\right)\delta^{k'}_{\, l'} = 0 \,.
\end{align}
This property admits a simple graphical proof (fig.~\ref{fig:proof}) \cite{BPSL}. Therefore, for vacuum diagrams, the subleading propagator can be safely ignored, and the topologies of the large-$N$ expansion of the U($N$) and SU($N$) theories coincide.

\subsection{Single-trace operators arising from nested commutators}\label{subsec:nest}

Generalizing the structure of $V_3$ and $V_4$, we observe that the above argument extends to connected correlators of single-trace operators in the SU($N$) theory, provided that the corresponding local vertices are proportional to a matrix product \cite{maltoni2003color} of $(T^a)^b_c= -if^{abc}$,
\begin{equation}\label{eq:thooft}
	(T^{a_2} \dots T^{a_{n-1}})^{a_1}_{a_n}=2\Tr(\lambda^{a_1}[\lambda^{a_2}, \dots [\lambda^{a_{n-1}},\lambda^{a_{n}}] \dots]) \,,
\end{equation} 
with $[\lambda^a,\lambda^b] = i f^{abc}\lambda^c$ (appendix~\ref{app:group}), which requires at least $3$-gluon operators in perturbation theory.\par
Indeed, the above vertices in the double-line representation have no $\frac{1}{N}$ corrections \cite{maltoni2003color}, and their contraction with the subleading propagator vanishes because they involve nested commutators.\par
We present the proof in two steps: first, there are no $\frac{1}{N}$ corrections; second, the contraction with the subleading propagator vanishes.\par
Given the completeness relation
\begin{equation}
	\sum_{a=1}^{N^2-1}(\lambda^a)^{i}_{\,j}(\lambda^a)^{k}_{\,l}=\frac{1}{2}\left(\delta^i_{\,l}\delta^k_{\,j}-\frac{1}{N}\delta^i_{\,j}\delta^k_{\,l}\right) \,,
\end{equation}
for a fixed $N\times N$ matrix $M$ we obtain
\begin{align}\label{eq:com1}
	\sum_{a=1}^{N^2-1}[\lambda^a,M]^{i}_{\,j}(\lambda^a)^{k}_{\,l} &=	\sum_{a=1}^{N^2-1}\left[(\lambda^a)^{i}_{\,m}M^{m}_{\,j}-M^{i}_{\,m}(\lambda^a)^{m}_{\,j}\right](\lambda^a)^{k}_{\,l}\nonumber\\
	&=	\sum_{a=1}^{N^2-1}\left[(\lambda^a)^{i}_{\,m}(\lambda^a)^{k}_{\,l}M^{m}_{\,j}-M^{i}_{\,m}(\lambda^a)^{m}_{\,j}(\lambda^a)^{k}_{\,l}\right]\nonumber\\
	&=\frac{1}{2}\left[\left(\delta^i_{\,l}\delta^k_{\,m}-\frac{1}{N}\delta^i_{\,m}\delta^k_{\,l}\right)M^{m}_{\,j}-M^{i}_{\,m}\left(\delta^m_{\,l}\delta^k_{\,j}-\frac{1}{N}\delta^m_{\,j}\delta^k_{\,l}\right)\right]\nonumber\\
	&= \frac{1}{2}\delta^{i}_{\,l}M^k_{\,j}- \frac{1}{2}M^i_{\,l}\delta^{k}_{\,j}-\frac{1}{2N}\left(M^{i}_{\,j}\delta^{k}_{\,l}-M^{i}_{\,j}\delta^{k}_{\,l} \right)\nonumber\\
	&= \frac{1}{2}\delta^{i}_{\,l}M^k_{\,j}- \frac{1}{2}M^i_{\,l}\delta^{k}_{\,j}\,.
\end{align}
Moreover, if $M$ is of order $N^0$ with no $\frac{1}{N}$ corrections, then so is the right-hand side of eq.~\eqref{eq:com1}. Similarly, for fixed $N\times N$ matrices $M$ and $O$, we have
\begin{align}\label{eq:com2}
	&\sum_{a=1}^{N^2-1}\Tr\left(\lambda^a[M,O]\right)(\lambda^a)^{i}_{\,j} =\sum_{a=1}^{N^2-1}(\lambda^a)^{n}_{\,m}[M,O]^{m}_{\,n}(\lambda^a)^{i}_{\,j} \nonumber\\
	&=\sum_{a=1}^{N^2-1}(\lambda^a)^{n}_{\,m}(\lambda^a)^{i}_{\,j} [M,O]^{m}_{\,n}=\frac{1}{2}\left(\delta^n_{\,j}\delta^i_{\,m}-\frac{1}{N}\delta^n_{\,m}\delta^i_{\,j}\right)	[M,O]^{m}_{\,n}\nonumber\\
	&=\frac{1}{2}[M,O]^i_{\,j}-\frac{1}{2N}\Tr\left([M,O]\right)\delta^{i}_{\,j}\nonumber\\
	&= \frac{1}{2} [M,O]^i_{\,j}\,,
\end{align}
where, again, if $M$ and $O$ are of order $N^0$ with no $\frac{1}{N}$ corrections, so is the right-hand side because the trace of the commutator vanishes. Now, representing the generalized vertices in the double-line representation, we get
\begin{align}\label{eq:vertexdoublea}
	&(T^{a_2} \dots T^{a_{n-1}})^{a_1}_{a_n}\,(\lambda^{a_1})^{i_1}_{\,j_1} \ldots (\lambda^{a_{n}})^{i_n}_{\,j_n} \nonumber\\
	&=2\Tr(\lambda^{a_1}[\lambda^{a_2}, \dots [\lambda^{a_{n-1}},\lambda^{a_{n}}] \dots])\,(\lambda^{a_1})^{i_1}_{\,j_1} \ldots (\lambda^{a_{n}})^{i_n}_{\,j_n}\nonumber\\
	&=[\lambda^{a_2}, \dots [\lambda^{a_{n-1}},\lambda^{a_{n}}] \dots]^{i_1}_{\,j_1}\,(\lambda^{a_2})^{i_2}_{\,j_2} \ldots (\lambda^{a_{n}})^{i_n}_{\,j_n}\,,
\end{align}
where in the last line we have employed eq.~\eqref{eq:com2}. Denoting
\begin{equation}\label{eq:Vn}
	\mathcal{V}_n{}^{i_1\,i_2\ldots i_n}_{j_1\,j_2\ldots j_n}=\sum_{a_2,\ldots,a_n=1}^{N^2-1}[\lambda^{a_2}, \dots [\lambda^{a_{n-1}},\lambda^{a_{n}}] \dots]^{i_1}_{\,j_1}\,(\lambda^{a_2})^{i_2}_{\,j_2} \ldots (\lambda^{a_{n}})^{i_n}_{\,j_n}\,,
\end{equation}
which is the right-hand side of eq.~\eqref{eq:vertexdoublea} (i.e., the double-line representation of the generalized vertex with $n$ legs), we prove by induction on $n\geq3$ that it is a sum of products of Kronecker deltas with $N$-independent coefficients.\par
For $n=3$, the vertex is $(T^{a_2})^{a_1}_{a_3}=-if^{a_2a_1a_3}=if^{a_1a_2a_3}$; in the double-line representation, this reads
\begin{align}\label{eq:vertexdoublea2}
	(T^{a_2} )^{a_1}_{a_3}\,(\lambda^{a_1})^{i_1}_{\,j_1}(\lambda^{a_2})^{i_2}_{\,j_2}   (\lambda^{a_{3}})^{i_3}_{\,j_3} 
	&=2\Tr(\lambda^{a_1}[\lambda^{a_2},\lambda^{a_3}])\,(\lambda^{a_1})^{i_1}_{\,j_1}(\lambda^{a_2})^{i_2}_{\,j_2}   (\lambda^{a_{3}})^{i_3}_{\,j_3}\nonumber\\
	&=[\lambda^{a_2},\lambda^{a_3}]^{i_1}_{\,j_1}\,(\lambda^{a_2})^{i_2}_{\,j_2}   (\lambda^{a_{3}})^{i_3}_{\,j_3}\,,
\end{align} 
which is eq.~\eqref{eq:com2} with $M=\lambda^{a_2}$ and $O=\lambda^{a_3}$. Thus, we eliminate $a_2$ by means of eq.~\eqref{eq:com1} and insert the completeness relation to evaluate the sum over $a_3$, obtaining
\begin{align}\label{eq:V3}
	\mathcal{V}_3{}^{i_1\,i_2\,i_3}_{j_1\,j_2\,j_3}&= \sum_{a_3=1}^{N^2-1}\frac{1}{2}\left[\delta^{i_1}_{\,j_2}(\lambda^{a_3})^{i_2}_{\,j_1}-(\lambda^{a_3})^{i_1}_{\,j_2}\delta^{i_2}_{\,j_1}\right](\lambda^{a_3})^{i_3}_{\,j_3}\nonumber\\
	&=\frac{1}{4}\left[\delta^{i_1}_{\,j_2}\left(\delta^{i_2}_{\,j_3}\delta^{i_3}_{\,j_1}-\frac{1}{N}\delta^{i_2}_{\,j_1}\delta^{i_3}_{\,j_3}\right)-\delta^{i_2}_{\,j_1}\left(\delta^{i_1}_{\,j_3}\delta^{i_3}_{\,j_2}-\frac{1}{N}\delta^{i_1}_{\,j_2}\delta^{i_3}_{\,j_3}\right)\right]\nonumber\\
	&=\frac{1}{4}\left(\delta^{i_1}_{\,j_2}\delta^{i_2}_{\,j_3}\delta^{i_3}_{\,j_1}-\delta^{i_1}_{\,j_3}\delta^{i_3}_{\,j_2}\delta^{i_2}_{\,j_1}\right) \,.
\end{align}
The two $\frac{1}{N}$ terms above cancel because the $u(1)$ projector drops out of the commutator. Hence, $f^{abc}(\lambda^a)^{i_1}_{\,j_1}(\lambda^b)^{i_2}_{\,j_2}(\lambda^c)^{i_3}_{\,j_3}$ is $N$-independent, as already shown in eq.~\eqref{vertices}.\par
Now, suppose that $\mathcal{V}_{n-1}$, with $n-1\geq3$, is $N$-independent. In eq.~\eqref{eq:Vn}, the nested commutator reads $[\lambda^{a_2},M]$, with $M=[\lambda^{a_3}, \dots [\lambda^{a_{n-1}},\lambda^{a_{n}}] \dots]$. Hence, eq.~\eqref{eq:com1} applies to the sum over $a_2$:
\begin{align}\label{eq:Vrec}
	\mathcal{V}_n{}^{i_1\,i_2\,i_3\ldots i_n}_{j_1\,j_2\,j_3\ldots j_n}&=\sum_{a_3,\ldots,a_n=1}^{N^2-1}\frac{1}{2}\left(\delta^{i_1}_{\,j_2}M^{i_2}_{\,j_1}-M^{i_1}_{\,j_2}\delta^{i_2}_{\,j_1}\right)(\lambda^{a_3})^{i_3}_{\,j_3} \ldots (\lambda^{a_{n}})^{i_n}_{\,j_n}\nonumber\\
	&=\frac{1}{2}\,\delta^{i_1}_{\,j_2}\,\mathcal{V}_{n-1}{}^{i_2\,i_3\ldots i_n}_{j_1\,j_3\ldots j_n}-\frac{1}{2}\,\delta^{i_2}_{\,j_1}\,\mathcal{V}_{n-1}{}^{i_1\,i_3\ldots i_n}_{j_2\,j_3\ldots j_n}\,,
\end{align}
since the remaining sums over $a_3,\ldots,a_n$ reproduce $\mathcal{V}_{n-1}$, with the indices of its first leg replaced by $(i_2,j_1)$ and $(i_1,j_2)$, respectively. By the induction hypothesis, the right-hand side is $N$-independent, and hence so is $\mathcal{V}_n$. This completes the induction for the first part: for every $n\geq3$, the generalized vertices of the SU($N$) theory in the double-line representation have no $\frac{1}{N}$ corrections and therefore coincide with the corresponding vertices of the U($N$) theory. Explicitly, eqs.~\eqref{eq:V3} and \eqref{eq:Vrec} show that $\mathcal{V}_n$ is a sum of $2^{n-2}$ products of $n$ Kronecker deltas.\par
We now prove the second part: the contraction of any leg labeled by $m$ in eq.~\eqref{eq:vertexdoublea} with the $u(1)$ projector arising from the subleading propagator -- which contracts $i_m$ with $j_m$ -- identically vanishes. Explicitly, contracting the projector with the first leg amounts to taking the trace of the nested commutator in eq.~\eqref{eq:vertexdoublea}:
\begin{align}\label{eq:Pproof1}
	&(T^{a_2} \dots T^{a_{n-1}})^{a_1}_{a_n}\,(\lambda^{a_1})^{i_1}_{\,j_1} \ldots (\lambda^{a_{n}})^{i_n}_{\,j_n}\, P^{i_1\,k}_{\,j_1\,l} \nonumber\\
	&=\frac{1}{N}\,\delta^{k}_{\,l}\,[\lambda^{a_2}, \dots [\lambda^{a_{n-1}},\lambda^{a_{n}}] \dots]^{i_1}_{\,j_1}\,\delta^{i_1}_{\,j_1}\,(\lambda^{a_2})^{i_2}_{\,j_2} \ldots (\lambda^{a_{n}})^{i_n}_{\,j_n} \nonumber\\
	&=\frac{1}{N}\,\delta^{k}_{\,l}\,\Tr\left([\lambda^{a_2}, \dots [\lambda^{a_{n-1}},\lambda^{a_{n}}] \dots]\right)(\lambda^{a_2})^{i_2}_{\,j_2} \ldots (\lambda^{a_{n}})^{i_n}_{\,j_n}=0\,,
\end{align}
which vanishes because the trace of a commutator is identically zero. For any other leg $2\le m\le n$, the projector closes the color loop of the $m$-th leg, $\delta^{i_m}_{\,j_m}(\lambda^{a_m})^{i_m}_{\,j_m}=\Tr\lambda^{a_m}$, so that
\begin{align}\label{eq:Pproof2}
	&(T^{a_2} \dots T^{a_{n-1}})^{a_1}_{a_n}\,(\lambda^{a_1})^{i_1}_{\,j_1} \ldots (\lambda^{a_{n}})^{i_n}_{\,j_n}\, P^{i_m\,k}_{\,j_m\,l}\nonumber\\
	&=\frac{1}{N}\,\delta^{k}_{\,l}\;(T^{a_2} \dots T^{a_{n-1}})^{a_1}_{a_n}\,\Tr(\lambda^{a_m})\prod_{r\neq m}(\lambda^{a_r})^{i_r}_{\,j_r}=0\,,
\end{align}
because the $su(N)$ generators are traceless (eq.~\eqref{eq:sun}).\par
Therefore, there exists an infinite -- though very special -- class of operators for which the 't Hooft topological expansion, consisting solely of smooth surfaces, applies in SU($N$) YM theory in the exact same way as in the U($N$) theory, since the extra U($1$) projection occurring in SU($N$) theory gives a vanishing contribution to their connected correlators.

	\section{Twist-$2$ operators in large-$N$ SU($N$) and renormalized U($N$) theories}\label{sec:s5}

We pointed out in \cite{BPSL} that the above proof does not apply to $2$-gluon operators, and specifically not to twist-$2$ operators. We thus investigate the new topologies that occur in both the SU($N$) and renormalized U($N$) theories.

\subsection{Large-$N$ expansion of twist-$2$ operators in the SU($N$) theory} 

\begin{figure}[h]
	\centering
	\includegraphics[width=0.4\linewidth]{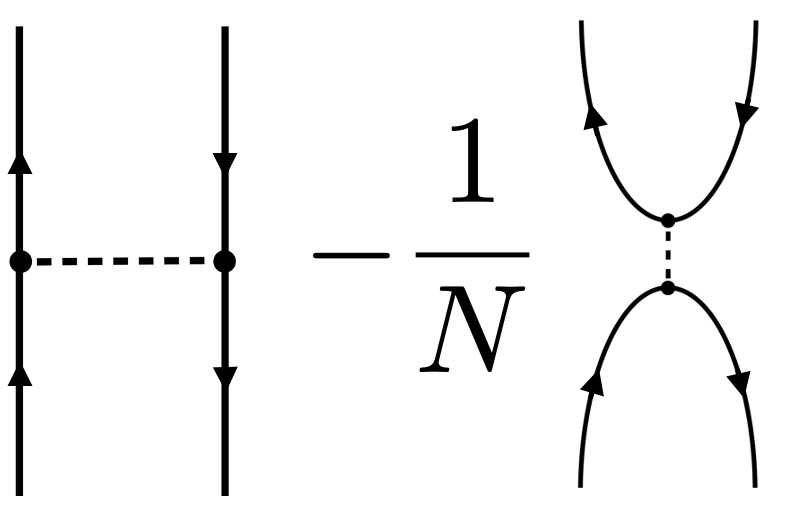}
	\caption{The $2$-gluon twist-$2$ vertex in the double-line representation. The dotted lines identify the punctures.}
	\label{fig:2vertex}
\end{figure}

\begin{figure}[h!]
	\centering
	\includegraphics[width=1\linewidth]{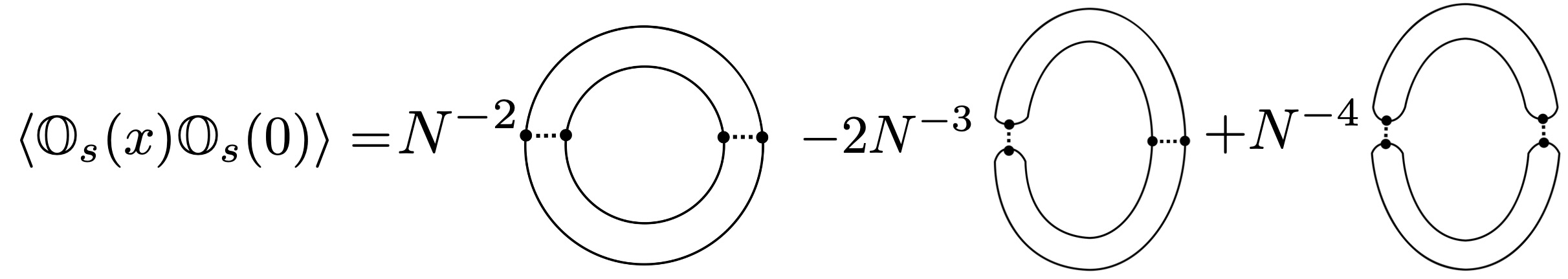}
	\caption{$2$-point correlators of twist-$2$ operators at leading perturbative order.}
	\label{fig:2pointcolor}
\end{figure}

Without loss of generality, we consider nonchiral twist-$2$ operators in the SU($N$) theory as an example, with even spin $s = 2,4,6,\ldots$ and maximal-spin component in the $p_+$ direction in the light-cone gauge \cite{BPS1,BPSpaper2}, analytically continued to Euclidean space-time (appendices~\ref{app:notation}, \ref{app:ymlightcone}, and \ref{app:twist2})
\begin{equation}
	\mathbb{O}_s=\Tr \bar{A}\mathcal{Y}_{s-2}^{\frac{5}{2}}(\overrightarrow{\partial}_z,\overleftarrow{\partial}_z) A=\frac{1}{2}\delta^{ab}\bar{A}^a\mathcal{Y}_{s-2}^{\frac{5}{2}}(\overrightarrow{\partial}_z,\overleftarrow{\partial}_z) A^a \,,
\end{equation}
where
\begin{align}
	\label{defY}
	\mathcal{Y}_{s-2}^{\frac{5}{2}}(\overrightarrow{\partial}_z,\overleftarrow{\partial}_z)=(-1)^{s+1}\overleftarrow{\partial}_z (\overrightarrow{\partial}_z+\overleftarrow{\partial}_z)^{s-2}C^{\frac{5}{2}}_{s-2}\left(\frac{\overrightarrow{\partial}_z-\overleftarrow{\partial}_z}{\overrightarrow{\partial}_z+\overleftarrow{\partial}_z}\right)\overrightarrow{\partial}_z \,,
\end{align}
with $C^{\frac{5}{2}}_{s-2}$ denoting the Gegenbauer polynomials (appendix~\ref{app:jacobi}). The color factor of the corresponding local vertex $\mathbf{V}_2$ involves $\delta^{ab}$ and contains a $\frac{1}{N}$ correction (fig.~\ref{fig:2vertex}) in its double-line representation, just as the gluon propagator does
\begin{equation}\label{eq:deltaab}(\mathbf{V}_2)^{i\, k}_{\, j\, l}=
	\delta^{ab}		(\lambda^a)^{i}_{\,j}(\lambda^b)^{k}_{\,l}=\frac{1}{2}\left(\delta^i_{\,l}\delta^k_{\,j}-\frac{1}{N}\delta^i_{\,j}\delta^k_{\,l}\right)\,,
\end{equation}
so that $\mathbb{O}_{s}$ reads
\begin{align}
	\label{eq:Oprojected}
	\mathbb{O}_{s} 
	&=2(\mathbf{V}_2)^{i\, k}_{\, j\, l}\,\,\bar{A}^{j}_{\, i}(x)\mathcal{Y}_{s-2}^{\frac{5}{2}}(\overrightarrow{\partial}_z,\overleftarrow{\partial}_z)A^{l}_{\, k}(x) \,.
\end{align}
Moreover,
\begin{equation}
	(\mathbf{V}_2)^{i\, k}_{\, j\, l}= \frac{1}{2} (I-P)^{i\, k}_{\, j\, l}\,,
\end{equation}
where the following properties hold (appendix~\ref{app:ymlightcone})
\begin{equation}
	P^2 = P,\qquad \qquad(I-P)P=0\,,\qquad\qquad\Tr (I-P)=N^2-1\,,
\end{equation}
from which it follows that
\begin{equation}\label{eq:propert}
	(I-P)^2=I-P = (I-P)I\,.
\end{equation}
Eq.~\eqref{eq:propert} shows that in the SU($N$) theory we can either take both the vertex and the propagator to be proportional to $I-P$, take the vertex proportional to $I-P$ and the propagator proportional to the identity $I$, or vice versa. 
Computationally, these choices are all equivalent, in the sense that they lead to the exact same final perturbative result; yet, topologically they are \emph{inequivalent}.
Thus, we choose to keep the vertex proportional to $I-P$ and the propagator proportional solely to $I$, so that the standard 't Hooft double-line representation of the propagator in the U($N$) theory also applies to the SU($N$) theory in our framework. This is by far the most convenient choice, owing both to its conservative nature and its simplicity. Indeed, the other two choices have a problematic physical interpretation, as they lead to new diagrams involving disks that are disconnected in color space but connected in space-time.
Hence, with our choice, we obtain for the $2$-point correlator, at lowest order in perturbation theory, the diagrammatic representation in fig.~\ref{fig:2pointcolor}.\par
From fig.~\ref{fig:2pointcolor}, it is apparent that in the SU($N$) theory for twist-$2$ operators there exist new geometrically planar contributions that are nevertheless subleading in the $\frac{1}{N}$ expansion, in stark contrast to the U($N$) case. Indeed, the first graph in fig.~\ref{fig:2pointcolor} represents the standard planar 't Hooft contribution, which gives rise to a $2$-punctured sphere once the boundaries of the two disks are glued together.\par
The remaining diagrams are geometrically planar but do not have the topology of a $2$-punctured sphere.
We would like to interpret them in terms of smooth closed Riemann surfaces, yet we cannot. However, a hint of what their topology should be comes from their large-$N$ weight, obtained by counting the number of closed color lines (each contributing a factor of $N$) and combining this with the weights from the definitions of the operators, the Wilsonian normalization of the action, and the leading and subleading vertices.\par
The total weight of the first diagram is $N^0$, which is precisely the weight of a $2$-punctured sphere, whereas the weight of the remaining diagrams -- despite being geometrically planar -- is $N^{-2}$, which in the standard 't Hooft expansion is the weight of a $2$-punctured smooth torus.
Yet, such a torus cannot arise from the above diagrams because they are geometrically planar.  \par
\begin{figure}[h!]
	\centering
	\includegraphics[width=1\linewidth]{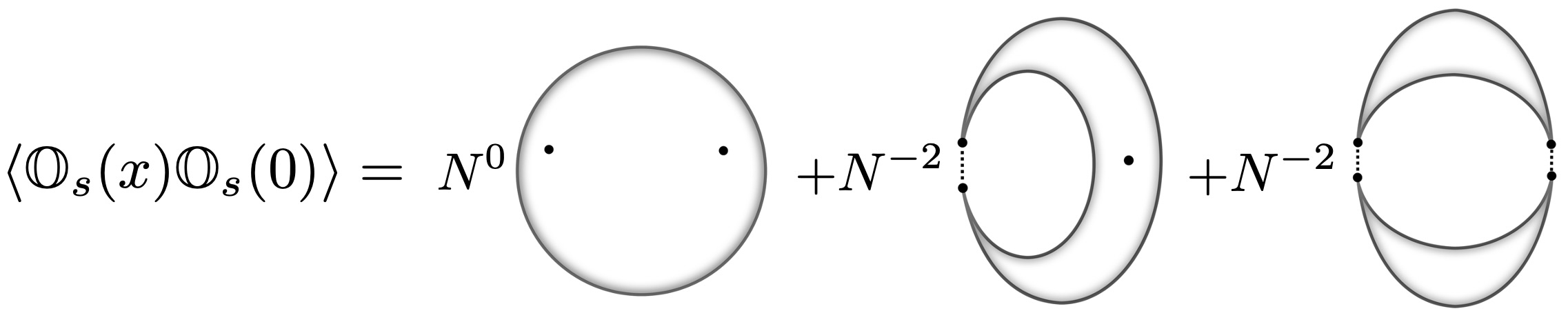}
	\caption{Topology and weights of the diagrams in fig.~\ref{fig:2pointcolor}.}
	\label{fig:2pointnewtopos}
\end{figure}

We provide a topological interpretation of the new diagrams that is consistent with both their weight and planarity \cite{BPSL}. We represent the punctured disks in the last two diagrams in fig.~\ref{fig:2pointcolor} as infinite strips and glue the opposite edges of the strips to obtain infinite cylinders, i.e., punctured spheres. As a result, we obtain the topologies in fig.~\ref{fig:2pointnewtopos}, which involve only punctured spheres, in accordance with the original planarity of the diagrams in fig.~\ref{fig:2pointcolor}. However, the last two diagrams in fig.~\ref{fig:2pointnewtopos} can be thought of as the normalization of a punctured torus with one pinch and of a doubly pinched torus, respectively. This interpretation implies the correct weights as well, provided that we view our diagrams as degenerations of smooth tori with pinches possibly colliding with punctures\footnote{Pinches colliding with punctures cannot arise in the Deligne-Mumford compactification of the moduli space of Riemann surfaces, a fact that has implications for the string interpretation \cite{BPSL}.}.\par
\begin{figure}[h!]
	\centering
	\includegraphics[width=1\linewidth]{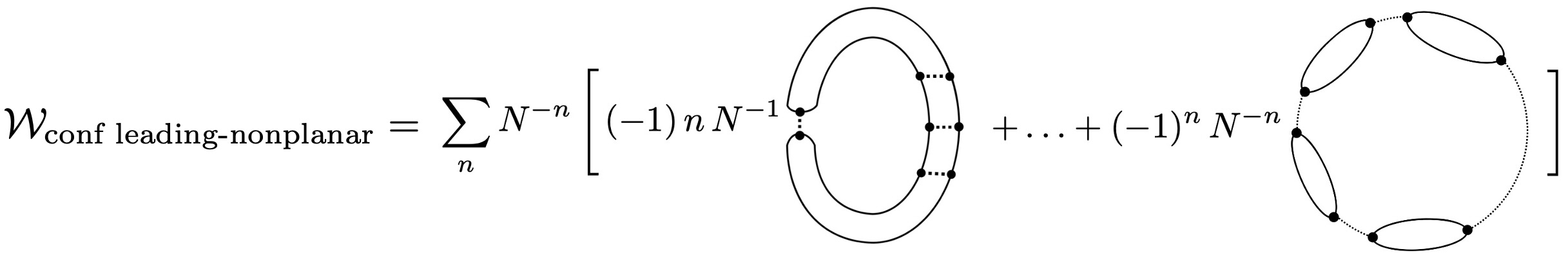}
	\caption{The refined perturbative expansion of the conformal generating functional at next-to-leading $\frac{1}{N}$ order: the square brackets contain the sum over $1 \leq p \leq n$ of possibly pinched annuli involving $n-p$ pairs of punctures and $p$ pinches arising from $p$ insertions of $P$, which give rise to the weights $N^{-p}$ compensated for by the color traces represented by loops. Some relative signs and combinatorics in the expansion of the second line of eq.~\eqref{Wconf1} with respect to the corresponding planar contributions in the first line are also displayed.}
	\label{fig:nonrenorm1}
\end{figure}
\begin{figure}[h!]
	\centering
	\includegraphics[width=1\linewidth]{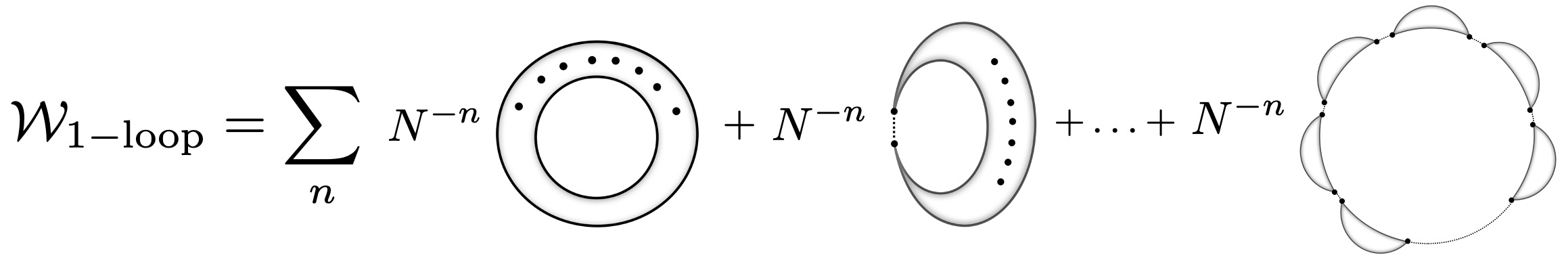}
	\caption{The refined topological expansion of the generating functional at next-to-leading $\frac{1}{N}$ order, including punctured spheres with $2$ punctures pairwise identified.}
	\label{fig:w1loop}
\end{figure}
\begin{figure}[h!]
	\centering
	\includegraphics[width=1\linewidth]{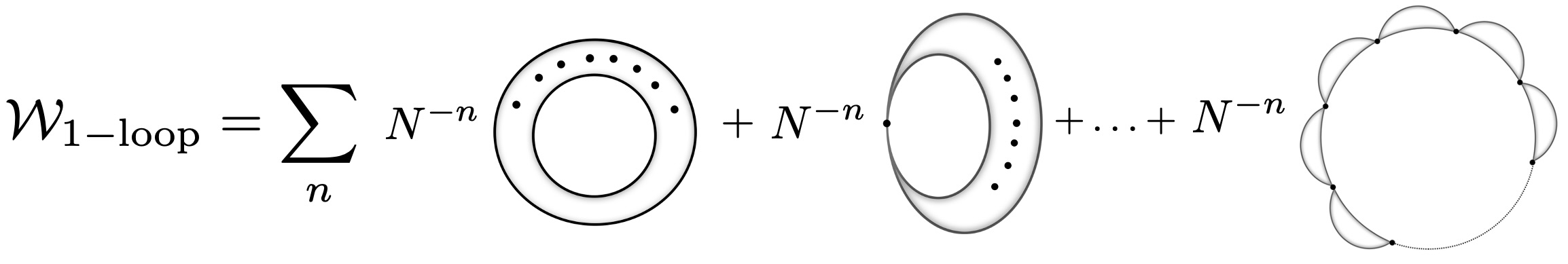}
	\caption{The new topologies in fig.~\ref{fig:w1loop} arise from the components of the normalization of $p$-pinched tori with $n-p$ punctures obtained by cutting the pinches.}
	\label{fig:w2loop}
\end{figure}

Collisions of pinches with punctures are inevitable with our choice for the twist-$2$ operator vertices, since the subleading corrections to the vertices carry the marked points associated with the local operators, which become the pinches after gluing. 
Similarly, for $n$-point correlators at lowest perturbative order, we obtain the contributions depicted in fig.~\ref{fig:nonrenorm1}, which correspond to the following calculation \cite{BPSL} of their generating functional 
in eq.~\eqref{eq:WUN} restricted to the SU($N$) sector
\begin{align}
	\label{Wconf0}
	\mathcal{W}^E_{\text{conf}}[J_{\mathbb{O}^E}] =-\log\Det\left(\mathcal{I}+\frac{1}{2}(I-P)\Delta^{-1}\frac{J_{\mathbb{O}^E_{s}}}{N}\otimes\mathcal{Y}_{s-2}^{E\frac{5}{2}}\right) \,,
\end{align}
where the term containing $I-P$ involves the SU($N$) propagator and $\mathcal{I}$ is the identity in both color and space-time.
The aforementioned equality follows by noticing that the color trace $\Tr$ in the loop expansion of eq.~\eqref{Wconf0} containing the insertion of $n$ sources $J_{\mathbb{O}^E_{s}}$ produces the overall factor 
\begin{align}
	\Tr(I-P)^n= \Tr(I-P)=N^2-1\,.
\end{align}
As for $2$-point correlators, in order to retain the 't Hooft double-line representation -- which involves only the leading propagator -- beyond the leading $\frac{1}{N}$ order of the SU($N$) theory as well, we transfer the $\frac{1}{N}$ dependence in eq.~\eqref{Wconf0} from the propagator to the local vertex for $2$-gluon operators (fig.~\ref{fig:2vertex}) via the identity
\begin{align}
	\label{Wconf01}
	&\mathcal{W}^E_{\text{conf}}[J_{\mathbb{O}^E}] \nonumber\\
	&=-\log\Det\left(\mathcal{I}+\frac{1}{2}(I-P)\Delta^{-1}\frac{J_{\mathbb{O}^E_{s}}}{N}\otimes\mathcal{Y}_{s-2}^{E\frac{5}{2}}\right)\nonumber\\
	&=-\log\Det\left(\mathcal{I}+\frac{1}{2}I\Delta^{-1}\frac{J_{\mathbb{O}^E_{s}}}{N}\otimes\mathcal{Y}_{s-2}^{E\frac{5}{2}}-\frac{1}{2}P\Delta^{-1}\frac{J_{\mathbb{O}^E_{s}}}{N}\otimes\mathcal{Y}_{s-2}^{E\frac{5}{2}}\right) \nonumber \\
	&=-\log\Det\Bigg[\Big(\mathcal{I}+\frac{1}{2}\Delta^{-1} I \frac{J_{\mathbb{O}^E_{s}}}{N}\otimes\mathcal{Y}_{s-2}^{E\frac{5}{2}}\Big) \nonumber \\
	&\,\,\,\,\,\,\,\,\, \left(\mathcal{I}-\frac{1}{2}\Big(\mathcal{I}+\frac{1}{2}\Delta^{-1} I \frac{J_{\mathbb{O}^E_{s}}}{N}\otimes\mathcal{Y}_{s-2}^{E\frac{5}{2}}\Big)^{-1}\Delta^{-1}P\frac{J_{\mathbb{O}^E_{s}}}{N}\otimes\mathcal{Y}_{s-2}^{E\frac{5}{2}}\right)\Bigg] \,,
\end{align}
so that
\begin{align}
	\label{Wconf1}
	\mathcal{W}^E_{\text{conf}}[J_{\mathbb{O}^E}]
	&=-\log\Det\left(\mathcal{I}+\frac{1}{2}\Delta^{-1} I \frac{J_{\mathbb{O}^E_{s}}}{N}\otimes\mathcal{Y}_{s-2}^{E\frac{5}{2}}\right)\nonumber\\
	&\quad-\log\Det\Bigg(\mathcal{I}-\frac{1}{2}\Big(\mathcal{I}+\frac{1}{2}\Delta^{-1} I \frac{J_{\mathbb{O}^E_{s}}}{N}\otimes\mathcal{Y}_{s-2}^{E\frac{5}{2}}\Big)^{-1} \Delta^{-1}P\frac{J_{\mathbb{O}^E_{s}}}{N}\otimes\mathcal{Y}_{s-2}^{E\frac{5}{2}}\Bigg) \,. \nonumber\\
\end{align}
The first $\log \Det$ above is the 't Hooft-planar contribution -- involving the leading local vertex (fig.~\ref{fig:2vertex}) -- while the second $\log \Det$ is the leading correction, which is still geometrically planar in the language of the present paper, yet non-'t Hooft-planar, involving at least one subleading local vertex (fig.~\ref{fig:2vertex}) carrying the factor of $P$.
Generalizing the new diagrams in fig.~\ref{fig:2pointcolor}, the subleading contributions due to the second $\log \Det$ in eq.~\eqref{Wconf1} involve the new topological sector (fig.~\ref{fig:nonrenorm1}).
This defines a refined topological expansion of the generating functional in which, in addition to the $n$-punctured tori, punctured spheres (fig.~\ref{fig:w1loop}) arise as the normalization of $p$-pinched and $(n-p)$-punctured tori (fig.~\ref{fig:w2loop}), with $1 \leq p \leq n$. 


	\subsection{Filling disks in the 't Hooft expansion versus nonrenormalization of pinched tori}

We emphasize a fundamental difference between the smooth tori in the 't Hooft expansion and our pinched tori.\par
The physical intuition behind the 't Hooft prescription of gluing disks in the double-line representation -- which generates smooth (punctured) Riemann surfaces from color loops -- is that the disks are ``filled in'' by an arbitrary number of gluon loops in Feynman diagrams with the same topology, arising from higher perturbative orders. For example, the sphere originates from an infinite sum of planar diagrams with an arbitrary number of gluon loops filling in the disks. Clearly, the same holds for all higher-genus surfaces.\par
The pinched tori are rather different. Since they arise from gluing oppositely oriented edges of strips involving at least two subleading vertices from fig.~\ref{fig:2vertex}, they possess a peculiar property: they do not receive any perturbative contribution from the interaction vertices of the action, since the projector $P$ makes any contraction with the action vertices vanish (fig.~\ref{fig:proof}). This means that the pinched tori are in fact ``empty'' and obey a perturbative nonrenormalization theorem. This property has a profound consequence: for correlators of twist-$2$ operators, the new pinched sector is one-loop exact\footnote{More properly, one-handle exact.} due to this nonrenormalization via decoupling from the interaction vertices (fig.~\ref{fig:proof}).

	\subsection{Mixing of pinched tori with smooth ones under perturbative renormalization of twist-$2$ operators in SU($N$) YM theory}

Moreover, the punctured smooth tori are suppressed in perturbation theory with respect to the pinched ones, since the smooth tori inevitably involve the action vertices that carry powers of the 't Hooft gauge coupling. 
Yet, the smooth tori must mix under renormalization with the pinched ones in order to provide the renormalization factors of the twist-$2$ operators due to their anomalous dimensions, since the pinched contributions are exactly conformal in perturbation theory.\par
Indeed, because of multiplicative renormalizability and scaling symmetry\footnote{Since $\gamma_{0\mathbb{O}_{s}}$ is diagonal, scaling symmetry holds generically to order $g^2$, whereas to the same order the additional symmetry under (collinear) conformal inversions is implemented in the conformal scheme \cite{Braun:2003rp}.}, the bare $2$-point correlators of twist-$2$ operators in $4-2\epsilon$ dimensions read, to order $g^2$ in perturbation theory,
\begin{align}
	\langle\mathbb{O}^B_{s}(x) \mathbb{O}^B_{s}(y)\rangle=&\left(1-\frac{1}{N^{2}}\right)C_s \frac{(x-y)_{+}^{2s}}{(|x-y|^{2})^{2s+2}} \nonumber \\
	&\biggl[1- \left(\frac{ \gamma_{0\mathbb{O}_{s}}g^2}{\epsilon}\right) +  \gamma_{0\mathbb{O}_{s}} g^2 \log\frac{1}{|x-y|^{2}\mu^{2}}+ \cdots \biggr]\,,\label{eq:renorm}
\end{align}
where the ellipsis stands for finite terms. As an example, we have verified eq.~\eqref{eq:renorm} for the $2$-point correlator of spin-$2$ unbalanced operators \cite{Galli}.
A similar expression holds for the correlators analytically continued to Euclidean space-time. 
Therefore, in the bare theory, the $2$-punctured smooth tori produce perturbative divergences,
\begin{align}
	\langle\mathbb{O}^B_{s}(x) \mathbb{O}^B_{s}(y)\rangle_{\text{smooth torus}}=&-\frac{1}{N^{2}} C_s \frac{(x-y)_{+}^{2s}}{(|x-y|^{2})^{2s+2}} \nonumber \\
	&\biggl[- \left(\frac{\gamma_{0\mathbb{O}_{s}}g^2}{\epsilon}\right) +  \gamma_{0\mathbb{O}_{s}} g^2 \log\frac{1}{|x-y|^{2}\mu^{2}}+ \cdots \biggr]\,,
\end{align}
which renormalize the lowest-order pinched contributions (fig.~\ref{fig:2pointcolor}) that we have interpreted as pinched tori (fig.~\ref{fig:2pointnewtopos}):
\begin{align}
	\langle\mathbb{O}^B_{s}(x) \mathbb{O}^B_{s}(y)\rangle_{\text{pinched tori}}=&\left(-\frac{2}{N^{2}}+\frac{1}{N^{2}}\right) C_s \frac{(x-y)_{+}^{2s}}{(|x-y|^{2})^{2s+2}} \nonumber \\
	&=-\frac{1}{N^{2}} C_s \frac{(x-y)_{+}^{2s}}{(|x-y|^{2})^{2s+2}} \,.
\end{align}
The above divergences are canceled by means of multiplicative renormalization (appendix~\ref{app:C3}),
\begin{equation}
	Z_{ss}^2=1+\frac{\gamma_{0\mathbb{O}_{s}} g^2}{ \epsilon} + \cdots\,,
\end{equation}
via counterterms that have precisely the space-time structure of pinched tori
\begin{align}
	-\left(\frac{ \gamma_{0\mathbb{O}_{s}}g^2}{\epsilon}\right)\frac{1}{N^{2}} C_s \frac{(x-y)_{+}^{2s}}{(|x-y|^{2})^{2s+2}}\,.
\end{align}
Hence, the smooth tori mix with the pinched ones under perturbative renormalization, so that their contributions cannot be separated in the renormalization-group (RG) improved correlators that we will discuss momentarily.

\subsection{Asymptotics of renormalized tori} 

After RG improvement, the leading non-'t Hooft-planar asymptotic generating functional reads \cite{BPSpaper2,BPSL}
\begin{align}
	\label{Wasnp}
	&\mathcal{W}^E_{\text{asym nonplanar}}[J_{\mathbb{O}^E},\lambda] \nonumber\\
	&=-\log\Det\Bigg(\mathcal{I}-\frac{1}{2}\Big(\mathcal{I}+ \frac{1}{2} \frac{Z_{\mathbb{O}_s}(\lambda)}{\lambda^{s+2}}  \Delta^{-1} I \frac{J_{\mathbb{O}^E_{s}}}{N}\otimes\mathcal{Y}_{s-2}^{E\frac{5}{2}}\Big)^{-1} \frac{Z_{\mathbb{O}_s}(\lambda)}{\lambda^{s+2}}\Delta^{-1}P\frac{J_{\mathbb{O}^E_{s}}}{N}\otimes\mathcal{Y}_{s-2}^{E\frac{5}{2}}\Bigg)\,.
\end{align}
We can view the leading-nonplanar RG-improved generating functional above as the resummation, via the Callan-Symanzik (CS) equation (appendix~\ref{app:C}), of the leading logarithms that rescale the conformal result arising from the pinched tori.\par
This result is derived as follows. Under a uniform rescaling of all coordinates by a common factor $\lambda \rightarrow 0$, the solution of the CS equation for the exact $n$-point correlators (eqs.~\eqref{csformuladiag} and \eqref{eq:expansion}) reads
\begin{align}
	&G^{(n)}_{j_1 \ldots j_n}(\lambda x_1,\ldots, \lambda x_n; \mu, g(\mu)) \nonumber \\
	&=Z_{\mathcal{O}_{j_1}}(\lambda) \ldots Z_{\mathcal{O}_{j_n}}(\lambda)\,\lambda^{-\sum_{i=1}^nD_{\mathcal{O}_{j_i}}}G^{(n)}_{j_1 \ldots j_n }( x_1, \ldots, x_n; \mu, g(\frac{\mu}{\lambda}))\,,
\end{align}
with 
\begin{align} 
	g^2(\frac{\mu}{\lambda}) \sim \dfrac{1}{\beta_0 \log(\frac{1}{\lambda^2})}\left(1-\frac{\beta_1}{\beta_0^2}\frac{\log\log(\frac{1}{\lambda^2})}{\log(\frac{1}{\lambda^2})}\right)
\end{align}
and
\begin{align}
	Z_{\mathcal{O}}(\lambda) = \left(\frac{g(\mu)}{g(\frac{\mu}{\lambda})}\right)^{\frac{\gamma_{0\mathcal{O}}}{\beta_0}} 
\end{align}
in the nonresonant renormalization scheme \cite{Bochicchio:2021geometry,S1}, with $N$-independent $\gamma_{0\mathcal{O}}$ (appendix~\ref{app:C}). \par
Then, as $\lambda \rightarrow 0$, in any renormalization scheme, the renormalized correlator on the right-hand side above admits the perturbative asymptotic expansion in terms of the renormalized coupling $g(\frac{\mu}{\lambda})$ at the scale $\frac{\mu}{\lambda}$
\begin{align}
	G^{(n)}_{j_1 \ldots j_n }( x_1, \ldots, x_n; \mu, g(\frac{\mu}{\lambda}))\,\sim\, &{\mathcal G}^{(n,0)}_{j_1 \ldots j_n }( x_1, \ldots, x_n; \mu)+g^2(\frac{\mu}{\lambda}) \, {\mathcal G}^{(n,2)}_{j_1 \ldots j_n }( x_1, \ldots, x_n; \mu)\nonumber\\
	&+ g^4(\frac{\mu}{\lambda}) \, {\mathcal G}^{(n,4)}_{j_1 \ldots j_n }( x_1, \ldots, x_n; \mu)
	+\cdots\,.
\end{align}
The leading asymptotics is therefore dominated by the first term -- provided that it is nonvanishing -- which is the conformal result employed in eq.~\eqref{Wasnp}.

\subsection{Renormalization of twist-$2$ operators in U($N$) YM theory}  

Since the interacting SU($N$) theory and the free U($1$) sector decouple, the twist-$2$ operators also split into two decoupled parts: the SU($N$) part acquires an anomalous dimension in general, whereas the U($1$) part does not renormalize. Consequently, the full U($N$) operator is not multiplicatively renormalizable; rather, its renormalization splits as
\begin{equation}\label{eq:fact}
	\mathbb{O}^{\text{U}(N)}_s\to (Z\mathbb{O})^{SU(N)}_s+\mathbb{O}_s^{U(1)}\,,
\end{equation}
where the two terms must be analyzed separately. Clearly, our considerations regarding the topology of the SU($N$) expansion apply to the first term, so that 't Hooft smooth topologies are incomplete for twist-$2$ operators in the \emph{renormalized} U($N$) theory as well.

	\section{Large-$N$ topology and renormalization in SU($N$) and U($N$) YM theories}\label{sec:s6}

\subsection{Quasi-partonic operators in the SU($N$) theory}

We first recall the definition of quasi-partonic operators \cite{BUKHVOSTOV1985601,Braun:2008ia} in YM theory. In the light-cone quantization of YM theory, only the transverse components of the gauge field are dynamical, and the corresponding field strength, $F_{+\perp}$, transforms at lowest order as a primary conformal operator under the collinear conformal subgroup $SL(2,\mathbb{R})$ of the conformal group $SO(4,2)$. Because of this property, $F_{+\perp}$ is classified as a ``good'' operator, whereas all remaining components are classified as ``bad''. More precisely, $F_{+\perp}$ carries canonical dimension $d=2$ and collinear-spin projection $s=1$ along the $+$ direction, hence collinear twist $\tau=d-s=1$ and collinear conformal spin $j=\frac{1}{2}(d+s)=\frac{3}{2}$, i.e., the minimal twist and maximal spin allowed for an elementary field in YM theory.\par
A quasi-partonic operator is a local gauge-invariant operator built exclusively out of the good field components $F_{+\perp}$ and their covariant derivatives $D_+$ along the light-cone direction, so that its collinear twist equals the number of its constituent fields, $\tau = n$, which is the minimal value for an $n$-field operator. Equivalently, for fixed collinear-spin projection $s$, the quasi-partonic operators carry the minimal canonical dimension $d=\tau+s$ among the operators built out of $n$ fields \cite{BUKHVOSTOV1985601,Braun:2003rp}.\par
 Since in a massless theory renormalization only mixes operators with the same canonical dimension and spin, it follows that quasi-partonic operators cannot mix with operators built out of more than $n$ fields (which have higher twist) at any order in perturbation theory, but they can mix with total derivatives of lower-spin operators with fewer than $n$ fields, which may include the descendants of the corresponding conformal primaries. Moreover, quasi-partonic operators of twist-$2$ may mix only among themselves, since they have minimal twist.\par
In the light-cone gauge $A_+=0$, $D_+=\partial_+$ and $F_{+\perp}=\partial_+A_\perp$ -- in the complex basis, $F=\partial_+A$ and $\bar F=\partial_+\bar A$ (appendix~\ref{app:ymlightcone}) -- so that an $n$-gluon quasi-partonic operator of collinear twist $\tau=n$ takes the general form
\begin{equation}\label{eq:qpgeneral}
	\mathcal{O}^{\tau}(x)=C_{a_1\ldots a_n}\,\mathcal{P}\big(\partial_+^{(1)},\ldots,\partial_+^{(n)}\big)\,\partial_+A_{\perp}^{a_1}(x_1)\cdots\partial_+A_{\perp}^{a_n}(x_n)\Big\vert_{x_1=\ldots=x_n=x}\,,
\end{equation}
where each transverse field $A_\perp$ is either $A$ or $\bar A$, $\partial_+^{(r)}$ acts on the coordinate $x_r$, $\mathcal{P}$ is a polynomial in the light-cone derivatives fixed by the collinear conformal properties (appendix~\ref{app:twist3}), and $C_{a_1\ldots a_n}$ is an invariant color tensor.\par
To order $g_{YM}^{2}$, the quasi-partonic operators of a given collinear twist close under renormalization~\cite{BUKHVOSTOV1985601}: they mix only among themselves, and the evolution kernel reduces to a sum of two-particle kernels acting pairwise on the $SL(2,\mathbb{R})$ modules carried by the constituent fields~\cite{BUKHVOSTOV1985601,Braun:2008ia},
\begin{equation}
	\mu\frac{d}{d\mu}\,\mathcal{O}^{\tau}
	\;=\; -\,\frac{\alpha_{s}}{2\pi}
	\sum_{1\le i<k\le n}\mathbb{H}_{ik}\;\mathcal{O}^{\tau} + \cdots \,,
	\label{eq:rge}
\end{equation}
with $\alpha_{s} = \frac{g^2_{YM}}{4\pi}$, where $\mathbb{H}_{ik}$ acts on two constituent primaries and the remaining $n-2$ constituents are spectators. This property does not survive beyond one loop \cite{BraunManashov2011, BUKHVOSTOV1985601,Braun:2008ia}: at higher loops, no symmetry prevents the quasi-partonic operators from mixing with operators of the same collinear spin built out of fewer fields, including bad operators.\par
For twist-$3$ operators, exceptions are $3$-gluon operators with color structure $d^{abc}$ in pure YM theory \cite{BUKHVOSTOV1985601}, which mix only among themselves to all orders in perturbation theory because they are odd under charge conjugation (section~\ref{sec:3gluon}), whereas all gauge-invariant operators with two gluons are even.\par
Moreover, bad operators of twist-$3$ reduce \cite{BraunManashovRohrwild2009,JiBelitsky2014}, by the equations of motion, to quasi-partonic operators and total derivatives of twist-$2$ operators, incidentally implying that the anomalous dimensions are not affected by the aforementioned mixing\footnote{Since anomalous dimensions may be obtained from zero-momentum transfer matrix elements, operators that are total derivatives do not contribute.}.\par
For twist-$4$, genuine non-quasi-partonic operators arise \cite{JiBelitsky2014}.\par
Conversely, even at one loop, operators containing bad components feed into the quasi-partonic sector through genuine $2\to 3$ kernels \cite{BraunManashovRohrwild2009,JiBelitsky2014}, so that their one-loop mixing matrix is block-triangular rather than block-diagonal.

\subsection{Renormalization of quasi-partonic operators in the U($N$) theory}

It is straightforward to see that all single-trace\footnote{Though they appear to be single-trace with respect to the $u(N)$ trace, they are not single-trace with respect to the $su(N)$ and $u(1)$ traces into which they decompose.} $n$-gluon quasi-partonic operators in the U($N$) theory can be decomposed as 
\begin{equation}
	\label{eq:UNdecomp}
	\mathcal{O}_n^{U(N)}=\mathcal{O}_n^{SU(N)}+\sum_{i=1}^{n-2} \mathcal{O}_n^i+\mathcal{O}_n^{U(1)}\,,
\end{equation}
where $\mathcal{O}_n^{SU(N)}$ depends only on the traceless combination $A-\frac{1}{N}\Tr A$, the operators\footnote{There is no contribution for $i=n-1$, since the corresponding operator $\Tr(A-\frac{1}{N}\Tr A)(\Tr A)^{n-1}$ vanishes because $\Tr (A-\frac{1}{N}\Tr A)=0$.} $\mathcal{O}^i$ depend only on $(\Tr A)^i$ and $A-\frac{1}{N}\Tr A$, and $\mathcal{O}_n^{U(1)}$ depends only on $\Tr A$. Schematically,
\begin{align}
	&\mathcal{O}_n^{SU(N)} \propto \Tr\left(A-\frac{1}{N}\Tr A\right)^n\nonumber\\
	&\mathcal{O}_n^i\propto \Tr\left(A-\frac{1}{N}\Tr A\right)^{n-i}\left(\Tr A\right)^i \nonumber\\
	&\mathcal{O}_n^{U(1)}\propto \left(\Tr A\right)^n \,,
\end{align}
with $n\geq 2$ and $1\leq i \leq n-2$.
They can be written more explicitly in terms of tensors valued in the $u(N)$ Lie algebra using eqs.~\eqref{eq:doubled} and \eqref{eq:trA0}:
\begin{align}
	&\mathcal{O}_n^{SU(N)} \propto \mathbb{T}^{SU(N)}_{a_1\ldots a_n} A^{a_1}\ldots A^{a_n}\nonumber\\
	&\mathcal{O}_n^i\propto\mathbb{T}^i_{a_1\ldots a_{n-i}} A^{a_1}\ldots  A^{a_{n-i}}(A^0)^i \nonumber\\
	&\mathcal{O}_n^{U(1)}\propto (A^0)^n\,.
\end{align}
Since the action factorizes according to eq.~\eqref{eq:compact}, so that the photon and gluons decouple, the above operators are orthogonal to all orders in perturbation theory:
\begin{align}
	\label{eq:ortogonal}
	&\langle \mathcal{O}^{SU(N)}_n\mathcal{O}^i_n\rangle = 0,\qquad\langle \mathcal{O}^{U(1)}_n\mathcal{O}^{i}_n\rangle =0,\qquad\langle \mathcal{O}^{SU(N)}_n\mathcal{O}^{U(1)}_n\rangle = 0,\qquad\langle \mathcal{O}^{i}_n\mathcal{O}^j_n\rangle \propto \delta^{ij}\,.
\end{align}
Indeed, $A^0$ is decoupled from the SU($N$) theory, and in connected $2$-point correlators the number of photons in each operator must match, assuming that tadpoles vanish.
Two observations are in order.\par
First, in general, the bare U($N$) operator $\mathcal{O}_n^{U(N)}$ is not multiplicatively renormalizable.\par
Second, operators with a fixed number of photons mix only with operators having the same number of photons.\par
Therefore, the mixing matrix for $\mathcal{O}_n^{U(N)}$ is block-diagonal with the last entry equal to $1$
\begin{equation}\label{eq:renormZ}
	[\mathcal{O}_n^{U(N)}]=
	\begin{pmatrix}
		Z^{SU(N)} &0 & 0 &\ldots &0&0 \\
		0 &\tilde{Z}^1 &0 &\ldots &0&0 \\
		0 & 0&\tilde{Z}^2  &\ldots &0&0 \\
		&\ldots&\ldots&\ldots\\
		0 & 0&0&\ldots& \tilde{Z}^{n-2} &0\\
		0 & 0&0&\ldots& 0 &1
	\end{pmatrix}\begin{pmatrix}
		\mathcal{O}^{SU(N)}  \\
		\mathcal{O}^1 \\
		\mathcal{O}^2 \\
		\vdots\\
		\mathcal{O}^{n-2} \\
		\mathcal{O}^{U(1)}
	\end{pmatrix}  \,,
\end{equation}
since $\mathcal{O}_n^{U(1)}$ does not renormalize because the U($1$) theory is free, while $Z^{SU(N)}$ and $\tilde{Z}^i$ are the renormalization mixing matrices for the corresponding operators. 
In the following sections, we will investigate only the renormalization in the first block of eq.~\eqref{eq:renormZ} in the U($N$) theory, which coincides with that of the SU($N$) theory. The analysis of the remaining blocks is straightforward but cumbersome; as it does not add anything essential to our topological classification, we omit it.

\section{Topologies in the large-$N$ expansion of correlators of $n$-gluon quasi-partonic operators in SU($N$) YM theory}\label{sec:s7}

\subsection{$2$-gluon operators in the adjoint representation}

We consider twist-$2$ operators in the adjoint representation. For example,
\begin{equation}\label{eq:OsAdj}
	\mathbb{O}^{\text{Adj}}_{s} = \Tr\, \bar F(x)\,(i\overrightarrow{D}_++i\overleftarrow{D}_+)^{s-2}\,C^{\frac{5}{2}}_{s-2}\left(\frac{\overrightarrow{D}_+-\overleftarrow{D}_+}{\overrightarrow{D}_++\overleftarrow{D}_+}\right)F(x)\,,
\end{equation}
where now $F=F^aT^a$ and $\bar F=\bar F^aT^a$ are the good components of the field strength in the adjoint representation (eq.~\eqref{OO}), and the same holds for $\tilde{\mathbb{O}}_{s}$, $\mathbb{S}_{s}$, and $\bar{\mathbb{S}}_{s}$.
Since
\begin{align}
	\label{adjoint}
	&\Tr T^{a}T^{b} = N \delta^{ab}\,,
\end{align}
with
\begin{align}
	(T^a)_{bc} = -i f^{abc} = -2\Tr \lambda^a\left[\lambda^b,\lambda^c\right] \,,
\end{align}
it follows that the color factor of the vertex in the adjoint representation is
\begin{equation}
	(\mathbf{F}_2)^{i\,k}_{j\,l} = N (\mathbf{V}_2)^{i\,k}_{j\,l} =	\frac{N}{2}\left(\delta^i_l\delta^k_j-\frac{1}{N}\delta^i_j\delta^k_l\right)\,,
\end{equation}
so that
\begin{equation}\label{eq:OsAdjfund}
	\mathbb{O}^{\text{Adj}}_{s} = 2N\, \mathbb{O}_{s}\,.
\end{equation}
Indeed, for $2$-gluon operators the color vertex is proportional to $\delta^{ab}$ in any representation.
As a consequence, the very same diagrammatic and topological expansion as for twist-$2$ operators in the fundamental representation applies (section~\ref{sec:s5}).

\subsection{$3$-gluon operators in the adjoint representation}

The $3$-gluon quasi-partonic operators in the adjoint representation read
\begin{equation}\label{eq:A3gaugeinv}
	\mathcal{A}_3 = \mathcal{P}\big(D^{(1)}_+,D^{(2)}_+,D^{(3)}_+\big)\,\Tr\left(F_{+\perp}\,F_{+\perp}\,F_{+\perp}\right),
\end{equation}
where, as in eq.~\eqref{eq:qpgeneral}, each good component $F_{+\perp}$ is either $F$ or $\bar F$ -- now valued in the adjoint representation -- $D^{(r)}_+$ acts covariantly on the $r$-th field, and $\mathcal{P}$ is the conformal polynomial (appendix~\ref{app:twist3}) built out of Jacobi polynomials (appendix~\ref{app:jacobi}). Since the trace is cyclic, the chirality assignments produce, up to parity, the two structures $\Tr(\bar F\,\bar F\,\bar F)$ and $\Tr(\bar F\,\bar F\,F)$, i.e., the operators $\mathbb{S}$ and $\mathbb{O}$ (appendix~\ref{app:twist3}), with $\Tr(\bar F\,F\,\bar F)$ reducing to $\Tr(\bar F\,\bar F\,F)$ upon relabeling the points. In the light-cone gauge, where $F=\partial_+A$, $\bar F=\partial_+\bar A$, and $D_+=\partial_+$, their color vertex reads
\begin{equation}\label{eq:A3gaugeinv2}
	\Tr(T^a T^b T^c)=i\,f^{ade}f^{bef}f^{cfd} = \frac{N}{2}if^{abc}\,.
\end{equation}
Hence, we conclude that they have no subleading corrections in the double-line representation
\begin{align}\label{eq:adjvertexdouble}
	(\mathbf{F}_3)^{i\,k\,m}_{j\,l\,n}&=\Tr(T^a T^b T^c)\,(\lambda^a)^{i}_{\,j}(\lambda^b)^{k}_{\,l}(\lambda^c)^{m}_{\,n}\nonumber\\
	&=N\,\Tr\left(\lambda^{a}\left[\lambda^{b},\lambda^{c}\right]\right)(\lambda^a)^{i}_{\,j}(\lambda^b)^{k}_{\,l}(\lambda^c)^{m}_{\,n}\nonumber\\
	&=\frac{N}{2}\,[\lambda^{b},\lambda^{c}]^{i}_{\,j}\,(\lambda^b)^{k}_{\,l}(\lambda^c)^{m}_{\,n}\nonumber\\
	&=\frac{N}{4}\left(\delta^{i}_{\,l}\,(\lambda^{c})^{k}_{\,j}-(\lambda^{c})^{i}_{\,l}\,\delta^{k}_{\,j}\right)(\lambda^c)^{m}_{\,n}\nonumber\\
	&=\frac{N}{8}\left(\delta^{i}_{\,l}\delta^{k}_{\,n}\delta^{m}_{\,j}-\delta^{i}_{\,n}\delta^{k}_{\,j}\delta^{m}_{\,l}\right),
\end{align}
where the second line follows from eq.~\eqref{eq:sun}, the third and fourth lines from eqs.~\eqref{eq:com2} and \eqref{eq:com1}, and in the last line the terms of order $\frac{1}{N}$ in the $su(N)$ completeness relation cancel out. Up to overall normalization, the operator vertex (fig.~\ref{fig:3vertex}) is therefore $N$-independent, and its color structure coincides with the $3$-gluon vertex $\mathcal{V}_3$ of the action in eq.~\eqref{vertices}. As a consequence, for these operators only the smooth 't Hooft topologies contribute in the large-$N$ expansion of the SU($N$) theory.

\subsection{$4$-gluon operators in the adjoint representation}

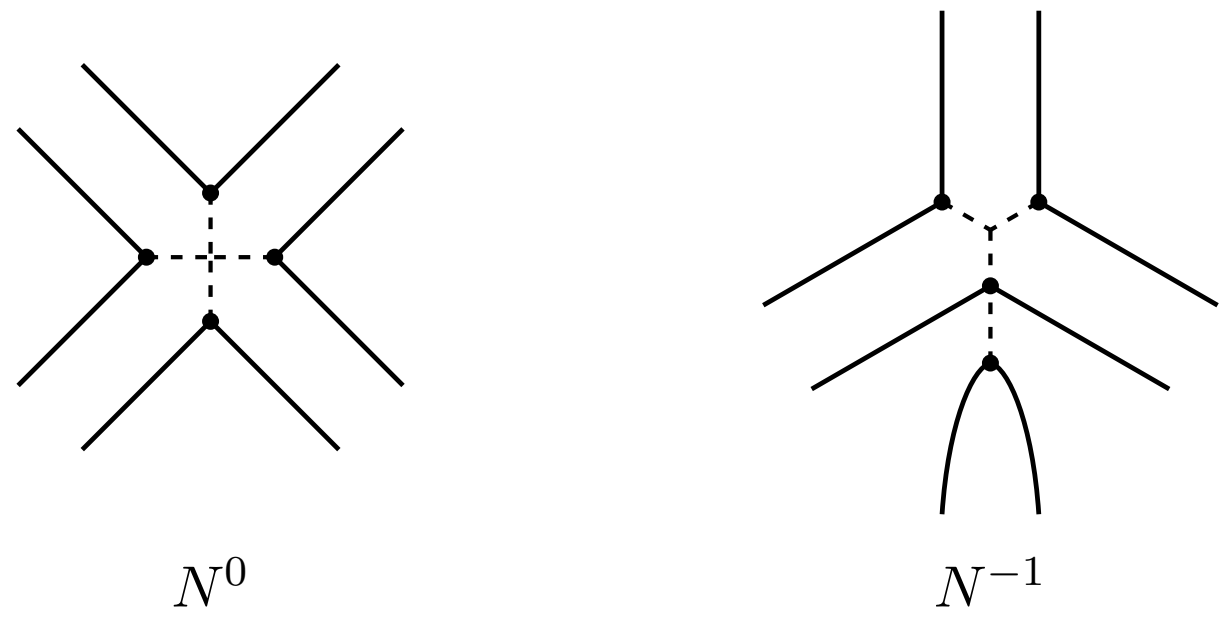
\begin{figure}[h!]
	\centering
	\begin{tikzpicture}[baseline={(0,-0.1)}]
		\path (-1.75,-2.55) rectangle (1.75,1.7);
		\foreach \p in {(0,0.4243),(-0.4243,0),(0,-0.4243),(0.4243,0)}{\draw[pinch] \p -- (0,0);}
		\draw[skel] (0.8485,1.2728) -- (0,0.4243) -- (-0.8485,1.2728);
		\draw[skel] (-1.2728,0.8485) -- (-0.4243,0) -- (-1.2728,-0.8485);
		\draw[skel] (-0.8485,-1.2728) -- (0,-0.4243) -- (0.8485,-1.2728);
		\draw[skel] (1.2728,-0.8485) -- (0.4243,0) -- (1.2728,0.8485);
		\fill (0,0.4243) circle (1.6pt);
		\fill (-0.4243,0) circle (1.6pt);
		\fill (0,-0.4243) circle (1.6pt);
		\fill (0.4243,0) circle (1.6pt);
		\node at (0,-2.15) {$N^0$};
	\end{tikzpicture}
	\qquad\qquad
	\begin{tikzpicture}[baseline={(0,-0.1)}]
		\path (-1.75,-2.55) rectangle (1.75,1.7);
		\begin{scope}[shift={(0,0.18)}]
			\draw[pinch] (0,-0.88) -- (0,0);
			\draw[pinch] (-0.32,0.1848) -- (0,0);
			\draw[pinch] (0.32,0.1848) -- (0,0);
			\draw[skel] (-0.32,1.45) -- (-0.32,0.1848) -- (-1.5023,-0.4979);
			\draw[skel] (-1.1823,-1.0521) -- (0,-0.3695) -- (1.1823,-1.0521);
			\draw[skel] (1.5023,-0.4979) -- (0.32,0.1848) -- (0.32,1.45);
			\fill (-0.32,0.1848) circle (1.6pt);
			\fill (0,-0.3695) circle (1.6pt);
			\fill (0.32,0.1848) circle (1.6pt);
			\ophairpin{270}{0.88}{1.0}
		\end{scope}
		\node at (0,-2.15) {$N^{-1}$};
	\end{tikzpicture}
	\caption{Operator vertex $(\mathbf{F}_4)^{i\,k\,m\,q}_{j\,l\,n\,p}$ in the double-line representation: the leading planar contribution (left) corresponding to the two terms in eq.~\eqref{eq:F4} and the subleading $\frac{1}{N}$ contribution corresponding to the eleven terms in eq.~\eqref{eq:F4}, up to permutations. The dots label multiple punctures corresponding to a single operator insertion that are later identified by 't Hooft's gluing prescription (dashed lines).}
	\label{fig:F4doubleline}
\end{figure}

\begin{figure}[h!]
		\centering
		\begin{tikzpicture}[scale=0.55,
			surf/.style={c969696, line width=1.2pt},
			ident/.style={line width=0.8pt, black, dash pattern=on 0.9pt off 1.4pt}]
			\begin{scope}[shift={(0.00,0.00)}]
				\begin{scope}[shift={(-2.900,0)}]
					\draw[skel] (1.707,0.281) -- (0.936,0.000) -- (1.707,-0.281);
					\draw[skel] (1.488,-0.882) -- (0.717,-0.601) -- (1.127,-1.312);
					\draw[skel] (1.488,0.882) -- (-0.468,0.170) -- (0.573,-1.632);
					\begin{scope}[shift={(60:1.200)}, rotate=-30]
						\draw[skel] (-0.320,0.500) .. controls (-0.280,0.325) and (-0.12,0.03) .. (0,0) .. controls (0.12,0.03) and (0.280,0.325) .. (0.320,0.500);
						\fill (0,0) circle (2.91pt);
						\draw[pinch] (0,0) -- (0,-1.200);
					\end{scope}
					\draw[pinch] (0.936,0.000) -- (0,0);
					\fill (0.936,0.000) circle (2.91pt);
					\draw[pinch] (0.717,-0.601) -- (0,0);
					\fill (0.717,-0.601) circle (2.91pt);
					\draw[pinch] (-0.468,0.170) -- (0,0);
					\fill (-0.468,0.170) circle (2.91pt);
				\end{scope}
				\begin{scope}[shift={(2.900,0)}, xscale=-1]
					\draw[skel] (1.127,1.312) -- (0.468,0.170) -- (1.707,-0.281);
					\draw[skel] (1.488,-0.882) -- (0.717,-0.601) -- (1.127,-1.312);
					\draw[skel] (0.573,1.632) -- (-0.370,0.000) -- (0.573,-1.632);
					\begin{scope}[shift={(20:1.200)}, rotate=-70]
						\draw[skel] (-0.320,0.500) .. controls (-0.280,0.325) and (-0.12,0.03) .. (0,0) .. controls (0.12,0.03) and (0.280,0.325) .. (0.320,0.500);
						\fill (0,0) circle (2.91pt);
						\draw[pinch] (0,0) -- (0,-1.200);
					\end{scope}
					\draw[pinch] (0.468,0.170) -- (0,0);
					\fill (0.468,0.170) circle (2.91pt);
					\draw[pinch] (0.717,-0.601) -- (0,0);
					\fill (0.717,-0.601) circle (2.91pt);
					\draw[pinch] (-0.370,0.000) -- (0,0);
					\fill (-0.370,0.000) circle (2.91pt);
				\end{scope}
				\draw[skel] (-2.327,1.632) arc[start angle=150.000, end angle=30.000, radius=2.687];
				\draw[skel] (-1.773,1.312) arc[start angle=150.000, end angle=30.000, radius=2.047];
				\draw[skel] (-1.412,0.882) arc[start angle=110.000, end angle=70.000, radius=4.128];
				\draw[skel] (-1.193,0.281) arc[start angle=110.000, end angle=70.000, radius=3.488];
				\draw[skel] (-1.193,-0.281) arc[start angle=-110.000, end angle=-70.000, radius=3.488];
				\draw[skel] (-1.412,-0.882) arc[start angle=-110.000, end angle=-70.000, radius=4.128];
				\draw[skel] (-1.773,-1.312) arc[start angle=-150.000, end angle=-30.000, radius=2.047];
				\draw[skel] (-2.327,-1.632) arc[start angle=-150.000, end angle=-30.000, radius=2.687];
			\end{scope}
			\begin{scope}[shift={(0.00,-5.20)}]
				\shade[inner color=white, outer color=gray!38, even odd rule] (0,0) ellipse (2.31 and 1.33) (-1.210,0) ellipse (1.100 and 0.700) (0.990,0) ellipse (1.100 and 0.700);
				\draw[surf] (0,0) ellipse (2.31 and 1.33);
				\draw[surf] (-1.210,0) ellipse (1.100 and 0.700);
				\draw[surf] (0.990,0) ellipse (1.100 and 0.700);
				\fill (-0.110,0) circle (3.64pt);
				\fill (-2.310,0) circle (3.64pt);
			\end{scope}
			\node at (0.00,-7.13) {$N^{-4}$};
			\begin{scope}[shift={(7.94,0.00)}]
				\begin{scope}[shift={(-2.900,0)}]
					\draw[skel] (1.707,0.281) -- (0.936,0.000) -- (1.707,-0.281);
					\draw[skel] (1.488,-0.882) -- (0.717,-0.601) -- (1.127,-1.312);
					\draw[skel] (1.488,0.882) -- (-0.468,0.170) -- (0.573,-1.632);
					\begin{scope}[shift={(60:1.200)}, rotate=-30]
						\draw[skel] (-0.320,0.500) .. controls (-0.280,0.325) and (-0.12,0.03) .. (0,0) .. controls (0.12,0.03) and (0.280,0.325) .. (0.320,0.500);
						\fill (0,0) circle (2.91pt);
						\draw[pinch] (0,0) -- (0,-1.200);
					\end{scope}
					\draw[pinch] (0.936,0.000) -- (0,0);
					\fill (0.936,0.000) circle (2.91pt);
					\draw[pinch] (0.717,-0.601) -- (0,0);
					\fill (0.717,-0.601) circle (2.91pt);
					\draw[pinch] (-0.468,0.170) -- (0,0);
					\fill (-0.468,0.170) circle (2.91pt);
				\end{scope}
				\begin{scope}[shift={(2.900,0)}, xscale=-1]
					\draw[skel] (1.127,1.312) -- (0.717,0.601) -- (1.488,0.882);
					\draw[skel] (1.707,0.281) -- (0.936,0.000) -- (1.707,-0.281);
					\draw[skel] (1.488,-0.882) -- (0.717,-0.601) -- (1.127,-1.312);
					\draw[skel] (0.573,1.632) -- (-0.370,0.000) -- (0.573,-1.632);
					\draw[pinch] (0.717,0.601) -- (0,0);
					\fill (0.717,0.601) circle (2.91pt);
					\draw[pinch] (0.936,0.000) -- (0,0);
					\fill (0.936,0.000) circle (2.91pt);
					\draw[pinch] (0.717,-0.601) -- (0,0);
					\fill (0.717,-0.601) circle (2.91pt);
					\draw[pinch] (-0.370,0.000) -- (0,0);
					\fill (-0.370,0.000) circle (2.91pt);
				\end{scope}
				
				\draw[skel] (-2.327,1.632) arc[start angle=150.000, end angle=30.000, radius=2.687];
				\draw[skel] (-1.773,1.312) arc[start angle=150.000, end angle=30.000, radius=2.047];
				\draw[skel] (-1.773,-1.312) arc[start angle=-150.000, end angle=-30.000, radius=2.047];
				\draw[skel] (-2.327,-1.632) arc[start angle=-150.000, end angle=-30.000, radius=2.687];
				\draw[skel] (-1.412,0.882) -- (-1.346,0.905) -- (-1.281,0.925) -- (-1.218,0.942) -- (-1.155,0.957) -- (-1.093,0.968) -- (-1.031,0.977) -- (-0.971,0.983) -- (-0.911,0.987) -- (-0.853,0.987) -- (-0.795,0.985) -- (-0.738,0.979) -- (-0.682,0.971) -- (-0.628,0.960) -- (-0.575,0.947) -- (-0.523,0.930) -- (-0.473,0.912) -- (-0.425,0.891) -- (-0.379,0.868) -- (-0.334,0.843) -- (-0.291,0.816) -- (-0.250,0.788) -- (-0.211,0.759) -- (-0.174,0.728) -- (-0.138,0.697) -- (-0.104,0.665) -- (-0.072,0.632) -- (-0.041,0.598) -- (-0.011,0.565) -- (0.018,0.530) -- (0.045,0.496) -- (0.072,0.462) -- (0.098,0.427) -- (0.122,0.393) -- (0.146,0.358) -- (0.170,0.324) -- (0.193,0.290) -- (0.215,0.256) -- (0.237,0.223) -- (0.258,0.190) -- (0.279,0.158) -- (0.300,0.126) -- (0.320,0.095) -- (0.340,0.064) -- (0.360,0.034) -- (0.380,0.005) -- (0.400,-0.023) -- (0.419,-0.050) -- (0.438,-0.076) -- (0.458,-0.101) -- (0.477,-0.124) -- (0.496,-0.147) -- (0.515,-0.168) -- (0.533,-0.188) -- (0.552,-0.207) -- (0.570,-0.224) -- (0.588,-0.240) -- (0.606,-0.255) -- (0.624,-0.268) -- (0.642,-0.281) -- (0.659,-0.292) -- (0.677,-0.302) -- (0.695,-0.310) -- (0.713,-0.318) -- (0.732,-0.325) -- (0.751,-0.331) -- (0.771,-0.336) -- (0.792,-0.341) -- (0.814,-0.344) -- (0.838,-0.346) -- (0.864,-0.347) -- (0.891,-0.347) -- (0.921,-0.345) -- (0.953,-0.342) -- (0.987,-0.337) -- (1.023,-0.330) -- (1.062,-0.321) -- (1.103,-0.310) -- (1.147,-0.297) -- (1.193,-0.281);
				\draw[skel] (-1.193,0.281) -- (-1.147,0.297) -- (-1.103,0.310) -- (-1.062,0.321) -- (-1.023,0.330) -- (-0.987,0.337) -- (-0.953,0.342) -- (-0.921,0.345) -- (-0.891,0.347) -- (-0.864,0.347) -- (-0.838,0.346) -- (-0.814,0.344) -- (-0.792,0.341) -- (-0.771,0.336) -- (-0.751,0.331) -- (-0.732,0.325) -- (-0.713,0.318) -- (-0.695,0.310) -- (-0.677,0.302) -- (-0.659,0.292) -- (-0.642,0.281) -- (-0.624,0.268) -- (-0.606,0.255) -- (-0.588,0.240) -- (-0.570,0.224) -- (-0.552,0.207) -- (-0.533,0.188) -- (-0.515,0.168) -- (-0.496,0.147) -- (-0.477,0.124) -- (-0.458,0.101) -- (-0.438,0.076) -- (-0.419,0.050) -- (-0.400,0.023) -- (-0.380,-0.005) -- (-0.360,-0.034) -- (-0.340,-0.064) -- (-0.320,-0.095) -- (-0.300,-0.126) -- (-0.279,-0.158) -- (-0.258,-0.190) -- (-0.237,-0.223) -- (-0.215,-0.256) -- (-0.193,-0.290) -- (-0.170,-0.324) -- (-0.146,-0.358) -- (-0.122,-0.393) -- (-0.098,-0.427) -- (-0.072,-0.462) -- (-0.045,-0.496) -- (-0.018,-0.530) -- (0.011,-0.565) -- (0.041,-0.598) -- (0.072,-0.632) -- (0.104,-0.665) -- (0.138,-0.697) -- (0.174,-0.728) -- (0.211,-0.759) -- (0.250,-0.788) -- (0.291,-0.816) -- (0.334,-0.843) -- (0.379,-0.868) -- (0.425,-0.891) -- (0.473,-0.912) -- (0.523,-0.930) -- (0.575,-0.947) -- (0.628,-0.960) -- (0.682,-0.971) -- (0.738,-0.979) -- (0.795,-0.985) -- (0.853,-0.987) -- (0.911,-0.987) -- (0.971,-0.983) -- (1.031,-0.977) -- (1.093,-0.968) -- (1.155,-0.957) -- (1.218,-0.942) -- (1.281,-0.925) -- (1.346,-0.905) -- (1.412,-0.882);
				\draw[skel] (-1.193,-0.281) -- (-1.147,-0.297) -- (-1.103,-0.310) -- (-1.062,-0.321) -- (-1.023,-0.330) -- (-0.987,-0.337) -- (-0.953,-0.342) -- (-0.921,-0.345) -- (-0.891,-0.347) -- (-0.864,-0.347) -- (-0.838,-0.346) -- (-0.814,-0.344) -- (-0.792,-0.341) -- (-0.771,-0.336) -- (-0.751,-0.331) -- (-0.732,-0.325) -- (-0.713,-0.318) -- (-0.695,-0.310) -- (-0.677,-0.302) -- (-0.659,-0.292) -- (-0.642,-0.281) -- (-0.624,-0.268) -- (-0.606,-0.255) -- (-0.588,-0.240) -- (-0.570,-0.224) -- (-0.552,-0.207) -- (-0.533,-0.188) -- (-0.515,-0.168) -- (-0.496,-0.147) -- (-0.477,-0.124) -- (-0.458,-0.101) -- (-0.438,-0.076) -- (-0.419,-0.050) -- (-0.400,-0.023) -- (-0.380,0.005) -- (-0.360,0.034) -- (-0.340,0.064) -- (-0.320,0.095) -- (-0.300,0.126) -- (-0.279,0.158) -- (-0.258,0.190) -- (-0.237,0.223) -- (-0.215,0.256) -- (-0.193,0.290) -- (-0.170,0.324) -- (-0.146,0.358) -- (-0.122,0.393) -- (-0.098,0.427) -- (-0.072,0.462) -- (-0.045,0.496) -- (-0.018,0.530) -- (0.011,0.565) -- (0.041,0.598) -- (0.072,0.632) -- (0.104,0.665) -- (0.138,0.697) -- (0.174,0.728) -- (0.211,0.759) -- (0.250,0.788) -- (0.291,0.816) -- (0.334,0.843) -- (0.379,0.868) -- (0.425,0.891) -- (0.473,0.912) -- (0.523,0.930) -- (0.575,0.947) -- (0.628,0.960) -- (0.682,0.971) -- (0.738,0.979) -- (0.795,0.985) -- (0.853,0.987) -- (0.911,0.987) -- (0.971,0.983) -- (1.031,0.977) -- (1.093,0.968) -- (1.155,0.957) -- (1.218,0.942) -- (1.281,0.925) -- (1.346,0.905) -- (1.412,0.882);
				\draw[skel] (-1.412,-0.882) -- (-1.346,-0.905) -- (-1.281,-0.925) -- (-1.218,-0.942) -- (-1.155,-0.957) -- (-1.093,-0.968) -- (-1.031,-0.977) -- (-0.971,-0.983) -- (-0.911,-0.987) -- (-0.853,-0.987) -- (-0.795,-0.985) -- (-0.738,-0.979) -- (-0.682,-0.971) -- (-0.628,-0.960) -- (-0.575,-0.947) -- (-0.523,-0.930) -- (-0.473,-0.912) -- (-0.425,-0.891) -- (-0.379,-0.868) -- (-0.334,-0.843) -- (-0.291,-0.816) -- (-0.250,-0.788) -- (-0.211,-0.759) -- (-0.174,-0.728) -- (-0.138,-0.697) -- (-0.104,-0.665) -- (-0.072,-0.632) -- (-0.041,-0.598) -- (-0.011,-0.565) -- (0.018,-0.530) -- (0.045,-0.496) -- (0.072,-0.462) -- (0.098,-0.427) -- (0.122,-0.393) -- (0.146,-0.358) -- (0.170,-0.324) -- (0.193,-0.290) -- (0.215,-0.256) -- (0.237,-0.223) -- (0.258,-0.190) -- (0.279,-0.158) -- (0.300,-0.126) -- (0.320,-0.095) -- (0.340,-0.064) -- (0.360,-0.034) -- (0.380,-0.005) -- (0.400,0.023) -- (0.419,0.050) -- (0.438,0.076) -- (0.458,0.101) -- (0.477,0.124) -- (0.496,0.147) -- (0.515,0.168) -- (0.533,0.188) -- (0.552,0.207) -- (0.570,0.224) -- (0.588,0.240) -- (0.606,0.255) -- (0.624,0.268) -- (0.642,0.281) -- (0.659,0.292) -- (0.677,0.302) -- (0.695,0.310) -- (0.713,0.318) -- (0.732,0.325) -- (0.751,0.331) -- (0.771,0.336) -- (0.792,0.341) -- (0.814,0.344) -- (0.838,0.346) -- (0.864,0.347) -- (0.891,0.347) -- (0.921,0.345) -- (0.953,0.342) -- (0.987,0.337) -- (1.023,0.330) -- (1.062,0.321) -- (1.103,0.310) -- (1.147,0.297) -- (1.193,0.281);
			\end{scope}
		\begin{scope}[shift={(7.94,-5.20)}]
				\shade[inner color=white, outer color=gray!38, even odd rule] (0,0) ellipse (2.31 and 1.33) (-1.421,0) ellipse (0.886 and 0.701) (0.934,0) ellipse (0.726 and 0.575);
				\draw[surf] (0,0) ellipse (2.31 and 1.33);
				\draw[surf] (-1.421,0) ellipse (0.886 and 0.701);
				\draw[surf] (0.934,0) ellipse (0.726 and 0.575);
				\fill (-2.306,0) circle (3.64pt);
				\fill (1.983,0) circle (3.64pt);
			\end{scope}
			\node at (7.94,-7.13) {$N^{-4}$};
		\end{tikzpicture}
		\caption{Pinched double tori arising in the SU($N$) theory for the $2$-point correlators of $4$-gluon operators in the adjoint representation that do not match the smooth topologies of the 't Hooft expansion of bare U($N$) YM theory.}
\label{fig:A4twopoint}
\end{figure}
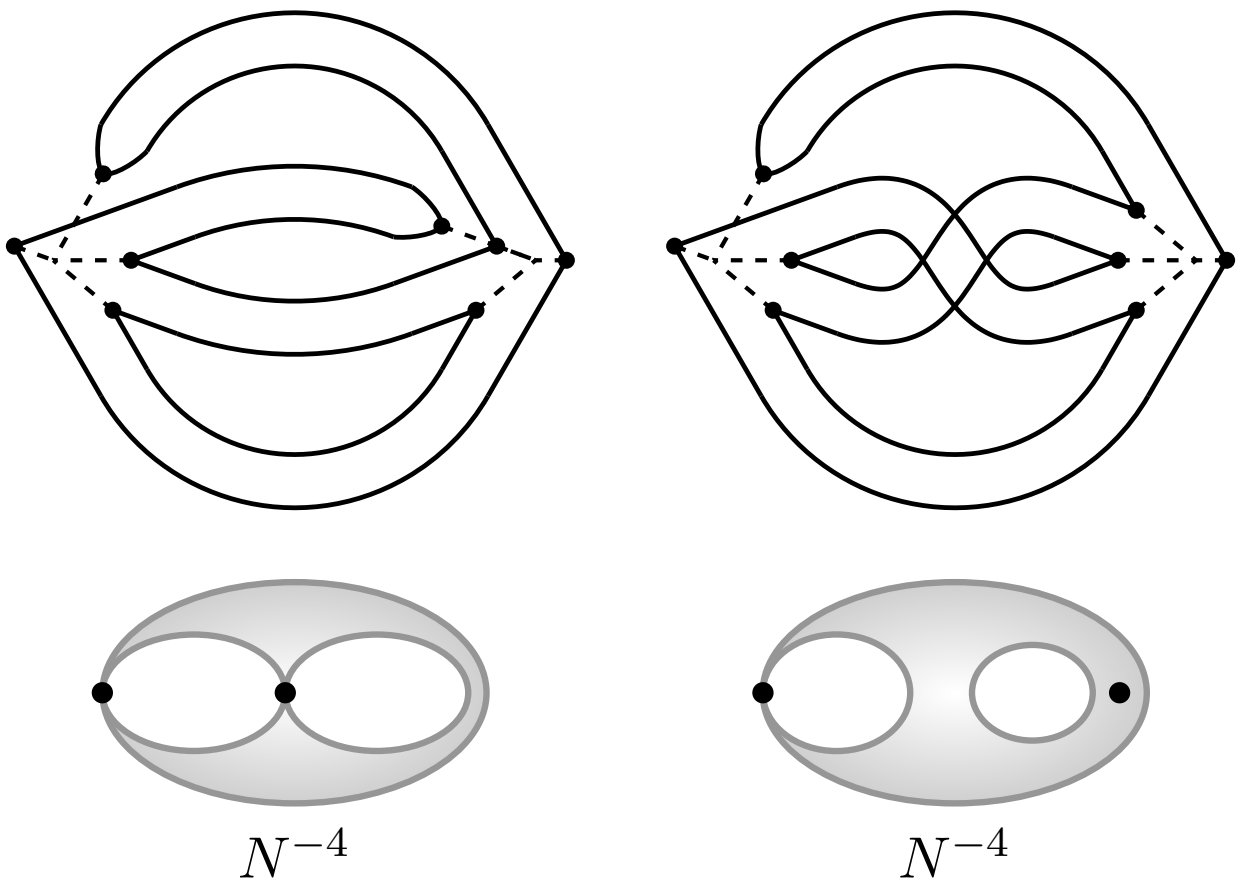

The $4$-gluon quasi-partonic operators have the structure
\begin{equation}\label{eq:A4gaugeinv}
	\mathcal{A}_4 = \mathcal{P}\big(D^{(1)}_+,\ldots,D^{(4)}_+\big)\,\Tr\left(F_{+\perp}\,F_{+\perp}\,F_{+\perp}\,F_{+\perp}\right),
\end{equation}
with each good component $F_{+\perp}$ being either $F$ or $\bar F$ in the adjoint representation, and the polynomial $\mathcal{P}$ suitably chosen so that the operators are conformal primaries with respect to the $SL(2,\mathbb{R})$ subgroup. Their color structure is given by
\begin{equation} \label{eq:trac4}
	\mathcal{A}_4 = \mathcal{P}\big(D^{(1)}_+,\ldots,D^{(4)}_+\big)\,\Tr (T^a T^b T^c T^d)F_{+\perp}^a F_{+\perp}^b F_{+\perp}^c F_{+\perp}^d\,,
\end{equation}
where
\begin{align}\label{eq:TrAdj4irr}
	\Tr(T^aT^bT^cT^d)=\;&\frac{2}{3}\left(\delta^{ab}\delta^{cd}+\delta^{ac}\delta^{bd}+\delta^{ad}\delta^{bc}\right)
	+\frac{N}{12}\left(d^{abe}d^{cde}+d^{ace}d^{bde}+d^{ade}d^{bce}\right)\nonumber\\
	&+\frac{N}{6}\left(f^{ade}f^{bce}-f^{abe}f^{cde}\right),
\end{align}
and the color vertex in the double-line representation is 
\begin{align}\label{eq:F4}
	(\mathbf{F}_4)^{i\,k\,m\,q}_{j\,l\,n\,p}=&\Tr (T^a T^b T^c T^d)(\lambda^a)^{i}_{j}(\lambda^b)^{k}_{l}(\lambda^c)^{m}_{n}(\lambda^d)^{q}_{p}\nonumber\\
	=&N\bigg(- \frac{1}{16N} \delta^q_{p} \delta_{l}^i \delta_{j}^m \delta_{n}^k 
	- \frac{1}{16N} \delta_{p}^q \delta_{l}^m \delta_{n}^i \delta_{j}^k 
	- \frac{1}{16N} \delta_{l}^q \delta_{p}^i \delta_{j}^k \delta_{n}^m \notag \\
	&+ \frac{1}{8N} \delta_{l}^q \delta_{p}^k \delta_{n}^i \delta_{j}^m 
	- \frac{1}{16N} \delta_{l}^q \delta_{p}^m \delta_{j}^i \delta_{n}^k 
	+ \frac{1}{16} \delta_{j}^q \delta_{l}^i \delta_{n}^k \delta_{p}^m  \notag \\
	&+ \frac{1}{8N} \delta_{j}^q \delta_{p}^i \delta_{l}^m \delta_{n}^k 
	- \frac{1}{16N} \delta_{j}^q \delta_{p}^k \delta_{l}^i \delta_{n}^m 
	- \frac{1}{16N} \delta_{j}^q \delta_{p}^m \delta_{l}^k \delta_{n}^i \notag \\
	&- \frac{1}{16N} \delta_{n}^q \delta_{p}^i \delta_{l}^k \delta_{j}^m 
	+ \frac{1}{16} \delta_{n}^q \delta_{p}^i \delta_{l}^m \delta_{j}^k 
	- \frac{1}{16N} \delta_{n}^q \delta_{p}^k \delta_{l}^m \delta_{j}^i \notag \\
	&+ \frac{1}{8N} \delta_{n}^q \delta_{p}^m \delta_{l}^i \delta_{j}^k\bigg)\,.
\end{align}
The contributions of order $N^0$ inside the parentheses correspond to the standard 't Hooft vertex, whereas the terms of order $\frac{1}{N}$ give rise to pinched surfaces (fig.~\ref{fig:F4doubleline}). The leading-order vertex (fig.~\ref{fig:F4doubleline}) gives rise to the standard smooth 't Hooft topologies: for $2$-point correlators, we obtain a $2$-punctured sphere, a $2$-punctured torus, a $2$-punctured double torus, and so on. However, the subleading vertex (fig.~\ref{fig:F4doubleline}), possibly contracted with itself or with the leading one, gives rise to a torus with $2$ pinches, a double torus with $2$ pinches, and so on (fig.~\ref{fig:A4twopoint}).\par
\par 

\subsection{$3$-gluon operators in the fundamental representation}\label{sec:3gluon}

The $3$-gluon quasi-partonic operators in the fundamental representation read
\begin{equation}\label{eq:O3gaugeinv}
	\mathcal{O}_3 = \mathcal{P}\big(D^{(1)}_+,D^{(2)}_+,D^{(3)}_+\big)\,\Tr\left(F_{+\perp}\,F_{+\perp}\,F_{+\perp}\right),
\end{equation}
where the good components of $F_{+\perp}$ -- $F$ or $\bar F$ -- are valued in the fundamental representation, $F_{+\perp}=F^a_{+\perp}\lambda^a$, with the same chirality assignments as below eq.~\eqref{eq:A3gaugeinv}. Their color structure is given by
\begin{equation}\label{eq:trac3}
	\mathcal{O}_3 = \mathcal{P}\big(D^{(1)}_+,D^{(2)}_+,D^{(3)}_+\big)\,\Tr(\lambda^a\lambda^b\lambda^c)\,F_{+\perp}^aF_{+\perp}^bF_{+\perp}^c\,,
\end{equation}
where
\begin{equation}\label{eq:Tr3fund}
	\Tr(\lambda^a\lambda^b\lambda^c)=\frac{1}{4}\left(d^{abc}+i f^{abc}\right),
\end{equation}
since
\begin{equation}
	\lambda^a\lambda^b=\frac{1}{2N}\delta^{ab}\,\mathbb{I}+\frac{1}{2}\left(d^{abc}+if^{abc}\right)\lambda^c\,,
\end{equation}
with $d^{abc}=2\Tr\left(\lambda^a\{\lambda^b,\lambda^c\}\right)$ being the totally symmetric symbol and $c$ running over the $su(N)$ generators. \par
The first term in eq.~\eqref{eq:Tr3fund} is totally symmetric and the second one is totally antisymmetric in the color indices. The two terms differ under charge conjugation $\Tr(\lambda^a\lambda^b\lambda^c)\to-\Tr(\lambda^c\lambda^b\lambda^a)$, so that the $d^{abc}$ term is odd and the $f^{abc}$ term is even. \par
\begin{figure}[h!]
	\centering
	\begin{tikzpicture}[baseline={(0,-0.1)}]
		\path (-1.7,-1.75) rectangle (1.7,1.75);
		\begin{scope}[shift={(0,-0.25)}]
			\draw[pinch] (-0.32,0.1848) -- (0,0);
			\draw[pinch] (0,-0.3695) -- (0,0);
			\draw[pinch] (0.32,0.1848) -- (0,0);
			\draw[skel] (-0.32,1.55) -- (-0.32,0.1848) -- (-1.5023,-0.4979);
			\draw[skel] (-1.1823,-1.0521) -- (0,-0.3695) -- (1.1823,-1.0521);
			\draw[skel] (1.5023,-0.4979) -- (0.32,0.1848) -- (0.32,1.55);
			\fill (-0.32,0.1848) circle (1.6pt);
			\fill (0,-0.3695) circle (1.6pt);
			\fill (0.32,0.1848) circle (1.6pt);
		\end{scope}
	\end{tikzpicture}
	$\;-\;\dfrac{2}{N}\;$
	\begin{tikzpicture}[baseline={(0,-0.1)}]
		\path (-1.7,-1.75) rectangle (1.7,1.75);
		\begin{scope}[shift={(0,-0.78)}]
			\draw[pinch] (0,0.85) -- (0,-0.45);
			\draw[skel] (-1.5,0.45) -- (1.5,0.45);
			\draw[skel] (-1.5,-0.45) -- (1.5,-0.45);
			\fill (0,0.45) circle (1.6pt);
			\fill (0,-0.45) circle (1.6pt);
			\ophairpin{90}{0.85}{1.15}
		\end{scope}
	\end{tikzpicture}
	$\;+\;\dfrac{4}{N^2}\;$
	\begin{tikzpicture}[baseline={(0,-0.1)}]
		\path (-1.7,-1.75) rectangle (1.7,1.75);
		\foreach \a in {90,210,330}{\draw[pinch] (\a:0.52) -- (0,0);}
		\ophairpin{90}{0.52}{1.05}
		\ophairpin{210}{0.52}{1.05}
		\ophairpin{330}{0.52}{1.05}
	\end{tikzpicture}
	\caption{Double-line representation of the operator vertex $(\mathbf{V}_3)^{i\,k\,m}_{j\,l\,n}$: one diagram is displayed for each color structure in eq.~\eqref{eq:D3SU}, with the remaining ones obtained by permutations of the legs. The single-pinched vertex has weight $\frac{1}{N}$, while the maximally pinched one has weight $\frac{1}{N^2}$.}
	\label{fig:D3doubleline}
\end{figure}
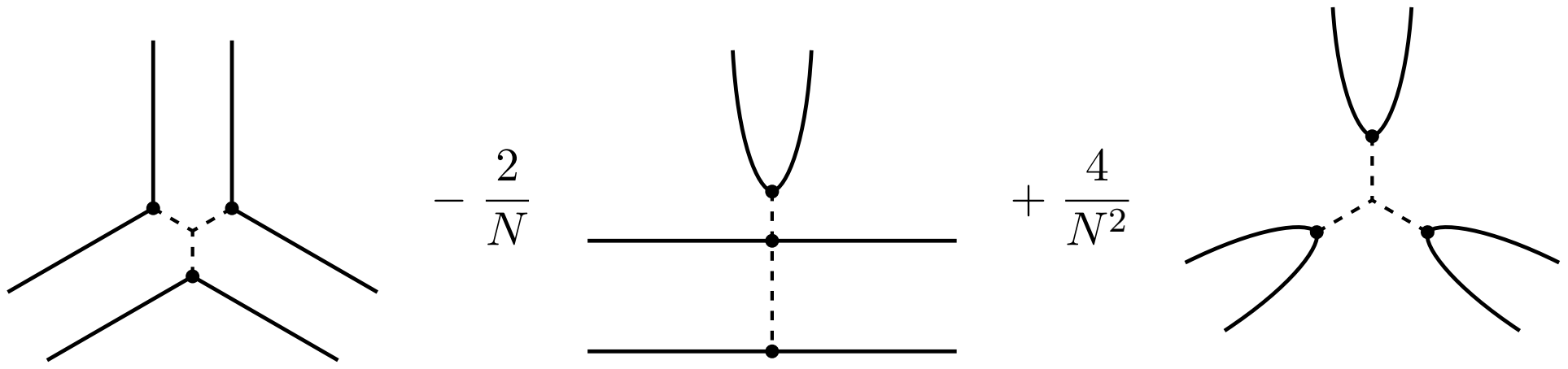

The latter reproduces the operators in the adjoint representation of the previous subsection, since $\Tr(T^aT^bT^c)=\frac{N}{2}if^{abc}$, whereas the former is new. Therefore, we consider only the operators
\begin{equation}\label{eq:O3d}
	\mathcal{O}_3 = \mathcal{P}\big(D^{(1)}_+,D^{(2)}_+,D^{(3)}_+\big)\,d^{abc}\,F_{+\perp}^aF_{+\perp}^bF_{+\perp}^c\,,
\end{equation}
so that they have the structure
\begin{equation}
	\mathcal{O}_3= \mathcal{P}(\partial^{\mu_1},\partial^{\mu_2},\partial^{\mu_3})\,(\mathbf{V}_3)^{i\,k\,m}_{j\,l\,n}\,(A_{\mu_1})^{j}_{\ i}(A_{\mu_2})^{l}_{\ k}(A_{\mu_3})^{n}_{\ m}\,,
\end{equation}
with the color vertex in the double-line representation given by
\begin{align}\label{eq:D3SU}
	(\mathbf{V}_3)^{i\,k\,m}_{j\,l\,n}=\frac{1}{4}\Big(&\delta^{i}_n\delta^{k}_j\delta^{m}_l+\delta^{i}_l\delta^{m}_j\delta^{k}_n-\frac{2}{N}\delta^{i}_j\delta^{m}_l\delta^{k}_n-\frac{2}{N}\delta^{i}_l\delta^{k}_j\delta^{m}_n-\frac{2}{N}\delta^{i}_n\delta^{m}_j\delta^{k}_l\nonumber\\
	&+\frac{4}{N^2}\delta^{i}_j\delta^{k}_l\delta^{m}_n\Big)\,.
\end{align}
Unlike the adjoint vertex in eq.~\eqref{eq:adjvertexdouble}, which is $N$-independent up to overall normalization, this vertex contains contributions of several orders. The terms of order $N^0$ inside the parentheses are standard 't Hooft vertices, whereas the contributions of order $\frac{1}{N}$ and $\frac{1}{N^2}$ give rise to pinched surfaces (fig.~\ref{fig:D3doubleline}).
The leading vertex gives rise to the standard smooth 't Hooft topologies. \par
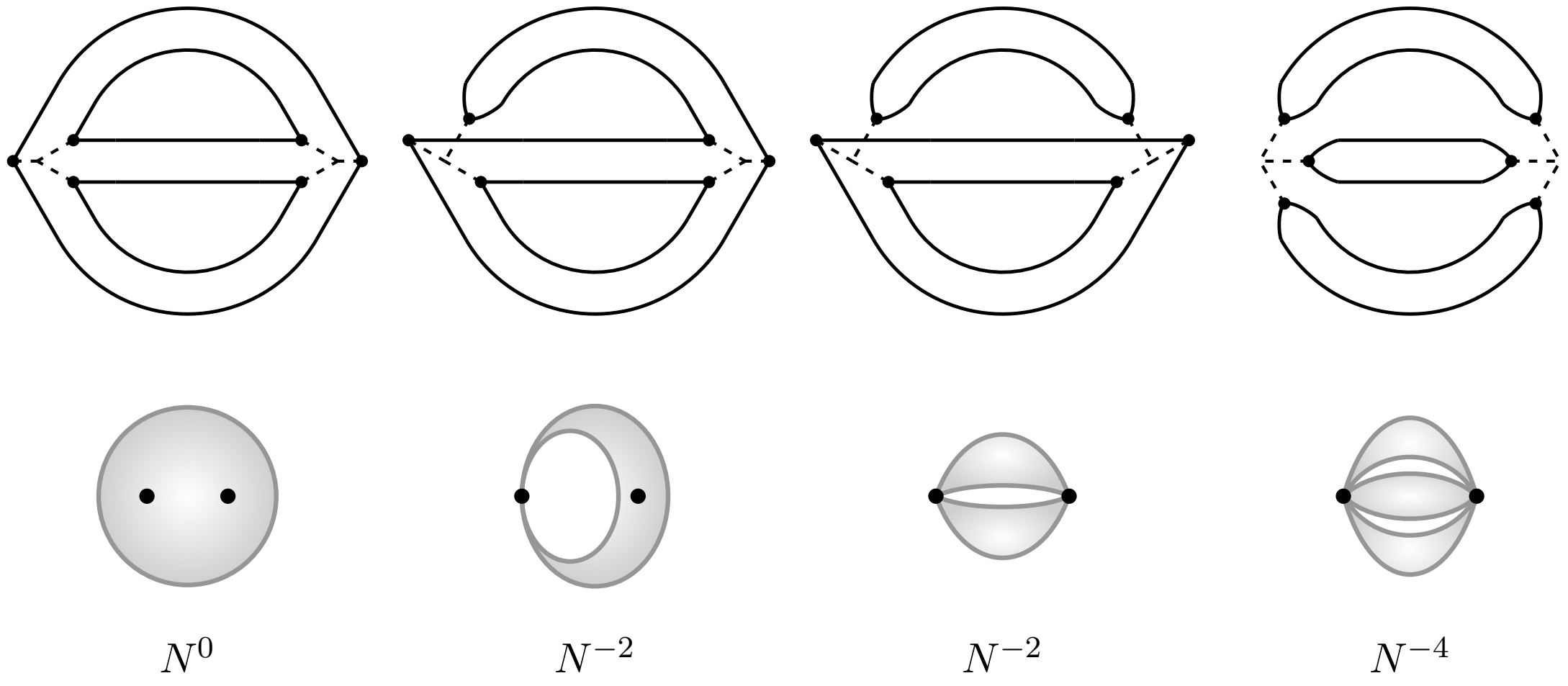
\begin{figure}[h!]
		\centering
		\begin{tikzpicture}[scale=0.6,
			surf/.style={c969696, line width=1.2pt}]
			\begin{scope}[shift={(0.00,0.00)}]
				\begin{scope}[shift={(-2.300,0)}]
					\draw[skel] (0.877,0.879) -- (0.554,0.320) -- (1.200,0.320);
					\draw[skel] (1.200,-0.320) -- (0.554,-0.320) -- (0.877,-0.879);
					\draw[skel] (0.323,1.199) -- (-0.370,0.000) -- (0.323,-1.199);
					\draw[pinch] (0.554,0.320) -- (0,0);
					\fill (0.554,0.320) circle (2.67pt);
					\draw[pinch] (0.554,-0.320) -- (0,0);
					\fill (0.554,-0.320) circle (2.67pt);
					\draw[pinch] (-0.370,0.000) -- (0,0);
					\fill (-0.370,0.000) circle (2.67pt);
				\end{scope}
				\begin{scope}[shift={(2.300,0)}, xscale=-1]
					\draw[skel] (0.877,0.879) -- (0.554,0.320) -- (1.200,0.320);
					\draw[skel] (1.200,-0.320) -- (0.554,-0.320) -- (0.877,-0.879);
					\draw[skel] (0.323,1.199) -- (-0.370,0.000) -- (0.323,-1.199);
					\draw[pinch] (0.554,0.320) -- (0,0);
					\fill (0.554,0.320) circle (2.67pt);
					\draw[pinch] (0.554,-0.320) -- (0,0);
					\fill (0.554,-0.320) circle (2.67pt);
					\draw[pinch] (-0.370,0.000) -- (0,0);
					\fill (-0.370,0.000) circle (2.67pt);
				\end{scope}
				\draw[skel] (-1.977,1.199) arc[start angle=150.000, end angle=30.000, radius=2.283];
				\draw[skel] (-1.423,0.879) arc[start angle=150.000, end angle=30.000, radius=1.643];
				\draw[skel] (-1.100,0.320) -- (1.100,0.320);
				\draw[skel] (-1.100,-0.320) -- (1.100,-0.320);
				\draw[skel] (-1.423,-0.879) arc[start angle=-150.000, end angle=-30.000, radius=1.643];
				\draw[skel] (-1.977,-1.199) arc[start angle=-150.000, end angle=-30.000, radius=2.283];
			\end{scope}
			\begin{scope}[shift={(6.24,0.00)}]
				\begin{scope}[shift={(-2.300,0)}]
					\draw[skel] (1.200,-0.320) -- (0.554,-0.320) -- (0.877,-0.879);
					\draw[skel] (1.200,0.320) -- (-0.554,0.320) -- (0.323,-1.199);
					\begin{scope}[shift={(60:0.750)}, rotate=-30]
						\draw[skel] (-0.320,0.450) .. controls (-0.280,0.292) and (-0.12,0.03) .. (0,0) .. controls (0.12,0.03) and (0.280,0.292) .. (0.320,0.450);
						\fill (0,0) circle (2.67pt);
						\draw[pinch] (0,0) -- (0,-0.750);
					\end{scope}
					\draw[pinch] (0.554,-0.320) -- (0,0);
					\fill (0.554,-0.320) circle (2.67pt);
					\draw[pinch] (-0.554,0.320) -- (0,0);
					\fill (-0.554,0.320) circle (2.67pt);
				\end{scope}
				\begin{scope}[shift={(2.300,0)}, xscale=-1]
					\draw[skel] (0.877,0.879) -- (0.554,0.320) -- (1.200,0.320);
					\draw[skel] (1.200,-0.320) -- (0.554,-0.320) -- (0.877,-0.879);
					\draw[skel] (0.323,1.199) -- (-0.370,0.000) -- (0.323,-1.199);
					\draw[pinch] (0.554,0.320) -- (0,0);
					\fill (0.554,0.320) circle (2.67pt);
					\draw[pinch] (0.554,-0.320) -- (0,0);
					\fill (0.554,-0.320) circle (2.67pt);
					\draw[pinch] (-0.370,0.000) -- (0,0);
					\fill (-0.370,0.000) circle (2.67pt);
				\end{scope}
				\draw[skel] (-1.977,1.199) arc[start angle=150.000, end angle=30.000, radius=2.283];
				\draw[skel] (-1.423,0.879) arc[start angle=150.000, end angle=30.000, radius=1.643];
				\draw[skel] (-1.100,0.320) -- (1.100,0.320);
				\draw[skel] (-1.100,-0.320) -- (1.100,-0.320);
				\draw[skel] (-1.423,-0.879) arc[start angle=-150.000, end angle=-30.000, radius=1.643];
				\draw[skel] (-1.977,-1.199) arc[start angle=-150.000, end angle=-30.000, radius=2.283];
			\end{scope}
			\begin{scope}[shift={(12.48,0.00)}]
				\begin{scope}[shift={(-2.300,0)}]
					\draw[skel] (1.200,-0.320) -- (0.554,-0.320) -- (0.877,-0.879);
					\draw[skel] (1.200,0.320) -- (-0.554,0.320) -- (0.323,-1.199);
					\begin{scope}[shift={(60:0.750)}, rotate=-30]
						\draw[skel] (-0.320,0.450) .. controls (-0.280,0.292) and (-0.12,0.03) .. (0,0) .. controls (0.12,0.03) and (0.280,0.292) .. (0.320,0.450);
						\fill (0,0) circle (2.67pt);
						\draw[pinch] (0,0) -- (0,-0.750);
					\end{scope}
					\draw[pinch] (0.554,-0.320) -- (0,0);
					\fill (0.554,-0.320) circle (2.67pt);
					\draw[pinch] (-0.554,0.320) -- (0,0);
					\fill (-0.554,0.320) circle (2.67pt);
				\end{scope}
				\begin{scope}[shift={(2.300,0)}, xscale=-1]
					\draw[skel] (1.200,-0.320) -- (0.554,-0.320) -- (0.877,-0.879);
					\draw[skel] (1.200,0.320) -- (-0.554,0.320) -- (0.323,-1.199);
					\begin{scope}[shift={(60:0.750)}, rotate=-30]
						\draw[skel] (-0.320,0.450) .. controls (-0.280,0.292) and (-0.12,0.03) .. (0,0) .. controls (0.12,0.03) and (0.280,0.292) .. (0.320,0.450);
						\fill (0,0) circle (2.67pt);
						\draw[pinch] (0,0) -- (0,-0.750);
					\end{scope}
					\draw[pinch] (0.554,-0.320) -- (0,0);
					\fill (0.554,-0.320) circle (2.67pt);
					\draw[pinch] (-0.554,0.320) -- (0,0);
					\fill (-0.554,0.320) circle (2.67pt);
				\end{scope}
				\draw[skel] (-1.977,1.199) arc[start angle=150.000, end angle=30.000, radius=2.283];
				\draw[skel] (-1.423,0.879) arc[start angle=150.000, end angle=30.000, radius=1.643];
				\draw[skel] (-1.100,0.320) -- (1.100,0.320);
				\draw[skel] (-1.100,-0.320) -- (1.100,-0.320);
				\draw[skel] (-1.423,-0.879) arc[start angle=-150.000, end angle=-30.000, radius=1.643];
				\draw[skel] (-1.977,-1.199) arc[start angle=-150.000, end angle=-30.000, radius=2.283];
			\end{scope}
			\begin{scope}[shift={(18.72,0.00)}]
				\begin{scope}[shift={(-2.300,0)}]
					\begin{scope}[shift={(60:0.750)}, rotate=-30]
						\draw[skel] (-0.320,0.450) .. controls (-0.280,0.292) and (-0.12,0.03) .. (0,0) .. controls (0.12,0.03) and (0.280,0.292) .. (0.320,0.450);
						\fill (0,0) circle (2.67pt);
						\draw[pinch] (0,0) -- (0,-0.750);
					\end{scope}
					\begin{scope}[shift={(0:0.750)}, rotate=-90]
						\draw[skel] (-0.320,0.450) .. controls (-0.280,0.292) and (-0.12,0.03) .. (0,0) .. controls (0.12,0.03) and (0.280,0.292) .. (0.320,0.450);
						\fill (0,0) circle (2.67pt);
						\draw[pinch] (0,0) -- (0,-0.750);
					\end{scope}
					\begin{scope}[shift={(-60:0.750)}, rotate=-150]
						\draw[skel] (-0.320,0.450) .. controls (-0.280,0.292) and (-0.12,0.03) .. (0,0) .. controls (0.12,0.03) and (0.280,0.292) .. (0.320,0.450);
						\fill (0,0) circle (2.67pt);
						\draw[pinch] (0,0) -- (0,-0.750);
					\end{scope}
				\end{scope}
				\begin{scope}[shift={(2.300,0)}, xscale=-1]
					\begin{scope}[shift={(60:0.750)}, rotate=-30]
						\draw[skel] (-0.320,0.450) .. controls (-0.280,0.292) and (-0.12,0.03) .. (0,0) .. controls (0.12,0.03) and (0.280,0.292) .. (0.320,0.450);
						\fill (0,0) circle (2.67pt);
						\draw[pinch] (0,0) -- (0,-0.750);
					\end{scope}
					\begin{scope}[shift={(0:0.750)}, rotate=-90]
						\draw[skel] (-0.320,0.450) .. controls (-0.280,0.292) and (-0.12,0.03) .. (0,0) .. controls (0.12,0.03) and (0.280,0.292) .. (0.320,0.450);
						\fill (0,0) circle (2.67pt);
						\draw[pinch] (0,0) -- (0,-0.750);
					\end{scope}
					\begin{scope}[shift={(-60:0.750)}, rotate=-150]
						\draw[skel] (-0.320,0.450) .. controls (-0.280,0.292) and (-0.12,0.03) .. (0,0) .. controls (0.12,0.03) and (0.280,0.292) .. (0.320,0.450);
						\fill (0,0) circle (2.67pt);
						\draw[pinch] (0,0) -- (0,-0.750);
					\end{scope}
				\end{scope}
				\draw[skel] (-1.977,1.199) arc[start angle=150.000, end angle=30.000, radius=2.283];
				\draw[skel] (-1.423,0.879) arc[start angle=150.000, end angle=30.000, radius=1.643];
				\draw[skel] (-1.100,0.320) -- (1.100,0.320);
				\draw[skel] (-1.100,-0.320) -- (1.100,-0.320);
				\draw[skel] (-1.423,-0.879) arc[start angle=-150.000, end angle=-30.000, radius=1.643];
				\draw[skel] (-1.977,-1.199) arc[start angle=-150.000, end angle=-30.000, radius=2.283];
			\end{scope}
			\begin{scope}[shift={(0.00,-5.13)}]
				\shade[inner color=white, outer color=gray!38] (0,0) circle (1.360);
				\draw[surf] (0,0) circle (1.360);
				\fill (-0.620,0) circle (3.33pt); \fill (0.620,0) circle (3.33pt);
			\end{scope}
			\node at (0.00,-7.56) {$N^{0}$};
			\begin{scope}[shift={(6.24,-5.13)}]
				\shade[inner color=white, outer color=gray!38, even odd rule] (0,0) ellipse (1.120 and 1.380) (-0.380,0) ellipse (0.740 and 1.000);
				\draw[surf] (0,0) ellipse (1.120 and 1.380);
				\draw[surf] (-0.380,0) ellipse (0.740 and 1.000);
				\fill (-1.120,0) circle (3.33pt); \fill (0.660,0) circle (3.33pt);
			\end{scope}
			\node at (6.24,-7.56) {$N^{-2}$};
			\begin{scope}[shift={(12.48,-5.13)}]
				\shade[inner color=white, outer color=gray!38] (-1.020,0) .. controls (-0.560,1.260) and (0.560,1.260) .. (1.020,0) .. controls (0.560,0.220) and (-0.560,0.220) .. cycle;
				\shade[inner color=white, outer color=gray!38] (-1.020,0) .. controls (-0.560,-0.220) and (0.560,-0.220) .. (1.020,0) .. controls (0.560,-1.260) and (-0.560,-1.260) .. cycle;
				\draw[surf] (-1.020,0) .. controls (-0.560,1.260) and (0.560,1.260) .. (1.020,0);
				\draw[surf] (-1.020,0) .. controls (-0.560,0.220) and (0.560,0.220) .. (1.020,0);
				\draw[surf] (-1.020,0) .. controls (-0.560,-0.220) and (0.560,-0.220) .. (1.020,0);
				\draw[surf] (-1.020,0) .. controls (-0.560,-1.260) and (0.560,-1.260) .. (1.020,0);
				\fill (-1.020,0) circle (3.33pt); \fill (1.020,0) circle (3.33pt);
			\end{scope}
			\node at (12.48,-7.56) {$N^{-2}$};
			\begin{scope}[shift={(18.72,-5.13)}]
				\shade[inner color=white, outer color=gray!38] (-1.020,0) .. controls (-0.560,1.600) and (0.560,1.600) .. (1.020,0) .. controls (0.560,0.800) and (-0.560,0.800) .. cycle;
				\shade[inner color=white, outer color=gray!38] (-1.020,0) .. controls (-0.560,0.460) and (0.560,0.460) .. (1.020,0) .. controls (0.560,-0.460) and (-0.560,-0.460) .. cycle;
				\shade[inner color=white, outer color=gray!38] (-1.020,0) .. controls (-0.560,-0.800) and (0.560,-0.800) .. (1.020,0) .. controls (0.560,-1.600) and (-0.560,-1.600) .. cycle;
				\draw[surf] (-1.020,0) .. controls (-0.560,1.600) and (0.560,1.600) .. (1.020,0);
				\draw[surf] (-1.020,0) .. controls (-0.560,0.800) and (0.560,0.800) .. (1.020,0);
				\draw[surf] (-1.020,0) .. controls (-0.560,0.460) and (0.560,0.460) .. (1.020,0);
				\draw[surf] (-1.020,0) .. controls (-0.560,-0.460) and (0.560,-0.460) .. (1.020,0);
				\draw[surf] (-1.020,0) .. controls (-0.560,-0.800) and (0.560,-0.800) .. (1.020,0);
				\draw[surf] (-1.020,0) .. controls (-0.560,-1.600) and (0.560,-1.600) .. (1.020,0);
				\fill (-1.020,0) circle (3.33pt); \fill (1.020,0) circle (3.33pt);
			\end{scope}
			\node at (18.72,-7.56) {$N^{-4}$};
		\end{tikzpicture}
	\caption{Planar contractions (i.e., without crossed propagators) of the $2$-point correlator of $3$-gluon operators in the fundamental representation at lowest order (top), and the corresponding surfaces with their weights (bottom): contractions of leading vertices give rise to a sphere with two punctures; contractions involving a single-pinched vertex give rise to a torus with a cycle pinched at a puncture; contractions of two single-pinched vertices via the same propagator give rise to a doubly pinched torus; and contractions of the maximally pinched vertices give rise to the maximally pinched double torus. Although the diagrams are planar, the pinched surfaces are degenerations of higher-genus ones.}
\label{fig:D3twopoint}
\label{fig:D3joined}
\end{figure}
\begin{figure}[h!]
		\centering
		\begin{tikzpicture}[scale=0.6,
			surf/.style={c969696, line width=1.2pt},
			ident/.style={line width=0.8pt, black, dash pattern=on 0.9pt off 1.4pt}]
			\begin{scope}[shift={(0.00,0.00)}]
				\begin{scope}[shift={(-2.300,0)}]
					\draw[skel] (0.877,0.879) -- (0.554,0.320) -- (1.200,0.320);
					\draw[skel] (1.200,-0.320) -- (0.554,-0.320) -- (0.877,-0.879);
					\draw[skel] (0.323,1.199) -- (-0.370,0.000) -- (0.323,-1.199);
					\draw[pinch] (0.554,0.320) -- (0,0);
					\fill (0.554,0.320) circle (2.67pt);
					\draw[pinch] (0.554,-0.320) -- (0,0);
					\fill (0.554,-0.320) circle (2.67pt);
					\draw[pinch] (-0.370,0.000) -- (0,0);
					\fill (-0.370,0.000) circle (2.67pt);
				\end{scope}
				\begin{scope}[shift={(2.300,0)}, xscale=-1]
					\draw[skel] (0.877,0.879) -- (0.554,0.320) -- (1.200,0.320);
					\draw[skel] (1.200,-0.320) -- (0.554,-0.320) -- (0.877,-0.879);
					\draw[skel] (0.323,1.199) -- (-0.370,0.000) -- (0.323,-1.199);
					\draw[pinch] (0.554,0.320) -- (0,0);
					\fill (0.554,0.320) circle (2.67pt);
					\draw[pinch] (0.554,-0.320) -- (0,0);
					\fill (0.554,-0.320) circle (2.67pt);
					\draw[pinch] (-0.370,0.000) -- (0,0);
					\fill (-0.370,0.000) circle (2.67pt);
				\end{scope}
				\draw[skel] (-1.977,1.199) arc[start angle=150.000, end angle=30.000, radius=2.283];
				\draw[skel] (-1.423,0.879) arc[start angle=150.000, end angle=30.000, radius=1.643];
				\draw[skel] (-1.100,0.320) -- (-1.039,0.319) -- (-0.978,0.315) -- (-0.918,0.309) -- (-0.859,0.300) -- (-0.800,0.289) -- (-0.743,0.276) -- (-0.686,0.262) -- (-0.630,0.245) -- (-0.575,0.227) -- (-0.521,0.207) -- (-0.468,0.185) -- (-0.416,0.162) -- (-0.365,0.138) -- (-0.314,0.113) -- (-0.265,0.087) -- (-0.216,0.059) -- (-0.168,0.031) -- (-0.120,0.002) -- (-0.074,-0.028) -- (-0.028,-0.059) -- (0.017,-0.090) -- (0.062,-0.121) -- (0.106,-0.154) -- (0.149,-0.186) -- (0.192,-0.219) -- (0.234,-0.252) -- (0.275,-0.285) -- (0.316,-0.319) -- (0.356,-0.352) -- (0.396,-0.385) -- (0.435,-0.419) -- (0.474,-0.452) -- (0.512,-0.485) -- (0.549,-0.517) -- (0.586,-0.549) -- (0.623,-0.581) -- (0.658,-0.612) -- (0.694,-0.643) -- (0.728,-0.673) -- (0.762,-0.702) -- (0.796,-0.730) -- (0.829,-0.758) -- (0.861,-0.785) -- (0.893,-0.810) -- (0.923,-0.835) -- (0.953,-0.858) -- (0.983,-0.880) -- (1.011,-0.900) -- (1.039,-0.920) -- (1.065,-0.938) -- (1.091,-0.954) -- (1.115,-0.969) -- (1.138,-0.982) -- (1.160,-0.993) -- (1.180,-1.003) -- (1.199,-1.011) -- (1.216,-1.018) -- (1.231,-1.023) -- (1.245,-1.027) -- (1.256,-1.030) -- (1.265,-1.031) -- (1.273,-1.032) -- (1.279,-1.032) -- (1.283,-1.032) -- (1.286,-1.031) -- (1.289,-1.031) -- (1.292,-1.030) -- (1.295,-1.029) -- (1.299,-1.027) -- (1.305,-1.024) -- (1.312,-1.019) -- (1.321,-1.012) -- (1.332,-1.003) -- (1.344,-0.990) -- (1.357,-0.975) -- (1.372,-0.957) -- (1.388,-0.934) -- (1.405,-0.909) -- (1.423,-0.879);
				\draw[skel] (-1.100,-0.320) -- (-1.066,-0.321) -- (-1.031,-0.323) -- (-0.997,-0.326) -- (-0.963,-0.331) -- (-0.928,-0.338) -- (-0.893,-0.346) -- (-0.858,-0.355) -- (-0.823,-0.365) -- (-0.788,-0.377) -- (-0.752,-0.390) -- (-0.717,-0.405) -- (-0.680,-0.420) -- (-0.644,-0.437) -- (-0.608,-0.456) -- (-0.571,-0.475) -- (-0.535,-0.496) -- (-0.498,-0.517) -- (-0.461,-0.540) -- (-0.424,-0.564) -- (-0.387,-0.588) -- (-0.350,-0.614) -- (-0.313,-0.640) -- (-0.276,-0.667) -- (-0.239,-0.695) -- (-0.202,-0.724) -- (-0.165,-0.753) -- (-0.128,-0.783) -- (-0.091,-0.813) -- (-0.054,-0.843) -- (-0.017,-0.874) -- (0.020,-0.905) -- (0.056,-0.937) -- (0.093,-0.968) -- (0.129,-1.000) -- (0.166,-1.032) -- (0.202,-1.063) -- (0.238,-1.095) -- (0.275,-1.127) -- (0.311,-1.158) -- (0.347,-1.189) -- (0.383,-1.220) -- (0.419,-1.250) -- (0.456,-1.280) -- (0.492,-1.309) -- (0.529,-1.338) -- (0.565,-1.367) -- (0.602,-1.394) -- (0.639,-1.421) -- (0.676,-1.447) -- (0.714,-1.473) -- (0.752,-1.497) -- (0.791,-1.521) -- (0.831,-1.543) -- (0.871,-1.564) -- (0.912,-1.584) -- (0.954,-1.603) -- (0.997,-1.620) -- (1.042,-1.635) -- (1.088,-1.647) -- (1.135,-1.658) -- (1.184,-1.666) -- (1.234,-1.671) -- (1.285,-1.672) -- (1.337,-1.669) -- (1.389,-1.663) -- (1.442,-1.652) -- (1.494,-1.637) -- (1.544,-1.618) -- (1.594,-1.595) -- (1.641,-1.568) -- (1.687,-1.538) -- (1.730,-1.504) -- (1.771,-1.468) -- (1.810,-1.429) -- (1.846,-1.388) -- (1.881,-1.344) -- (1.915,-1.298) -- (1.947,-1.250) -- (1.977,-1.199);
				\draw[skel] (-1.423,-0.879) -- (-1.405,-0.909) -- (-1.388,-0.934) -- (-1.372,-0.957) -- (-1.357,-0.975) -- (-1.344,-0.990) -- (-1.332,-1.003) -- (-1.321,-1.012) -- (-1.312,-1.019) -- (-1.305,-1.024) -- (-1.299,-1.027) -- (-1.295,-1.029) -- (-1.292,-1.030) -- (-1.289,-1.031) -- (-1.286,-1.031) -- (-1.283,-1.032) -- (-1.279,-1.032) -- (-1.273,-1.032) -- (-1.265,-1.031) -- (-1.256,-1.030) -- (-1.245,-1.027) -- (-1.231,-1.023) -- (-1.216,-1.018) -- (-1.199,-1.011) -- (-1.180,-1.003) -- (-1.160,-0.993) -- (-1.138,-0.982) -- (-1.115,-0.969) -- (-1.091,-0.954) -- (-1.065,-0.938) -- (-1.039,-0.920) -- (-1.011,-0.900) -- (-0.983,-0.880) -- (-0.953,-0.858) -- (-0.923,-0.835) -- (-0.893,-0.810) -- (-0.861,-0.785) -- (-0.829,-0.758) -- (-0.796,-0.730) -- (-0.762,-0.702) -- (-0.728,-0.673) -- (-0.694,-0.643) -- (-0.658,-0.612) -- (-0.623,-0.581) -- (-0.586,-0.549) -- (-0.549,-0.517) -- (-0.512,-0.485) -- (-0.474,-0.452) -- (-0.435,-0.419) -- (-0.396,-0.385) -- (-0.356,-0.352) -- (-0.316,-0.319) -- (-0.275,-0.285) -- (-0.234,-0.252) -- (-0.192,-0.219) -- (-0.149,-0.186) -- (-0.106,-0.154) -- (-0.062,-0.121) -- (-0.017,-0.090) -- (0.028,-0.059) -- (0.074,-0.028) -- (0.120,0.002) -- (0.168,0.031) -- (0.216,0.059) -- (0.265,0.087) -- (0.314,0.113) -- (0.365,0.138) -- (0.416,0.162) -- (0.468,0.185) -- (0.521,0.207) -- (0.575,0.227) -- (0.630,0.245) -- (0.686,0.262) -- (0.743,0.276) -- (0.800,0.289) -- (0.859,0.300) -- (0.918,0.309) -- (0.978,0.315) -- (1.039,0.319) -- (1.100,0.320);
				\draw[skel] (-1.977,-1.199) -- (-1.947,-1.250) -- (-1.915,-1.298) -- (-1.881,-1.344) -- (-1.846,-1.388) -- (-1.810,-1.429) -- (-1.771,-1.468) -- (-1.730,-1.504) -- (-1.687,-1.538) -- (-1.641,-1.568) -- (-1.594,-1.595) -- (-1.544,-1.618) -- (-1.494,-1.637) -- (-1.442,-1.652) -- (-1.389,-1.663) -- (-1.337,-1.669) -- (-1.285,-1.672) -- (-1.234,-1.671) -- (-1.184,-1.666) -- (-1.135,-1.658) -- (-1.088,-1.647) -- (-1.042,-1.635) -- (-0.997,-1.620) -- (-0.954,-1.603) -- (-0.912,-1.584) -- (-0.871,-1.564) -- (-0.831,-1.543) -- (-0.791,-1.521) -- (-0.752,-1.497) -- (-0.714,-1.473) -- (-0.676,-1.447) -- (-0.639,-1.421) -- (-0.602,-1.394) -- (-0.565,-1.367) -- (-0.529,-1.338) -- (-0.492,-1.309) -- (-0.456,-1.280) -- (-0.419,-1.250) -- (-0.383,-1.220) -- (-0.347,-1.189) -- (-0.311,-1.158) -- (-0.275,-1.127) -- (-0.238,-1.095) -- (-0.202,-1.063) -- (-0.166,-1.032) -- (-0.129,-1.000) -- (-0.093,-0.968) -- (-0.056,-0.937) -- (-0.020,-0.905) -- (0.017,-0.874) -- (0.054,-0.843) -- (0.091,-0.813) -- (0.128,-0.783) -- (0.165,-0.753) -- (0.202,-0.724) -- (0.239,-0.695) -- (0.276,-0.667) -- (0.313,-0.640) -- (0.350,-0.614) -- (0.387,-0.588) -- (0.424,-0.564) -- (0.461,-0.540) -- (0.498,-0.517) -- (0.535,-0.496) -- (0.571,-0.475) -- (0.608,-0.456) -- (0.644,-0.437) -- (0.680,-0.420) -- (0.717,-0.405) -- (0.752,-0.390) -- (0.788,-0.377) -- (0.823,-0.365) -- (0.858,-0.355) -- (0.893,-0.346) -- (0.928,-0.338) -- (0.963,-0.331) -- (0.997,-0.326) -- (1.031,-0.323) -- (1.066,-0.321) -- (1.100,-0.320);
			\end{scope}
			\begin{scope}[shift={(0.00,-4.40)}]
				\shade[inner color=white, outer color=gray!38, even odd rule] (0,0) ellipse (1.81 and 1.16) (0,0) ellipse (0.73 and 0.44);
				\draw[surf] (0,0) ellipse (1.81 and 1.16);
				\draw[surf] (0,0) ellipse (0.73 and 0.44);
				\fill (-1.270,0) circle (3.33pt);
				\fill (1.270,0) circle (3.33pt);
			\end{scope}
			\node at (0.00,-6.16) {$N^{-2}$};
			\begin{scope}[shift={(6.74,0.00)}]
				\begin{scope}[shift={(-2.300,0)}]
					\draw[skel] (1.200,-0.320) -- (0.554,-0.320) -- (0.877,-0.879);
					\draw[skel] (1.200,0.320) -- (-0.554,0.320) -- (0.323,-1.199);
					\begin{scope}[shift={(60:0.750)}, rotate=-30]
						\draw[skel] (-0.320,0.450) .. controls (-0.280,0.292) and (-0.12,0.03) .. (0,0) .. controls (0.12,0.03) and (0.280,0.292) .. (0.320,0.450);
						\fill (0,0) circle (2.67pt);
						\draw[pinch] (0,0) -- (0,-0.750);
					\end{scope}
					\draw[pinch] (0.554,-0.320) -- (0,0);
					\fill (0.554,-0.320) circle (2.67pt);
					\draw[pinch] (-0.554,0.320) -- (0,0);
					\fill (-0.554,0.320) circle (2.67pt);
				\end{scope}
				\begin{scope}[shift={(2.300,0)}, xscale=-1]
					\draw[skel] (1.200,-0.320) -- (0.554,-0.320) -- (0.877,-0.879);
					\draw[skel] (1.200,0.320) -- (-0.554,0.320) -- (0.323,-1.199);
					\begin{scope}[shift={(60:0.750)}, rotate=-30]
						\draw[skel] (-0.320,0.450) .. controls (-0.280,0.292) and (-0.12,0.03) .. (0,0) .. controls (0.12,0.03) and (0.280,0.292) .. (0.320,0.450);
						\fill (0,0) circle (2.67pt);
						\draw[pinch] (0,0) -- (0,-0.750);
					\end{scope}
					\draw[pinch] (0.554,-0.320) -- (0,0);
					\fill (0.554,-0.320) circle (2.67pt);
					\draw[pinch] (-0.554,0.320) -- (0,0);
					\fill (-0.554,0.320) circle (2.67pt);
				\end{scope}
				\draw[skel] (-1.423,-0.879) arc[start angle=-150.000, end angle=-30.000, radius=1.643];
				\draw[skel] (-1.977,-1.199) arc[start angle=-150.000, end angle=-30.000, radius=2.283];
				\draw[skel] (-1.977,1.199) -- (-1.947,1.250) -- (-1.915,1.298) -- (-1.881,1.344) -- (-1.846,1.388) -- (-1.810,1.429) -- (-1.771,1.468) -- (-1.730,1.504) -- (-1.687,1.538) -- (-1.641,1.568) -- (-1.594,1.595) -- (-1.544,1.618) -- (-1.494,1.637) -- (-1.442,1.652) -- (-1.389,1.663) -- (-1.337,1.669) -- (-1.285,1.672) -- (-1.234,1.671) -- (-1.184,1.666) -- (-1.135,1.658) -- (-1.088,1.647) -- (-1.042,1.635) -- (-0.997,1.620) -- (-0.954,1.603) -- (-0.912,1.584) -- (-0.871,1.564) -- (-0.831,1.543) -- (-0.791,1.521) -- (-0.752,1.497) -- (-0.714,1.473) -- (-0.676,1.447) -- (-0.639,1.421) -- (-0.602,1.394) -- (-0.565,1.367) -- (-0.529,1.338) -- (-0.492,1.309) -- (-0.456,1.280) -- (-0.419,1.250) -- (-0.383,1.220) -- (-0.347,1.189) -- (-0.311,1.158) -- (-0.275,1.127) -- (-0.238,1.095) -- (-0.202,1.063) -- (-0.166,1.032) -- (-0.129,1.000) -- (-0.093,0.968) -- (-0.056,0.937) -- (-0.020,0.905) -- (0.017,0.874) -- (0.054,0.843) -- (0.091,0.813) -- (0.128,0.783) -- (0.165,0.753) -- (0.202,0.724) -- (0.239,0.695) -- (0.276,0.667) -- (0.313,0.640) -- (0.350,0.614) -- (0.387,0.588) -- (0.424,0.564) -- (0.461,0.540) -- (0.498,0.517) -- (0.535,0.496) -- (0.571,0.475) -- (0.608,0.456) -- (0.644,0.437) -- (0.680,0.420) -- (0.717,0.405) -- (0.752,0.390) -- (0.788,0.377) -- (0.823,0.365) -- (0.858,0.355) -- (0.893,0.346) -- (0.928,0.338) -- (0.963,0.331) -- (0.997,0.326) -- (1.031,0.323) -- (1.066,0.321) -- (1.100,0.320);
				\draw[skel] (-1.423,0.879) -- (-1.405,0.909) -- (-1.388,0.934) -- (-1.372,0.957) -- (-1.357,0.975) -- (-1.344,0.990) -- (-1.332,1.003) -- (-1.321,1.012) -- (-1.312,1.019) -- (-1.305,1.024) -- (-1.299,1.027) -- (-1.295,1.029) -- (-1.292,1.030) -- (-1.289,1.031) -- (-1.286,1.031) -- (-1.283,1.032) -- (-1.279,1.032) -- (-1.273,1.032) -- (-1.265,1.031) -- (-1.256,1.030) -- (-1.245,1.027) -- (-1.231,1.023) -- (-1.216,1.018) -- (-1.199,1.011) -- (-1.180,1.003) -- (-1.160,0.993) -- (-1.138,0.982) -- (-1.115,0.969) -- (-1.091,0.954) -- (-1.065,0.938) -- (-1.039,0.920) -- (-1.011,0.900) -- (-0.983,0.880) -- (-0.953,0.858) -- (-0.923,0.835) -- (-0.893,0.810) -- (-0.861,0.785) -- (-0.829,0.758) -- (-0.796,0.730) -- (-0.762,0.702) -- (-0.728,0.673) -- (-0.694,0.643) -- (-0.658,0.612) -- (-0.623,0.581) -- (-0.586,0.549) -- (-0.549,0.517) -- (-0.512,0.485) -- (-0.474,0.452) -- (-0.435,0.419) -- (-0.396,0.385) -- (-0.356,0.352) -- (-0.316,0.319) -- (-0.275,0.285) -- (-0.234,0.252) -- (-0.192,0.219) -- (-0.149,0.186) -- (-0.106,0.154) -- (-0.062,0.121) -- (-0.017,0.090) -- (0.028,0.059) -- (0.074,0.028) -- (0.120,-0.002) -- (0.168,-0.031) -- (0.216,-0.059) -- (0.265,-0.087) -- (0.314,-0.113) -- (0.365,-0.138) -- (0.416,-0.162) -- (0.468,-0.185) -- (0.521,-0.207) -- (0.575,-0.227) -- (0.630,-0.245) -- (0.686,-0.262) -- (0.743,-0.276) -- (0.800,-0.289) -- (0.859,-0.300) -- (0.918,-0.309) -- (0.978,-0.315) -- (1.039,-0.319) -- (1.100,-0.320);
				\draw[skel] (-1.100,0.320) -- (-1.066,0.321) -- (-1.031,0.323) -- (-0.997,0.326) -- (-0.963,0.331) -- (-0.928,0.338) -- (-0.893,0.346) -- (-0.858,0.355) -- (-0.823,0.365) -- (-0.788,0.377) -- (-0.752,0.390) -- (-0.717,0.405) -- (-0.680,0.420) -- (-0.644,0.437) -- (-0.608,0.456) -- (-0.571,0.475) -- (-0.535,0.496) -- (-0.498,0.517) -- (-0.461,0.540) -- (-0.424,0.564) -- (-0.387,0.588) -- (-0.350,0.614) -- (-0.313,0.640) -- (-0.276,0.667) -- (-0.239,0.695) -- (-0.202,0.724) -- (-0.165,0.753) -- (-0.128,0.783) -- (-0.091,0.813) -- (-0.054,0.843) -- (-0.017,0.874) -- (0.020,0.905) -- (0.056,0.937) -- (0.093,0.968) -- (0.129,1.000) -- (0.166,1.032) -- (0.202,1.063) -- (0.238,1.095) -- (0.275,1.127) -- (0.311,1.158) -- (0.347,1.189) -- (0.383,1.220) -- (0.419,1.250) -- (0.456,1.280) -- (0.492,1.309) -- (0.529,1.338) -- (0.565,1.367) -- (0.602,1.394) -- (0.639,1.421) -- (0.676,1.447) -- (0.714,1.473) -- (0.752,1.497) -- (0.791,1.521) -- (0.831,1.543) -- (0.871,1.564) -- (0.912,1.584) -- (0.954,1.603) -- (0.997,1.620) -- (1.042,1.635) -- (1.088,1.647) -- (1.135,1.658) -- (1.184,1.666) -- (1.234,1.671) -- (1.285,1.672) -- (1.337,1.669) -- (1.389,1.663) -- (1.442,1.652) -- (1.494,1.637) -- (1.544,1.618) -- (1.594,1.595) -- (1.641,1.568) -- (1.687,1.538) -- (1.730,1.504) -- (1.771,1.468) -- (1.810,1.429) -- (1.846,1.388) -- (1.881,1.344) -- (1.915,1.298) -- (1.947,1.250) -- (1.977,1.199);
				\draw[skel] (-1.100,-0.320) -- (-1.039,-0.319) -- (-0.978,-0.315) -- (-0.918,-0.309) -- (-0.859,-0.300) -- (-0.800,-0.289) -- (-0.743,-0.276) -- (-0.686,-0.262) -- (-0.630,-0.245) -- (-0.575,-0.227) -- (-0.521,-0.207) -- (-0.468,-0.185) -- (-0.416,-0.162) -- (-0.365,-0.138) -- (-0.314,-0.113) -- (-0.265,-0.087) -- (-0.216,-0.059) -- (-0.168,-0.031) -- (-0.120,-0.002) -- (-0.074,0.028) -- (-0.028,0.059) -- (0.017,0.090) -- (0.062,0.121) -- (0.106,0.154) -- (0.149,0.186) -- (0.192,0.219) -- (0.234,0.252) -- (0.275,0.285) -- (0.316,0.319) -- (0.356,0.352) -- (0.396,0.385) -- (0.435,0.419) -- (0.474,0.452) -- (0.512,0.485) -- (0.549,0.517) -- (0.586,0.549) -- (0.623,0.581) -- (0.658,0.612) -- (0.694,0.643) -- (0.728,0.673) -- (0.762,0.702) -- (0.796,0.730) -- (0.829,0.758) -- (0.861,0.785) -- (0.893,0.810) -- (0.923,0.835) -- (0.953,0.858) -- (0.983,0.880) -- (1.011,0.900) -- (1.039,0.920) -- (1.065,0.938) -- (1.091,0.954) -- (1.115,0.969) -- (1.138,0.982) -- (1.160,0.993) -- (1.180,1.003) -- (1.199,1.011) -- (1.216,1.018) -- (1.231,1.023) -- (1.245,1.027) -- (1.256,1.030) -- (1.265,1.031) -- (1.273,1.032) -- (1.279,1.032) -- (1.283,1.032) -- (1.286,1.031) -- (1.289,1.031) -- (1.292,1.030) -- (1.295,1.029) -- (1.299,1.027) -- (1.305,1.024) -- (1.312,1.019) -- (1.321,1.012) -- (1.332,1.003) -- (1.344,0.990) -- (1.357,0.975) -- (1.372,0.957) -- (1.388,0.934) -- (1.405,0.909) -- (1.423,0.879);
			\end{scope}
			\begin{scope}[shift={(6.74,-4.40)}]
				\shade[inner color=white, outer color=gray!38, even odd rule] (0,0) ellipse (1.81 and 1.04) (-1.721,0.321) .. controls (-0.574,0.855) and (0.574,0.855) .. (1.721,0.321) .. controls (0.574,0.028) and (-0.574,0.028) .. (-1.721,0.321) -- cycle (0,-0.450) ellipse (0.800 and 0.280);
				\draw[surf] (0,0) ellipse (1.81 and 1.04);
				\draw[surf, line join=round] (-1.721,0.321) .. controls (-0.574,0.855) and (0.574,0.855) .. (1.721,0.321) .. controls (0.574,0.028) and (-0.574,0.028) .. (-1.721,0.321) -- cycle;
				\draw[surf] (0,-0.450) ellipse (0.800 and 0.280);
				\fill (-1.721,0.321) circle (3.33pt);
				\fill (1.721,0.321) circle (3.33pt);
			\end{scope}
			\node at (6.74,-6.16) {$N^{-4}$};
			
		\end{tikzpicture}
		latex
\caption{Nonplanar contractions via two crossed propagators: contractions of the leading vertices give rise to a smooth torus of weight $N^{-2}$, whereas contractions involving two single-pinched vertices give rise to a double torus with two pinched cycles of weight $N^{-4}$.}
\label{fig:D3nonplanar}
\end{figure}
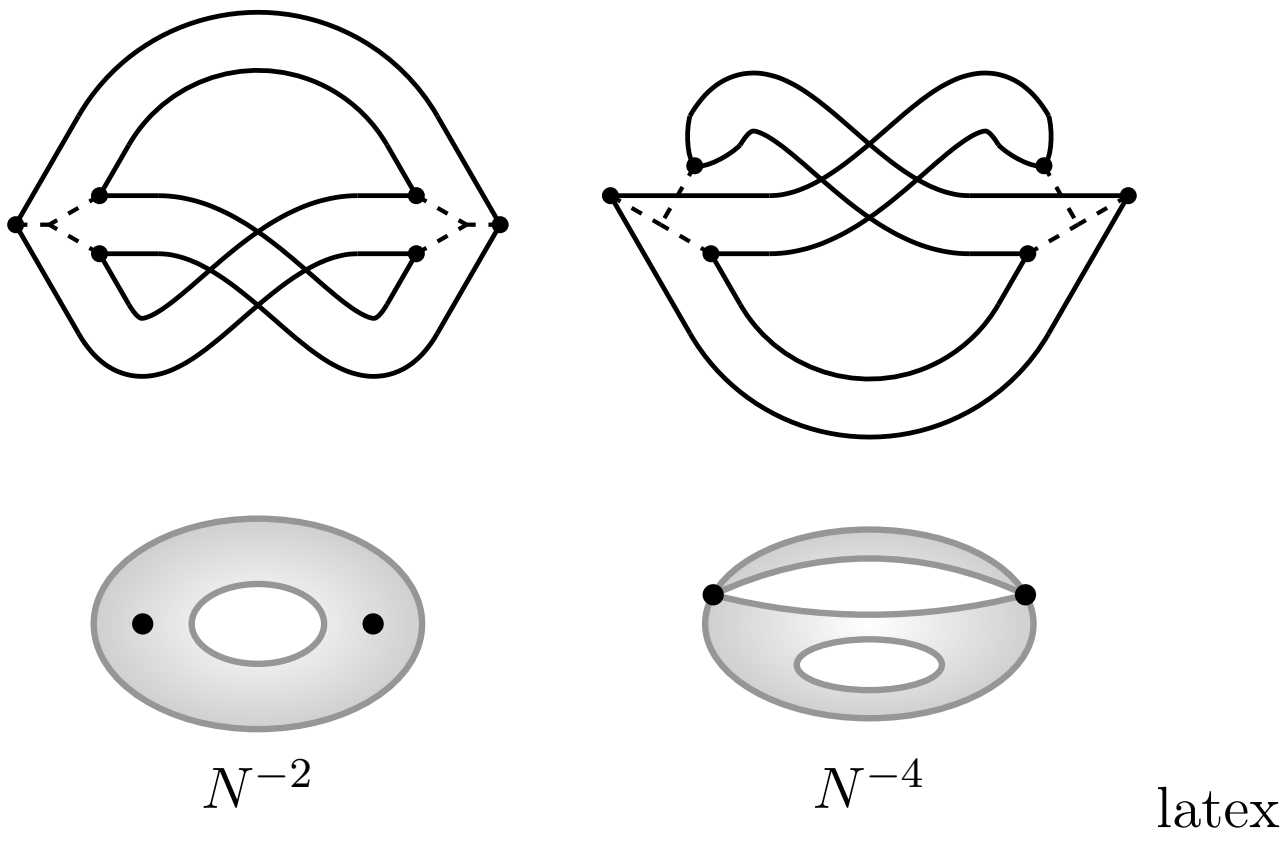

For the $2$-point correlator at lowest order, we obtain a $2$-punctured sphere and a $2$-punctured smooth torus (fig.~\ref{fig:D3nonplanar}, left). Moreover, the planar contractions -- i.e., the contractions without crossed propagators -- of the subleading vertices, either with themselves or with the leading one, give rise to pinched surfaces (fig.~\ref{fig:D3twopoint}): a $1$-pinched torus and a $2$-pinched torus of weight $N^{-2}$, and a maximally pinched double torus of weight $N^{-4}$.\par
Nonplanar contractions of the subleading vertices give rise to a $2$-pinched double torus (fig.~\ref{fig:D3nonplanar}, right).

\subsection{$4$-gluon operators in the fundamental representation}

The $4$-gluon quasi-partonic operators in the fundamental representation read
\begin{equation}\label{eq:O4gaugeinv}
	\mathcal{O}_4 = \mathcal{P}\big(D^{(1)}_+,\ldots,D^{(4)}_+\big)\,\Tr\left(F_{+\perp}\,F_{+\perp}\,F_{+\perp}\,F_{+\perp}\right),
\end{equation}
with each good component $F_{+\perp}$ being either $F$ or $\bar F$ in the fundamental representation. \par
Their color structure is given by
\begin{equation}\label{eq:trac4fund}
	\mathcal{O}_4 = \mathcal{P}\big(D^{(1)}_+,\ldots,D^{(4)}_+\big)\,\Tr(\lambda^a\lambda^b\lambda^c\lambda^d)\,F_{+\perp}^aF_{+\perp}^bF_{+\perp}^cF_{+\perp}^d\,,
\end{equation}
where, by iterating the product of two $\lambda$ matrices,
\begin{align}\label{eq:TrFund4irr}
	\Tr(\lambda^a\lambda^b\lambda^c\lambda^d)=\;&\frac{1}{12N}\left(\delta^{ab}\delta^{cd}+\delta^{ac}\delta^{bd}+\delta^{ad}\delta^{bc}\right)
	+\frac{1}{24}\left(d^{abe}d^{cde}+d^{ace}d^{bde}+d^{ade}d^{bce}\right)\nonumber\\
	&+\frac{1}{12}\left(f^{ade}f^{bce}-f^{abe}f^{cde}\right)+\frac{i}{8}\left(d^{abe}f^{cde}+f^{abe}d^{cde}\right)\,.
\end{align}
Each term in parentheses in eq.~\eqref{eq:TrFund4irr} is irreducible under permutations of the color indices. \par
The term
\begin{equation}
f^{ade}f^{bce}-f^{abe}f^{cde}
 \end{equation}
is a combination of nested commutators, since
\begin{equation}\label{eq:ff}
	f^{abe}f^{cde}=-2\Tr\big(\lambda^a[\lambda^b,[\lambda^c,\lambda^d]]\big)\,,
\end{equation}
so that in the double-line representation it has no $\frac{1}{N}$ corrections by eq.~\eqref{eq:vertexdoublea} and, therefore, gives rise only to 't Hooft smooth topologies. \par
The last term, with 
\begin{equation}
d^{abe}f^{cde}=-2i\Tr\big(\{\lambda^a,\lambda^b\}[\lambda^c,\lambda^d]\big) \,,
\end{equation}
which does not occur in the adjoint representation, is odd under charge conjugation, so that it is the difference of two oriented vertices
\begin{equation}\label{eq:df4}
	i\left(d^{abe}f^{cde}+f^{abe}d^{cde}\right)=4\left[\Tr(\lambda^a\lambda^b\lambda^c\lambda^d)-\Tr(\lambda^d\lambda^c\lambda^b\lambda^a)\right]\,.
\end{equation}'
\par
In the double-line representation
\begin{equation}
(\mathbf{V}^{\rm odd}_4)^{i\,k\,m\,q}_{j\,l\,n\,p}=i\left(d^{abe}f^{cde}+f^{abe}d^{cde}\right)(\lambda^a)^{i}_{j}(\lambda^b)^{k}_{l}(\lambda^c)^{m}_{n}(\lambda^d)^{q}_{p}
\end{equation}
reads (fig.~\ref{fig:F4doubleline})
\begin{align}
	(\mathbf{V}^{\rm odd}_4)^{i\,k\,m\,q}_{j\,l\,n\,p}=\frac{1}{4}\Big(&\delta^i_l\delta^k_n\delta^m_p\delta^q_j-\delta^i_p\delta^q_n\delta^m_l\delta^k_j\nonumber\\
	&-\frac{1}{N}\Big[\delta^i_j\big(\delta^k_n\delta^m_p\delta^q_l-\delta^k_p\delta^q_n\delta^m_l\big)
	+\delta^k_l\big(\delta^i_n\delta^m_p\delta^q_j-\delta^i_p\delta^q_n\delta^m_j\big)\nonumber\\
	&\qquad+\delta^m_n\big(\delta^i_l\delta^k_p\delta^q_j-\delta^i_p\delta^q_l\delta^k_j\big)
	+\delta^q_p\big(\delta^i_l\delta^k_n\delta^m_j-\delta^i_n\delta^m_l\delta^k_j\big)\Big]\Big)\,.
\end{align}
Finally, the first two terms in eq.~\eqref{eq:TrFund4irr} reconstruct the trace of the totally symmetric vertex
\begin{align}\label{eq:trsym4}
	\Tr(\lambda^{(a}\lambda^b\lambda^c\lambda^{d)})=\frac{1}{12N}\left(\delta^{ab}\delta^{cd}+\delta^{ac}\delta^{bd}+\delta^{ad}\delta^{bc}\right)
	+\frac{1}{24}\left(d^{abe}d^{cde}+d^{ace}d^{bde}+d^{ade}d^{bce}\right)\,.
\end{align}
As for $3$-gluon operators, where we considered only $d^{abc}$, we restrict our attention to 
\begin{equation}\label{eq:O4d}
	\mathcal{O}_4 = \mathcal{P}\big(D^{(1)}_+,\ldots,D^{(4)}_+\big)\,\Tr(\lambda^{(a}\lambda^b\lambda^c\lambda^{d)})\,F_{+\perp}^aF_{+\perp}^bF_{+\perp}^cF_{+\perp}^d\,,
\end{equation}
which in the double-line representation,
\begin{equation}
 (\mathbf{V}^{\rm sym}_4)^{i\,k\,m\,q}_{j\,l\,n\,p}=\Tr(\lambda^{(a}\lambda^b\lambda^c\lambda^{d)})(\lambda^a)^{i}_{j}(\lambda^b)^{k}_{l}(\lambda^c)^{m}_{n}(\lambda^d)^{q}_{p} \,,
 \end{equation}
reads (fig.~\ref{fig:D4doubleline})
\begin{align}\label{eq:D4SU}
	(\mathbf{V}^{\rm sym}_4)^{i\,k\,m\,q}_{j\,l\,n\,p}=\frac{1}{96}\Big(&\delta^i_l\delta^k_n\delta^m_p\delta^q_j+\delta^i_l\delta^k_p\delta^q_n\delta^m_j+\delta^i_n\delta^m_l\delta^k_p\delta^q_j+\delta^i_n\delta^m_p\delta^q_l\delta^k_j+\delta^i_p\delta^q_l\delta^k_n\delta^m_j+\delta^i_p\delta^q_n\delta^m_l\delta^k_j\nonumber\\
	&-\frac{3}{N}\Big[\delta^i_j\big(\delta^k_n\delta^m_p\delta^q_l+\delta^k_p\delta^q_n\delta^m_l\big)
	+\delta^k_l\big(\delta^i_n\delta^m_p\delta^q_j+\delta^i_p\delta^q_n\delta^m_j\big)\nonumber\\
	&\qquad+\delta^m_n\big(\delta^i_l\delta^k_p\delta^q_j+\delta^i_p\delta^q_l\delta^k_j\big)
	+\delta^q_p\big(\delta^i_l\delta^k_n\delta^m_j+\delta^i_n\delta^m_l\delta^k_j\big)\Big]\nonumber\\
	&+\frac{6}{N^2}\big(\delta^i_j\delta^k_l\delta^m_p\delta^q_n+\delta^i_j\delta^m_n\delta^k_p\delta^q_l+\delta^i_j\delta^q_p\delta^k_n\delta^m_l\nonumber\\
	&\qquad+\delta^k_l\delta^m_n\delta^i_p\delta^q_j+\delta^k_l\delta^q_p\delta^i_n\delta^m_j+\delta^m_n\delta^q_p\delta^i_l\delta^k_j\big)
	-\frac{18}{N^3}\,\delta^i_j\delta^k_l\delta^m_n\delta^q_p\Big)\,.
\end{align}
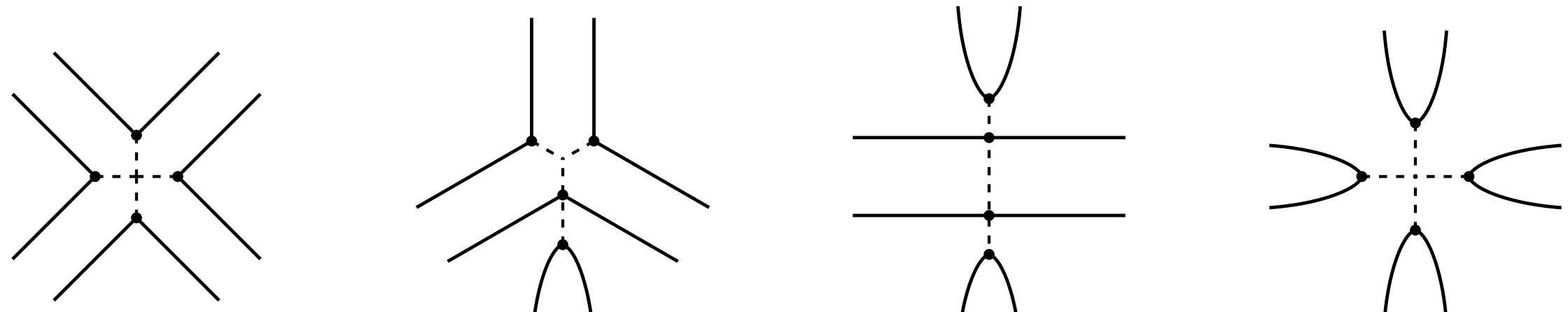
\begin{figure}[h!]
	\centering
	\begin{tikzpicture}[baseline={(0,-0.1)}, scale=0.92]
		\path (-1.65,-2.5) rectangle (1.65,1.75);
		\foreach \p in {(0,0.4243),(-0.4243,0),(0,-0.4243),(0.4243,0)}{\draw[pinch] \p -- (0,0);}
		\draw[skel] (0.8485,1.2728) -- (0,0.4243) -- (-0.8485,1.2728);
		\draw[skel] (-1.2728,0.8485) -- (-0.4243,0) -- (-1.2728,-0.8485);
		\draw[skel] (-0.8485,-1.2728) -- (0,-0.4243) -- (0.8485,-1.2728);
		\draw[skel] (1.2728,-0.8485) -- (0.4243,0) -- (1.2728,0.8485);
		\fill (0,0.4243) circle (1.6pt);
		\fill (-0.4243,0) circle (1.6pt);
		\fill (0,-0.4243) circle (1.6pt);
		\fill (0.4243,0) circle (1.6pt);
		\node at (0,-2.1) {$N^0$};
	\end{tikzpicture}\hfill
	\begin{tikzpicture}[baseline={(0,-0.1)}, scale=0.92]
		\path (-1.65,-2.5) rectangle (1.65,1.75);
		\begin{scope}[shift={(0,0.18)}]
			\draw[pinch] (0,-0.88) -- (0,0);
			\draw[pinch] (-0.32,0.1848) -- (0,0);
			\draw[pinch] (0.32,0.1848) -- (0,0);
			\draw[skel] (-0.32,1.45) -- (-0.32,0.1848) -- (-1.5023,-0.4979);
			\draw[skel] (-1.1823,-1.0521) -- (0,-0.3695) -- (1.1823,-1.0521);
			\draw[skel] (1.5023,-0.4979) -- (0.32,0.1848) -- (0.32,1.45);
			\fill (-0.32,0.1848) circle (1.6pt);
			\fill (0,-0.3695) circle (1.6pt);
			\fill (0.32,0.1848) circle (1.6pt);
			\ophairpin{270}{0.88}{1.0}
		\end{scope}
		\node at (0,-2.1) {$N^{-1}$};
	\end{tikzpicture}\hfill
	\begin{tikzpicture}[baseline={(0,-0.1)}, scale=0.92]
		\path (-1.65,-2.5) rectangle (1.65,1.75);
		\draw[pinch] (0,0.8) -- (0,-0.8);
		\draw[skel] (-1.4,0.4) -- (1.4,0.4);
		\draw[skel] (-1.4,-0.4) -- (1.4,-0.4);
		\fill (0,0.4) circle (1.6pt);
		\fill (0,-0.4) circle (1.6pt);
		\ophairpin{90}{0.8}{0.95}
		\ophairpin{270}{0.8}{0.95}
		\node at (0,-2.1) {$N^{-2}$};
	\end{tikzpicture}\hfill
	\begin{tikzpicture}[baseline={(0,-0.1)}, scale=0.92]
		\path (-1.65,-2.5) rectangle (1.65,1.75);
		\foreach \a in {0,90,180,270}{\draw[pinch] (\a:0.55) -- (0,0);}
		\ophairpin{0}{0.55}{0.95}
		\ophairpin{90}{0.55}{0.95}
		\ophairpin{180}{0.55}{0.95}
		\ophairpin{270}{0.55}{0.95}
		\node at (0,-2.1) {$N^{-3}$};
	\end{tikzpicture}
	\caption{Double-line representation of the operator vertex $(\mathbf{V}^{\rm sym}_4)^{i\,k\,m\,q}_{j\,l\,n\,p}$:  one diagram is shown for each $\frac{1}{N}$ order, involving the extra single-, double-, and maximally-pinched vertices, with the remaining contributions obtained by permutations of the legs.}
	\label{fig:D4doubleline}
\end{figure}
As an example, we list some of the topologies occurring in the planar contractions of the vertices in fig.~\ref{fig:D4doubleline}. \par
The leading vertex gives rise to the standard smooth 't Hooft topologies of the $2$-point correlator, i.e., a $2$-punctured sphere and a $2$-punctured torus.\par
Planar contractions of the subleading vertices, either with themselves or with the leading one, give rise to pinched surfaces (fig.~\ref{fig:D4twopoint}): $1$- and $2$-pinched tori of weight $N^{-2}$, a maximally pinched double torus of weight $N^{-4}$, and a maximally pinched triple torus of weight $N^{-6}$.\par
Nonplanar contractions of the subleading vertices give rise to a $2$-pinched double torus (fig.~\ref{fig:D4nonplanar}).
\begin{figure}[h!]
		\centering
		\begin{tikzpicture}[scale=0.40,
			surf/.style={c969696, line width=1.2pt}]
			\begin{scope}[shift={(0.00,0.00)}]
				\begin{scope}[shift={(-2.900,0)}]
					\draw[skel] (1.127,1.312) -- (0.717,0.601) -- (1.488,0.882);
					\draw[skel] (1.707,0.281) -- (0.936,0.000) -- (1.707,-0.281);
					\draw[skel] (1.488,-0.882) -- (0.717,-0.601) -- (1.127,-1.312);
					\draw[skel] (0.573,1.632) -- (-0.370,0.000) -- (0.573,-1.632);
					\draw[pinch] (0.717,0.601) -- (0,0);
					\fill (0.717,0.601) circle (4.00pt);
					\draw[pinch] (0.936,0.000) -- (0,0);
					\fill (0.936,0.000) circle (4.00pt);
					\draw[pinch] (0.717,-0.601) -- (0,0);
					\fill (0.717,-0.601) circle (4.00pt);
					\draw[pinch] (-0.370,0.000) -- (0,0);
					\fill (-0.370,0.000) circle (4.00pt);
				\end{scope}
				\begin{scope}[shift={(2.900,0)}, xscale=-1]
					\draw[skel] (1.127,1.312) -- (0.717,0.601) -- (1.488,0.882);
					\draw[skel] (1.707,0.281) -- (0.936,0.000) -- (1.707,-0.281);
					\draw[skel] (1.488,-0.882) -- (0.717,-0.601) -- (1.127,-1.312);
					\draw[skel] (0.573,1.632) -- (-0.370,0.000) -- (0.573,-1.632);
					\draw[pinch] (0.717,0.601) -- (0,0);
					\fill (0.717,0.601) circle (4.00pt);
					\draw[pinch] (0.936,0.000) -- (0,0);
					\fill (0.936,0.000) circle (4.00pt);
					\draw[pinch] (0.717,-0.601) -- (0,0);
					\fill (0.717,-0.601) circle (4.00pt);
					\draw[pinch] (-0.370,0.000) -- (0,0);
					\fill (-0.370,0.000) circle (4.00pt);
				\end{scope}
				\draw[skel] (-2.327,1.632) arc[start angle=150.000, end angle=30.000, radius=2.687];
				\draw[skel] (-1.773,1.312) arc[start angle=150.000, end angle=30.000, radius=2.047];
				\draw[skel] (-1.412,0.882) arc[start angle=110.000, end angle=70.000, radius=4.128];
				\draw[skel] (-1.193,0.281) arc[start angle=110.000, end angle=70.000, radius=3.488];
				\draw[skel] (-1.193,-0.281) arc[start angle=-110.000, end angle=-70.000, radius=3.488];
				\draw[skel] (-1.412,-0.882) arc[start angle=-110.000, end angle=-70.000, radius=4.128];
				\draw[skel] (-1.773,-1.312) arc[start angle=-150.000, end angle=-30.000, radius=2.047];
				\draw[skel] (-2.327,-1.632) arc[start angle=-150.000, end angle=-30.000, radius=2.687];
			\end{scope}
			\begin{scope}[shift={(7.44,0.00)}]
				\begin{scope}[shift={(-2.900,0)}]
					\draw[skel] (1.707,0.281) -- (0.936,0.000) -- (1.707,-0.281);
					\draw[skel] (1.488,-0.882) -- (0.717,-0.601) -- (1.127,-1.312);
					\draw[skel] (1.488,0.882) -- (-0.468,0.170) -- (0.573,-1.632);
					\begin{scope}[shift={(60:1.200)}, rotate=-30]
						\draw[skel] (-0.320,0.500) .. controls (-0.280,0.325) and (-0.12,0.03) .. (0,0) .. controls (0.12,0.03) and (0.280,0.325) .. (0.320,0.500);
						\fill (0,0) circle (4.00pt);
						\draw[pinch] (0,0) -- (0,-1.200);
					\end{scope}
					\draw[pinch] (0.936,0.000) -- (0,0);
					\fill (0.936,0.000) circle (4.00pt);
					\draw[pinch] (0.717,-0.601) -- (0,0);
					\fill (0.717,-0.601) circle (4.00pt);
					\draw[pinch] (-0.468,0.170) -- (0,0);
					\fill (-0.468,0.170) circle (4.00pt);
				\end{scope}
				\begin{scope}[shift={(2.900,0)}, xscale=-1]
					\draw[skel] (1.127,1.312) -- (0.717,0.601) -- (1.488,0.882);
					\draw[skel] (1.707,0.281) -- (0.936,0.000) -- (1.707,-0.281);
					\draw[skel] (1.488,-0.882) -- (0.717,-0.601) -- (1.127,-1.312);
					\draw[skel] (0.573,1.632) -- (-0.370,0.000) -- (0.573,-1.632);
					\draw[pinch] (0.717,0.601) -- (0,0);
					\fill (0.717,0.601) circle (4.00pt);
					\draw[pinch] (0.936,0.000) -- (0,0);
					\fill (0.936,0.000) circle (4.00pt);
					\draw[pinch] (0.717,-0.601) -- (0,0);
					\fill (0.717,-0.601) circle (4.00pt);
					\draw[pinch] (-0.370,0.000) -- (0,0);
					\fill (-0.370,0.000) circle (4.00pt);
				\end{scope}
				\draw[skel] (-2.327,1.632) arc[start angle=150.000, end angle=30.000, radius=2.687];
				\draw[skel] (-1.773,1.312) arc[start angle=150.000, end angle=30.000, radius=2.047];
				\draw[skel] (-1.412,0.882) arc[start angle=110.000, end angle=70.000, radius=4.128];
				\draw[skel] (-1.193,0.281) arc[start angle=110.000, end angle=70.000, radius=3.488];
				\draw[skel] (-1.193,-0.281) arc[start angle=-110.000, end angle=-70.000, radius=3.488];
				\draw[skel] (-1.412,-0.882) arc[start angle=-110.000, end angle=-70.000, radius=4.128];
				\draw[skel] (-1.773,-1.312) arc[start angle=-150.000, end angle=-30.000, radius=2.047];
				\draw[skel] (-2.327,-1.632) arc[start angle=-150.000, end angle=-30.000, radius=2.687];
			\end{scope}
			\begin{scope}[shift={(14.88,0.00)}]
				\begin{scope}[shift={(-2.900,0)}]
					\draw[skel] (1.707,0.281) -- (0.936,0.000) -- (1.707,-0.281);
					\draw[skel] (1.488,-0.882) -- (0.717,-0.601) -- (1.127,-1.312);
					\draw[skel] (1.488,0.882) -- (-0.468,0.170) -- (0.573,-1.632);
					\begin{scope}[shift={(60:1.200)}, rotate=-30]
						\draw[skel] (-0.320,0.500) .. controls (-0.280,0.325) and (-0.12,0.03) .. (0,0) .. controls (0.12,0.03) and (0.280,0.325) .. (0.320,0.500);
						\fill (0,0) circle (4.00pt);
						\draw[pinch] (0,0) -- (0,-1.200);
					\end{scope}
					\draw[pinch] (0.936,0.000) -- (0,0);
					\fill (0.936,0.000) circle (4.00pt);
					\draw[pinch] (0.717,-0.601) -- (0,0);
					\fill (0.717,-0.601) circle (4.00pt);
					\draw[pinch] (-0.468,0.170) -- (0,0);
					\fill (-0.468,0.170) circle (4.00pt);
				\end{scope}
				\begin{scope}[shift={(2.900,0)}, xscale=-1]
					\draw[skel] (1.707,0.281) -- (0.936,0.000) -- (1.707,-0.281);
					\draw[skel] (1.488,-0.882) -- (0.717,-0.601) -- (1.127,-1.312);
					\draw[skel] (1.488,0.882) -- (-0.468,0.170) -- (0.573,-1.632);
					\begin{scope}[shift={(60:1.200)}, rotate=-30]
						\draw[skel] (-0.320,0.500) .. controls (-0.280,0.325) and (-0.12,0.03) .. (0,0) .. controls (0.12,0.03) and (0.280,0.325) .. (0.320,0.500);
						\fill (0,0) circle (4.00pt);
						\draw[pinch] (0,0) -- (0,-1.200);
					\end{scope}
					\draw[pinch] (0.936,0.000) -- (0,0);
					\fill (0.936,0.000) circle (4.00pt);
					\draw[pinch] (0.717,-0.601) -- (0,0);
					\fill (0.717,-0.601) circle (4.00pt);
					\draw[pinch] (-0.468,0.170) -- (0,0);
					\fill (-0.468,0.170) circle (4.00pt);
				\end{scope}
				\draw[skel] (-2.327,1.632) arc[start angle=150.000, end angle=30.000, radius=2.687];
				\draw[skel] (-1.773,1.312) arc[start angle=150.000, end angle=30.000, radius=2.047];
				\draw[skel] (-1.412,0.882) arc[start angle=110.000, end angle=70.000, radius=4.128];
				\draw[skel] (-1.193,0.281) arc[start angle=110.000, end angle=70.000, radius=3.488];
				\draw[skel] (-1.193,-0.281) arc[start angle=-110.000, end angle=-70.000, radius=3.488];
				\draw[skel] (-1.412,-0.882) arc[start angle=-110.000, end angle=-70.000, radius=4.128];
				\draw[skel] (-1.773,-1.312) arc[start angle=-150.000, end angle=-30.000, radius=2.047];
				\draw[skel] (-2.327,-1.632) arc[start angle=-150.000, end angle=-30.000, radius=2.687];
			\end{scope}
			\begin{scope}[shift={(22.32,0.00)}]
				\begin{scope}[shift={(-2.900,0)}]
					\draw[skel] (1.488,-0.882) -- (0.717,-0.601) -- (1.127,-1.312);
					\draw[skel] (1.707,-0.281) -- (-0.717,0.601) -- (0.573,-1.632);
					\begin{scope}[shift={(60:1.200)}, rotate=-30]
						\draw[skel] (-0.320,0.500) .. controls (-0.280,0.325) and (-0.12,0.03) .. (0,0) .. controls (0.12,0.03) and (0.280,0.325) .. (0.320,0.500);
						\fill (0,0) circle (4.00pt);
						\draw[pinch] (0,0) -- (0,-1.200);
					\end{scope}
					\begin{scope}[shift={(20:1.200)}, rotate=-70]
						\draw[skel] (-0.320,0.500) .. controls (-0.280,0.325) and (-0.12,0.03) .. (0,0) .. controls (0.12,0.03) and (0.280,0.325) .. (0.320,0.500);
						\fill (0,0) circle (4.00pt);
						\draw[pinch] (0,0) -- (0,-1.200);
					\end{scope}
					\draw[pinch] (0.717,-0.601) -- (0,0);
					\fill (0.717,-0.601) circle (4.00pt);
					\draw[pinch] (-0.717,0.601) -- (0,0);
					\fill (-0.717,0.601) circle (4.00pt);
				\end{scope}
				\begin{scope}[shift={(2.900,0)}, xscale=-1]
					\draw[skel] (1.488,-0.882) -- (0.717,-0.601) -- (1.127,-1.312);
					\draw[skel] (1.707,-0.281) -- (-0.717,0.601) -- (0.573,-1.632);
					\begin{scope}[shift={(60:1.200)}, rotate=-30]
						\draw[skel] (-0.320,0.500) .. controls (-0.280,0.325) and (-0.12,0.03) .. (0,0) .. controls (0.12,0.03) and (0.280,0.325) .. (0.320,0.500);
						\fill (0,0) circle (4.00pt);
						\draw[pinch] (0,0) -- (0,-1.200);
					\end{scope}
					\begin{scope}[shift={(20:1.200)}, rotate=-70]
						\draw[skel] (-0.320,0.500) .. controls (-0.280,0.325) and (-0.12,0.03) .. (0,0) .. controls (0.12,0.03) and (0.280,0.325) .. (0.320,0.500);
						\fill (0,0) circle (4.00pt);
						\draw[pinch] (0,0) -- (0,-1.200);
					\end{scope}
					\draw[pinch] (0.717,-0.601) -- (0,0);
					\fill (0.717,-0.601) circle (4.00pt);
					\draw[pinch] (-0.717,0.601) -- (0,0);
					\fill (-0.717,0.601) circle (4.00pt);
				\end{scope}
				\draw[skel] (-2.327,1.632) arc[start angle=150.000, end angle=30.000, radius=2.687];
				\draw[skel] (-1.773,1.312) arc[start angle=150.000, end angle=30.000, radius=2.047];
				\draw[skel] (-1.412,0.882) arc[start angle=110.000, end angle=70.000, radius=4.128];
				\draw[skel] (-1.193,0.281) arc[start angle=110.000, end angle=70.000, radius=3.488];
				\draw[skel] (-1.193,-0.281) arc[start angle=-110.000, end angle=-70.000, radius=3.488];
				\draw[skel] (-1.412,-0.882) arc[start angle=-110.000, end angle=-70.000, radius=4.128];
				\draw[skel] (-1.773,-1.312) arc[start angle=-150.000, end angle=-30.000, radius=2.047];
				\draw[skel] (-2.327,-1.632) arc[start angle=-150.000, end angle=-30.000, radius=2.687];
			\end{scope}
			\begin{scope}[shift={(29.76,0.00)}]
				\begin{scope}[shift={(-2.900,0)}]
					\begin{scope}[shift={(60:1.200)}, rotate=-30]
						\draw[skel] (-0.320,0.500) .. controls (-0.280,0.325) and (-0.12,0.03) .. (0,0) .. controls (0.12,0.03) and (0.280,0.325) .. (0.320,0.500);
						\fill (0,0) circle (4.00pt);
						\draw[pinch] (0,0) -- (0,-1.200);
					\end{scope}
					\begin{scope}[shift={(20:1.200)}, rotate=-70]
						\draw[skel] (-0.320,0.500) .. controls (-0.280,0.325) and (-0.12,0.03) .. (0,0) .. controls (0.12,0.03) and (0.280,0.325) .. (0.320,0.500);
						\fill (0,0) circle (4.00pt);
						\draw[pinch] (0,0) -- (0,-1.200);
					\end{scope}
					\begin{scope}[shift={(-20:1.200)}, rotate=-110]
						\draw[skel] (-0.320,0.500) .. controls (-0.280,0.325) and (-0.12,0.03) .. (0,0) .. controls (0.12,0.03) and (0.280,0.325) .. (0.320,0.500);
						\fill (0,0) circle (4.00pt);
						\draw[pinch] (0,0) -- (0,-1.200);
					\end{scope}
					\begin{scope}[shift={(-60:1.200)}, rotate=-150]
						\draw[skel] (-0.320,0.500) .. controls (-0.280,0.325) and (-0.12,0.03) .. (0,0) .. controls (0.12,0.03) and (0.280,0.325) .. (0.320,0.500);
						\fill (0,0) circle (4.00pt);
						\draw[pinch] (0,0) -- (0,-1.200);
					\end{scope}
				\end{scope}
				\begin{scope}[shift={(2.900,0)}, xscale=-1]
					\begin{scope}[shift={(60:1.200)}, rotate=-30]
						\draw[skel] (-0.320,0.500) .. controls (-0.280,0.325) and (-0.12,0.03) .. (0,0) .. controls (0.12,0.03) and (0.280,0.325) .. (0.320,0.500);
						\fill (0,0) circle (4.00pt);
						\draw[pinch] (0,0) -- (0,-1.200);
					\end{scope}
					\begin{scope}[shift={(20:1.200)}, rotate=-70]
						\draw[skel] (-0.320,0.500) .. controls (-0.280,0.325) and (-0.12,0.03) .. (0,0) .. controls (0.12,0.03) and (0.280,0.325) .. (0.320,0.500);
						\fill (0,0) circle (4.00pt);
						\draw[pinch] (0,0) -- (0,-1.200);
					\end{scope}
					\begin{scope}[shift={(-20:1.200)}, rotate=-110]
						\draw[skel] (-0.320,0.500) .. controls (-0.280,0.325) and (-0.12,0.03) .. (0,0) .. controls (0.12,0.03) and (0.280,0.325) .. (0.320,0.500);
						\fill (0,0) circle (4.00pt);
						\draw[pinch] (0,0) -- (0,-1.200);
					\end{scope}
					\begin{scope}[shift={(-60:1.200)}, rotate=-150]
						\draw[skel] (-0.320,0.500) .. controls (-0.280,0.325) and (-0.12,0.03) .. (0,0) .. controls (0.12,0.03) and (0.280,0.325) .. (0.320,0.500);
						\fill (0,0) circle (4.00pt);
						\draw[pinch] (0,0) -- (0,-1.200);
					\end{scope}
				\end{scope}
				\draw[skel] (-2.327,1.632) arc[start angle=150.000, end angle=30.000, radius=2.687];
				\draw[skel] (-1.773,1.312) arc[start angle=150.000, end angle=30.000, radius=2.047];
				\draw[skel] (-1.412,0.882) arc[start angle=110.000, end angle=70.000, radius=4.128];
				\draw[skel] (-1.193,0.281) arc[start angle=110.000, end angle=70.000, radius=3.488];
				\draw[skel] (-1.193,-0.281) arc[start angle=-110.000, end angle=-70.000, radius=3.488];
				\draw[skel] (-1.412,-0.882) arc[start angle=-110.000, end angle=-70.000, radius=4.128];
				\draw[skel] (-1.773,-1.312) arc[start angle=-150.000, end angle=-30.000, radius=2.047];
				\draw[skel] (-2.327,-1.632) arc[start angle=-150.000, end angle=-30.000, radius=2.687];
			\end{scope}
			\begin{scope}[shift={(0.00,-7.14)}]
				\shade[inner color=white, outer color=gray!38] (0,0) circle (1.768);
				\draw[surf] (0,0) circle (1.768);
				\fill (-0.806,0) circle (5.00pt); \fill (0.806,0) circle (5.00pt);
			\end{scope}
			\node at (0.00,-10.96) {$N^{0}$};
			\begin{scope}[shift={(7.44,-7.14)}]
				\shade[inner color=white, outer color=gray!38, even odd rule] (0,0) ellipse (1.456 and 1.794) (-0.494,0) ellipse (0.962 and 1.300);
				\draw[surf] (0,0) ellipse (1.456 and 1.794);
				\draw[surf] (-0.494,0) ellipse (0.962 and 1.300);
				\fill (-1.456,0) circle (5.00pt); \fill (0.858,0) circle (5.00pt);
			\end{scope}
			\node at (7.44,-10.96) {$N^{-2}$};
			\begin{scope}[shift={(14.88,-7.14)}]
				\shade[inner color=white, outer color=gray!38] (-1.326,0) .. controls (-0.728,1.638) and (0.728,1.638) .. (1.326,0) .. controls (0.728,0.286) and (-0.728,0.286) .. cycle;
				\shade[inner color=white, outer color=gray!38] (-1.326,0) .. controls (-0.728,-0.286) and (0.728,-0.286) .. (1.326,0) .. controls (0.728,-1.638) and (-0.728,-1.638) .. cycle;
				\draw[surf] (-1.326,0) .. controls (-0.728,1.638) and (0.728,1.638) .. (1.326,0);
				\draw[surf] (-1.326,0) .. controls (-0.728,0.286) and (0.728,0.286) .. (1.326,0);
				\draw[surf] (-1.326,0) .. controls (-0.728,-0.286) and (0.728,-0.286) .. (1.326,0);
				\draw[surf] (-1.326,0) .. controls (-0.728,-1.638) and (0.728,-1.638) .. (1.326,0);
				\fill (-1.326,0) circle (5.00pt); \fill (1.326,0) circle (5.00pt);
			\end{scope}
			\node at (14.88,-10.96) {$N^{-2}$};
			\begin{scope}[shift={(22.32,-7.14)}]
				\shade[inner color=white, outer color=gray!38] (-1.326,0) .. controls (-0.728,2.080) and (0.728,2.080) .. (1.326,0) .. controls (0.728,1.040) and (-0.728,1.040) .. cycle;
				\shade[inner color=white, outer color=gray!38] (-1.326,0) .. controls (-0.728,0.598) and (0.728,0.598) .. (1.326,0) .. controls (0.728,-0.598) and (-0.728,-0.598) .. cycle;
				\shade[inner color=white, outer color=gray!38] (-1.326,0) .. controls (-0.728,-1.040) and (0.728,-1.040) .. (1.326,0) .. controls (0.728,-2.080) and (-0.728,-2.080) .. cycle;
				\draw[surf] (-1.326,0) .. controls (-0.728,2.080) and (0.728,2.080) .. (1.326,0);
				\draw[surf] (-1.326,0) .. controls (-0.728,1.040) and (0.728,1.040) .. (1.326,0);
				\draw[surf] (-1.326,0) .. controls (-0.728,0.598) and (0.728,0.598) .. (1.326,0);
				\draw[surf] (-1.326,0) .. controls (-0.728,-0.598) and (0.728,-0.598) .. (1.326,0);
				\draw[surf] (-1.326,0) .. controls (-0.728,-1.040) and (0.728,-1.040) .. (1.326,0);
				\draw[surf] (-1.326,0) .. controls (-0.728,-2.080) and (0.728,-2.080) .. (1.326,0);
				\fill (-1.326,0) circle (5.00pt); \fill (1.326,0) circle (5.00pt);
			\end{scope}
			\node at (22.32,-10.96) {$N^{-4}$};
			\begin{scope}[shift={(29.76,-7.14)}]
				\shade[inner color=white, outer color=gray!38] (-1.326,0) .. controls (-0.728,2.522) and (0.728,2.522) .. (1.326,0) .. controls (0.728,1.534) and (-0.728,1.534) .. cycle;
				\shade[inner color=white, outer color=gray!38] (-1.326,0) .. controls (-0.728,1.144) and (0.728,1.144) .. (1.326,0) .. controls (0.728,0.208) and (-0.728,0.208) .. cycle;
				\shade[inner color=white, outer color=gray!38] (-1.326,0) .. controls (-0.728,-0.208) and (0.728,-0.208) .. (1.326,0) .. controls (0.728,-1.144) and (-0.728,-1.144) .. cycle;
				\shade[inner color=white, outer color=gray!38] (-1.326,0) .. controls (-0.728,-1.534) and (0.728,-1.534) .. (1.326,0) .. controls (0.728,-2.522) and (-0.728,-2.522) .. cycle;
				\draw[surf] (-1.326,0) .. controls (-0.728,2.522) and (0.728,2.522) .. (1.326,0);
				\draw[surf] (-1.326,0) .. controls (-0.728,1.534) and (0.728,1.534) .. (1.326,0);
				\draw[surf] (-1.326,0) .. controls (-0.728,1.144) and (0.728,1.144) .. (1.326,0);
				\draw[surf] (-1.326,0) .. controls (-0.728,0.208) and (0.728,0.208) .. (1.326,0);
				\draw[surf] (-1.326,0) .. controls (-0.728,-0.208) and (0.728,-0.208) .. (1.326,0);
				\draw[surf] (-1.326,0) .. controls (-0.728,-1.144) and (0.728,-1.144) .. (1.326,0);
				\draw[surf] (-1.326,0) .. controls (-0.728,-1.534) and (0.728,-1.534) .. (1.326,0);
				\draw[surf] (-1.326,0) .. controls (-0.728,-2.522) and (0.728,-2.522) .. (1.326,0);
				\fill (-1.326,0) circle (5.00pt); \fill (1.326,0) circle (5.00pt);
			\end{scope}
			\node at (29.76,-10.96) {$N^{-6}$};
		\end{tikzpicture}
		\caption{Planar contractions for the $2$-point correlator of $4$-gluon operators in the fundamental representation at lowest order (top), and the corresponding surfaces with their weights (bottom): contractions of leading vertices give rise to a sphere with two punctures; contractions involving a single-pinched vertex give rise to a torus with a cycle pinched at a puncture; contractions of two single-pinched vertices via the same propagator give rise to a doubly pinched torus; contractions of two double-pinched vertices give rise to a maximally pinched double torus; and contractions of maximally pinched vertices give rise to a maximally pinched triple torus. Although the diagrams are planar, the pinched surfaces are degenerations of higher-genus ones.}
				\label{fig:D4twopoint}
		\label{fig:D4joined}
	\end{figure}
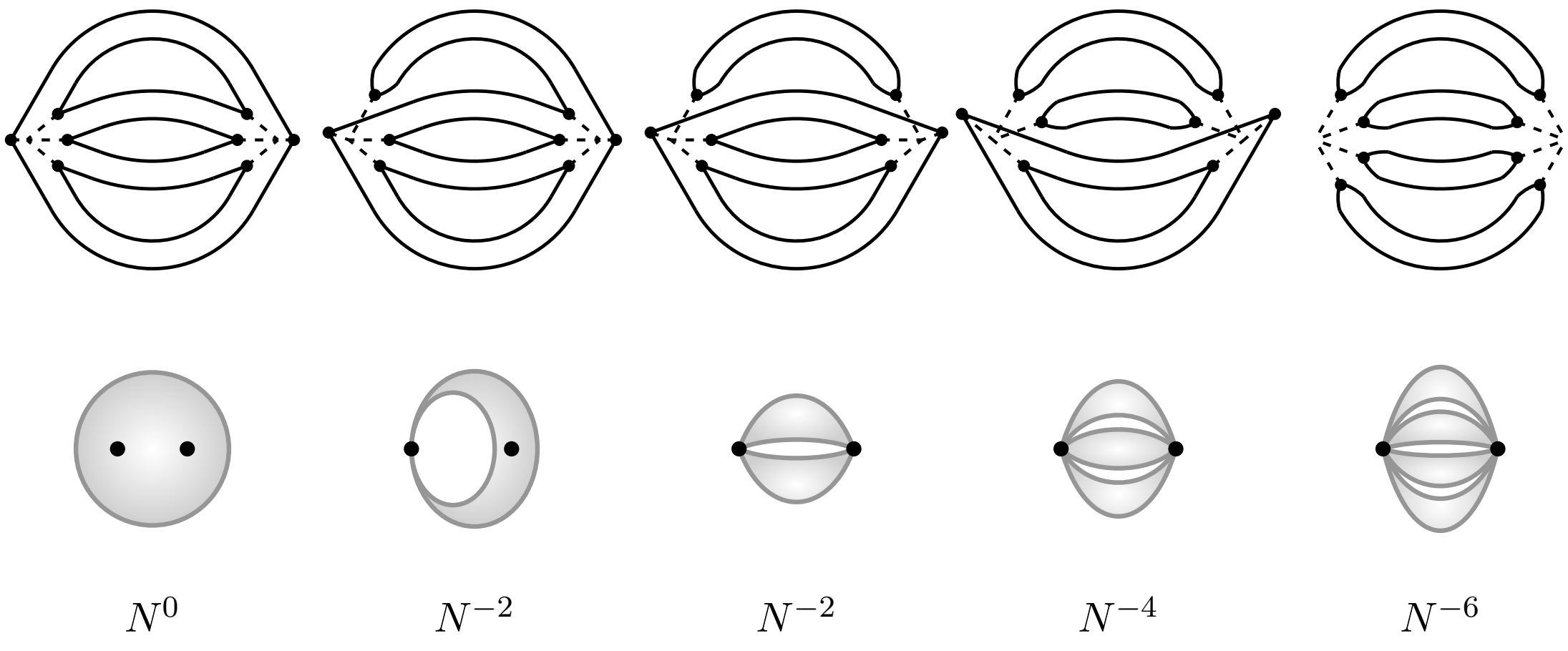
	\begin{figure}[h!]
		\centering
		\begin{tikzpicture}[scale=0.55,
			surf/.style={c969696, line width=1.2pt},
			ident/.style={line width=0.8pt, black, dash pattern=on 0.9pt off 1.4pt}]
			\begin{scope}[shift={(0.00,0.00)}]
				\begin{scope}[shift={(-2.900,0)}]
					\draw[skel] (1.707,0.281) -- (0.936,0.000) -- (1.707,-0.281);
					\draw[skel] (1.488,-0.882) -- (0.717,-0.601) -- (1.127,-1.312);
					\draw[skel] (1.488,0.882) -- (-0.468,0.170) -- (0.573,-1.632);
					\begin{scope}[shift={(60:1.200)}, rotate=-30]
						\draw[skel] (-0.320,0.500) .. controls (-0.280,0.325) and (-0.12,0.03) .. (0,0) .. controls (0.12,0.03) and (0.280,0.325) .. (0.320,0.500);
						\fill (0,0) circle (2.91pt);
						\draw[pinch] (0,0) -- (0,-1.200);
					\end{scope}
					\draw[pinch] (0.936,0.000) -- (0,0);
					\fill (0.936,0.000) circle (2.91pt);
					\draw[pinch] (0.717,-0.601) -- (0,0);
					\fill (0.717,-0.601) circle (2.91pt);
					\draw[pinch] (-0.468,0.170) -- (0,0);
					\fill (-0.468,0.170) circle (2.91pt);
				\end{scope}
				\begin{scope}[shift={(2.900,0)}, xscale=-1]
					\draw[skel] (1.127,1.312) -- (0.717,0.601) -- (1.488,0.882);
					\draw[skel] (1.707,0.281) -- (0.936,0.000) -- (1.707,-0.281);
					\draw[skel] (1.488,-0.882) -- (0.717,-0.601) -- (1.127,-1.312);
					\draw[skel] (0.573,1.632) -- (-0.370,0.000) -- (0.573,-1.632);
					\draw[pinch] (0.717,0.601) -- (0,0);
					\fill (0.717,0.601) circle (2.91pt);
					\draw[pinch] (0.936,0.000) -- (0,0);
					\fill (0.936,0.000) circle (2.91pt);
					\draw[pinch] (0.717,-0.601) -- (0,0);
					\fill (0.717,-0.601) circle (2.91pt);
					\draw[pinch] (-0.370,0.000) -- (0,0);
					\fill (-0.370,0.000) circle (2.91pt);
				\end{scope}
				\draw[skel] (-2.327,1.632) arc[start angle=150.000, end angle=30.000, radius=2.687];
				\draw[skel] (-1.773,1.312) arc[start angle=150.000, end angle=30.000, radius=2.047];
				\draw[skel] (-1.773,-1.312) arc[start angle=-150.000, end angle=-30.000, radius=2.047];
				\draw[skel] (-2.327,-1.632) arc[start angle=-150.000, end angle=-30.000, radius=2.687];
				\draw[skel] (-1.412,0.882) -- (-1.346,0.905) -- (-1.281,0.925) -- (-1.218,0.942) -- (-1.155,0.957) -- (-1.093,0.968) -- (-1.031,0.977) -- (-0.971,0.983) -- (-0.911,0.987) -- (-0.853,0.987) -- (-0.795,0.985) -- (-0.738,0.979) -- (-0.682,0.971) -- (-0.628,0.960) -- (-0.575,0.947) -- (-0.523,0.930) -- (-0.473,0.912) -- (-0.425,0.891) -- (-0.379,0.868) -- (-0.334,0.843) -- (-0.291,0.816) -- (-0.250,0.788) -- (-0.211,0.759) -- (-0.174,0.728) -- (-0.138,0.697) -- (-0.104,0.665) -- (-0.072,0.632) -- (-0.041,0.598) -- (-0.011,0.565) -- (0.018,0.530) -- (0.045,0.496) -- (0.072,0.462) -- (0.098,0.427) -- (0.122,0.393) -- (0.146,0.358) -- (0.170,0.324) -- (0.193,0.290) -- (0.215,0.256) -- (0.237,0.223) -- (0.258,0.190) -- (0.279,0.158) -- (0.300,0.126) -- (0.320,0.095) -- (0.340,0.064) -- (0.360,0.034) -- (0.380,0.005) -- (0.400,-0.023) -- (0.419,-0.050) -- (0.438,-0.076) -- (0.458,-0.101) -- (0.477,-0.124) -- (0.496,-0.147) -- (0.515,-0.168) -- (0.533,-0.188) -- (0.552,-0.207) -- (0.570,-0.224) -- (0.588,-0.240) -- (0.606,-0.255) -- (0.624,-0.268) -- (0.642,-0.281) -- (0.659,-0.292) -- (0.677,-0.302) -- (0.695,-0.310) -- (0.713,-0.318) -- (0.732,-0.325) -- (0.751,-0.331) -- (0.771,-0.336) -- (0.792,-0.341) -- (0.814,-0.344) -- (0.838,-0.346) -- (0.864,-0.347) -- (0.891,-0.347) -- (0.921,-0.345) -- (0.953,-0.342) -- (0.987,-0.337) -- (1.023,-0.330) -- (1.062,-0.321) -- (1.103,-0.310) -- (1.147,-0.297) -- (1.193,-0.281);
				\draw[skel] (-1.193,0.281) -- (-1.147,0.297) -- (-1.103,0.310) -- (-1.062,0.321) -- (-1.023,0.330) -- (-0.987,0.337) -- (-0.953,0.342) -- (-0.921,0.345) -- (-0.891,0.347) -- (-0.864,0.347) -- (-0.838,0.346) -- (-0.814,0.344) -- (-0.792,0.341) -- (-0.771,0.336) -- (-0.751,0.331) -- (-0.732,0.325) -- (-0.713,0.318) -- (-0.695,0.310) -- (-0.677,0.302) -- (-0.659,0.292) -- (-0.642,0.281) -- (-0.624,0.268) -- (-0.606,0.255) -- (-0.588,0.240) -- (-0.570,0.224) -- (-0.552,0.207) -- (-0.533,0.188) -- (-0.515,0.168) -- (-0.496,0.147) -- (-0.477,0.124) -- (-0.458,0.101) -- (-0.438,0.076) -- (-0.419,0.050) -- (-0.400,0.023) -- (-0.380,-0.005) -- (-0.360,-0.034) -- (-0.340,-0.064) -- (-0.320,-0.095) -- (-0.300,-0.126) -- (-0.279,-0.158) -- (-0.258,-0.190) -- (-0.237,-0.223) -- (-0.215,-0.256) -- (-0.193,-0.290) -- (-0.170,-0.324) -- (-0.146,-0.358) -- (-0.122,-0.393) -- (-0.098,-0.427) -- (-0.072,-0.462) -- (-0.045,-0.496) -- (-0.018,-0.530) -- (0.011,-0.565) -- (0.041,-0.598) -- (0.072,-0.632) -- (0.104,-0.665) -- (0.138,-0.697) -- (0.174,-0.728) -- (0.211,-0.759) -- (0.250,-0.788) -- (0.291,-0.816) -- (0.334,-0.843) -- (0.379,-0.868) -- (0.425,-0.891) -- (0.473,-0.912) -- (0.523,-0.930) -- (0.575,-0.947) -- (0.628,-0.960) -- (0.682,-0.971) -- (0.738,-0.979) -- (0.795,-0.985) -- (0.853,-0.987) -- (0.911,-0.987) -- (0.971,-0.983) -- (1.031,-0.977) -- (1.093,-0.968) -- (1.155,-0.957) -- (1.218,-0.942) -- (1.281,-0.925) -- (1.346,-0.905) -- (1.412,-0.882);
				\draw[skel] (-1.193,-0.281) -- (-1.147,-0.297) -- (-1.103,-0.310) -- (-1.062,-0.321) -- (-1.023,-0.330) -- (-0.987,-0.337) -- (-0.953,-0.342) -- (-0.921,-0.345) -- (-0.891,-0.347) -- (-0.864,-0.347) -- (-0.838,-0.346) -- (-0.814,-0.344) -- (-0.792,-0.341) -- (-0.771,-0.336) -- (-0.751,-0.331) -- (-0.732,-0.325) -- (-0.713,-0.318) -- (-0.695,-0.310) -- (-0.677,-0.302) -- (-0.659,-0.292) -- (-0.642,-0.281) -- (-0.624,-0.268) -- (-0.606,-0.255) -- (-0.588,-0.240) -- (-0.570,-0.224) -- (-0.552,-0.207) -- (-0.533,-0.188) -- (-0.515,-0.168) -- (-0.496,-0.147) -- (-0.477,-0.124) -- (-0.458,-0.101) -- (-0.438,-0.076) -- (-0.419,-0.050) -- (-0.400,-0.023) -- (-0.380,0.005) -- (-0.360,0.034) -- (-0.340,0.064) -- (-0.320,0.095) -- (-0.300,0.126) -- (-0.279,0.158) -- (-0.258,0.190) -- (-0.237,0.223) -- (-0.215,0.256) -- (-0.193,0.290) -- (-0.170,0.324) -- (-0.146,0.358) -- (-0.122,0.393) -- (-0.098,0.427) -- (-0.072,0.462) -- (-0.045,0.496) -- (-0.018,0.530) -- (0.011,0.565) -- (0.041,0.598) -- (0.072,0.632) -- (0.104,0.665) -- (0.138,0.697) -- (0.174,0.728) -- (0.211,0.759) -- (0.250,0.788) -- (0.291,0.816) -- (0.334,0.843) -- (0.379,0.868) -- (0.425,0.891) -- (0.473,0.912) -- (0.523,0.930) -- (0.575,0.947) -- (0.628,0.960) -- (0.682,0.971) -- (0.738,0.979) -- (0.795,0.985) -- (0.853,0.987) -- (0.911,0.987) -- (0.971,0.983) -- (1.031,0.977) -- (1.093,0.968) -- (1.155,0.957) -- (1.218,0.942) -- (1.281,0.925) -- (1.346,0.905) -- (1.412,0.882);
				\draw[skel] (-1.412,-0.882) -- (-1.346,-0.905) -- (-1.281,-0.925) -- (-1.218,-0.942) -- (-1.155,-0.957) -- (-1.093,-0.968) -- (-1.031,-0.977) -- (-0.971,-0.983) -- (-0.911,-0.987) -- (-0.853,-0.987) -- (-0.795,-0.985) -- (-0.738,-0.979) -- (-0.682,-0.971) -- (-0.628,-0.960) -- (-0.575,-0.947) -- (-0.523,-0.930) -- (-0.473,-0.912) -- (-0.425,-0.891) -- (-0.379,-0.868) -- (-0.334,-0.843) -- (-0.291,-0.816) -- (-0.250,-0.788) -- (-0.211,-0.759) -- (-0.174,-0.728) -- (-0.138,-0.697) -- (-0.104,-0.665) -- (-0.072,-0.632) -- (-0.041,-0.598) -- (-0.011,-0.565) -- (0.018,-0.530) -- (0.045,-0.496) -- (0.072,-0.462) -- (0.098,-0.427) -- (0.122,-0.393) -- (0.146,-0.358) -- (0.170,-0.324) -- (0.193,-0.290) -- (0.215,-0.256) -- (0.237,-0.223) -- (0.258,-0.190) -- (0.279,-0.158) -- (0.300,-0.126) -- (0.320,-0.095) -- (0.340,-0.064) -- (0.360,-0.034) -- (0.380,-0.005) -- (0.400,0.023) -- (0.419,0.050) -- (0.438,0.076) -- (0.458,0.101) -- (0.477,0.124) -- (0.496,0.147) -- (0.515,0.168) -- (0.533,0.188) -- (0.552,0.207) -- (0.570,0.224) -- (0.588,0.240) -- (0.606,0.255) -- (0.624,0.268) -- (0.642,0.281) -- (0.659,0.292) -- (0.677,0.302) -- (0.695,0.310) -- (0.713,0.318) -- (0.732,0.325) -- (0.751,0.331) -- (0.771,0.336) -- (0.792,0.341) -- (0.814,0.344) -- (0.838,0.346) -- (0.864,0.347) -- (0.891,0.347) -- (0.921,0.345) -- (0.953,0.342) -- (0.987,0.337) -- (1.023,0.330) -- (1.062,0.321) -- (1.103,0.310) -- (1.147,0.297) -- (1.193,0.281);
			\end{scope}
			\begin{scope}[shift={(0.00,-5.88)}]
				\shade[inner color=white, outer color=gray!38, even odd rule] (0,0) ellipse (2.31 and 1.33) (-1.421,0) ellipse (0.886 and 0.701) (0.934,0) ellipse (0.726 and 0.575);
				\draw[surf] (0,0) ellipse (2.31 and 1.33);
				\draw[surf] (-1.421,0) ellipse (0.886 and 0.701);
				\draw[surf] (0.934,0) ellipse (0.726 and 0.575);
				\fill (-2.306,0) circle (3.64pt);
				\fill (1.983,0) circle (3.64pt);
			\end{scope}
			\node at (0.00,-8.49) {$N^{-4}$};
			\begin{scope}[shift={(7.94,0.00)}]
				\begin{scope}[shift={(-2.900,0)}]
					\draw[skel] (1.707,0.281) -- (0.936,0.000) -- (1.707,-0.281);
					\draw[skel] (1.488,-0.882) -- (0.717,-0.601) -- (1.127,-1.312);
					\draw[skel] (1.488,0.882) -- (-0.468,0.170) -- (0.573,-1.632);
					\begin{scope}[shift={(60:1.200)}, rotate=-30]
						\draw[skel] (-0.320,0.500) .. controls (-0.280,0.325) and (-0.12,0.03) .. (0,0) .. controls (0.12,0.03) and (0.280,0.325) .. (0.320,0.500);
						\fill (0,0) circle (2.91pt);
						\draw[pinch] (0,0) -- (0,-1.200);
					\end{scope}
					\draw[pinch] (0.936,0.000) -- (0,0);
					\fill (0.936,0.000) circle (2.91pt);
					\draw[pinch] (0.717,-0.601) -- (0,0);
					\fill (0.717,-0.601) circle (2.91pt);
					\draw[pinch] (-0.468,0.170) -- (0,0);
					\fill (-0.468,0.170) circle (2.91pt);
				\end{scope}
				\begin{scope}[shift={(2.900,0)}, xscale=-1]
					\draw[skel] (1.707,0.281) -- (0.936,0.000) -- (1.707,-0.281);
					\draw[skel] (1.488,-0.882) -- (0.717,-0.601) -- (1.127,-1.312);
					\draw[skel] (1.488,0.882) -- (-0.468,0.170) -- (0.573,-1.632);
					\begin{scope}[shift={(60:1.200)}, rotate=-30]
						\draw[skel] (-0.320,0.500) .. controls (-0.280,0.325) and (-0.12,0.03) .. (0,0) .. controls (0.12,0.03) and (0.280,0.325) .. (0.320,0.500);
						\fill (0,0) circle (2.91pt);
						\draw[pinch] (0,0) -- (0,-1.200);
					\end{scope}
					\draw[pinch] (0.936,0.000) -- (0,0);
					\fill (0.936,0.000) circle (2.91pt);
					\draw[pinch] (0.717,-0.601) -- (0,0);
					\fill (0.717,-0.601) circle (2.91pt);
					\draw[pinch] (-0.468,0.170) -- (0,0);
					\fill (-0.468,0.170) circle (2.91pt);
				\end{scope}
				\draw[skel] (-2.327,1.632) arc[start angle=150.000, end angle=30.000, radius=2.687];
				\draw[skel] (-1.773,1.312) arc[start angle=150.000, end angle=30.000, radius=2.047];
				\draw[skel] (-1.773,-1.312) arc[start angle=-150.000, end angle=-30.000, radius=2.047];
				\draw[skel] (-2.327,-1.632) arc[start angle=-150.000, end angle=-30.000, radius=2.687];
				\draw[skel] (-1.412,0.882) -- (-1.346,0.905) -- (-1.281,0.925) -- (-1.218,0.942) -- (-1.155,0.957) -- (-1.093,0.968) -- (-1.031,0.977) -- (-0.971,0.983) -- (-0.911,0.987) -- (-0.853,0.987) -- (-0.795,0.985) -- (-0.738,0.979) -- (-0.682,0.971) -- (-0.628,0.960) -- (-0.575,0.947) -- (-0.523,0.930) -- (-0.473,0.912) -- (-0.425,0.891) -- (-0.379,0.868) -- (-0.334,0.843) -- (-0.291,0.816) -- (-0.250,0.788) -- (-0.211,0.759) -- (-0.174,0.728) -- (-0.138,0.697) -- (-0.104,0.665) -- (-0.072,0.632) -- (-0.041,0.598) -- (-0.011,0.565) -- (0.018,0.530) -- (0.045,0.496) -- (0.072,0.462) -- (0.098,0.427) -- (0.122,0.393) -- (0.146,0.358) -- (0.170,0.324) -- (0.193,0.290) -- (0.215,0.256) -- (0.237,0.223) -- (0.258,0.190) -- (0.279,0.158) -- (0.300,0.126) -- (0.320,0.095) -- (0.340,0.064) -- (0.360,0.034) -- (0.380,0.005) -- (0.400,-0.023) -- (0.419,-0.050) -- (0.438,-0.076) -- (0.458,-0.101) -- (0.477,-0.124) -- (0.496,-0.147) -- (0.515,-0.168) -- (0.533,-0.188) -- (0.552,-0.207) -- (0.570,-0.224) -- (0.588,-0.240) -- (0.606,-0.255) -- (0.624,-0.268) -- (0.642,-0.281) -- (0.659,-0.292) -- (0.677,-0.302) -- (0.695,-0.310) -- (0.713,-0.318) -- (0.732,-0.325) -- (0.751,-0.331) -- (0.771,-0.336) -- (0.792,-0.341) -- (0.814,-0.344) -- (0.838,-0.346) -- (0.864,-0.347) -- (0.891,-0.347) -- (0.921,-0.345) -- (0.953,-0.342) -- (0.987,-0.337) -- (1.023,-0.330) -- (1.062,-0.321) -- (1.103,-0.310) -- (1.147,-0.297) -- (1.193,-0.281);
				\draw[skel] (-1.193,0.281) -- (-1.147,0.297) -- (-1.103,0.310) -- (-1.062,0.321) -- (-1.023,0.330) -- (-0.987,0.337) -- (-0.953,0.342) -- (-0.921,0.345) -- (-0.891,0.347) -- (-0.864,0.347) -- (-0.838,0.346) -- (-0.814,0.344) -- (-0.792,0.341) -- (-0.771,0.336) -- (-0.751,0.331) -- (-0.732,0.325) -- (-0.713,0.318) -- (-0.695,0.310) -- (-0.677,0.302) -- (-0.659,0.292) -- (-0.642,0.281) -- (-0.624,0.268) -- (-0.606,0.255) -- (-0.588,0.240) -- (-0.570,0.224) -- (-0.552,0.207) -- (-0.533,0.188) -- (-0.515,0.168) -- (-0.496,0.147) -- (-0.477,0.124) -- (-0.458,0.101) -- (-0.438,0.076) -- (-0.419,0.050) -- (-0.400,0.023) -- (-0.380,-0.005) -- (-0.360,-0.034) -- (-0.340,-0.064) -- (-0.320,-0.095) -- (-0.300,-0.126) -- (-0.279,-0.158) -- (-0.258,-0.190) -- (-0.237,-0.223) -- (-0.215,-0.256) -- (-0.193,-0.290) -- (-0.170,-0.324) -- (-0.146,-0.358) -- (-0.122,-0.393) -- (-0.098,-0.427) -- (-0.072,-0.462) -- (-0.045,-0.496) -- (-0.018,-0.530) -- (0.011,-0.565) -- (0.041,-0.598) -- (0.072,-0.632) -- (0.104,-0.665) -- (0.138,-0.697) -- (0.174,-0.728) -- (0.211,-0.759) -- (0.250,-0.788) -- (0.291,-0.816) -- (0.334,-0.843) -- (0.379,-0.868) -- (0.425,-0.891) -- (0.473,-0.912) -- (0.523,-0.930) -- (0.575,-0.947) -- (0.628,-0.960) -- (0.682,-0.971) -- (0.738,-0.979) -- (0.795,-0.985) -- (0.853,-0.987) -- (0.911,-0.987) -- (0.971,-0.983) -- (1.031,-0.977) -- (1.093,-0.968) -- (1.155,-0.957) -- (1.218,-0.942) -- (1.281,-0.925) -- (1.346,-0.905) -- (1.412,-0.882);
				\draw[skel] (-1.193,-0.281) -- (-1.147,-0.297) -- (-1.103,-0.310) -- (-1.062,-0.321) -- (-1.023,-0.330) -- (-0.987,-0.337) -- (-0.953,-0.342) -- (-0.921,-0.345) -- (-0.891,-0.347) -- (-0.864,-0.347) -- (-0.838,-0.346) -- (-0.814,-0.344) -- (-0.792,-0.341) -- (-0.771,-0.336) -- (-0.751,-0.331) -- (-0.732,-0.325) -- (-0.713,-0.318) -- (-0.695,-0.310) -- (-0.677,-0.302) -- (-0.659,-0.292) -- (-0.642,-0.281) -- (-0.624,-0.268) -- (-0.606,-0.255) -- (-0.588,-0.240) -- (-0.570,-0.224) -- (-0.552,-0.207) -- (-0.533,-0.188) -- (-0.515,-0.168) -- (-0.496,-0.147) -- (-0.477,-0.124) -- (-0.458,-0.101) -- (-0.438,-0.076) -- (-0.419,-0.050) -- (-0.400,-0.023) -- (-0.380,0.005) -- (-0.360,0.034) -- (-0.340,0.064) -- (-0.320,0.095) -- (-0.300,0.126) -- (-0.279,0.158) -- (-0.258,0.190) -- (-0.237,0.223) -- (-0.215,0.256) -- (-0.193,0.290) -- (-0.170,0.324) -- (-0.146,0.358) -- (-0.122,0.393) -- (-0.098,0.427) -- (-0.072,0.462) -- (-0.045,0.496) -- (-0.018,0.530) -- (0.011,0.565) -- (0.041,0.598) -- (0.072,0.632) -- (0.104,0.665) -- (0.138,0.697) -- (0.174,0.728) -- (0.211,0.759) -- (0.250,0.788) -- (0.291,0.816) -- (0.334,0.843) -- (0.379,0.868) -- (0.425,0.891) -- (0.473,0.912) -- (0.523,0.930) -- (0.575,0.947) -- (0.628,0.960) -- (0.682,0.971) -- (0.738,0.979) -- (0.795,0.985) -- (0.853,0.987) -- (0.911,0.987) -- (0.971,0.983) -- (1.031,0.977) -- (1.093,0.968) -- (1.155,0.957) -- (1.218,0.942) -- (1.281,0.925) -- (1.346,0.905) -- (1.412,0.882);
				\draw[skel] (-1.412,-0.882) -- (-1.346,-0.905) -- (-1.281,-0.925) -- (-1.218,-0.942) -- (-1.155,-0.957) -- (-1.093,-0.968) -- (-1.031,-0.977) -- (-0.971,-0.983) -- (-0.911,-0.987) -- (-0.853,-0.987) -- (-0.795,-0.985) -- (-0.738,-0.979) -- (-0.682,-0.971) -- (-0.628,-0.960) -- (-0.575,-0.947) -- (-0.523,-0.930) -- (-0.473,-0.912) -- (-0.425,-0.891) -- (-0.379,-0.868) -- (-0.334,-0.843) -- (-0.291,-0.816) -- (-0.250,-0.788) -- (-0.211,-0.759) -- (-0.174,-0.728) -- (-0.138,-0.697) -- (-0.104,-0.665) -- (-0.072,-0.632) -- (-0.041,-0.598) -- (-0.011,-0.565) -- (0.018,-0.530) -- (0.045,-0.496) -- (0.072,-0.462) -- (0.098,-0.427) -- (0.122,-0.393) -- (0.146,-0.358) -- (0.170,-0.324) -- (0.193,-0.290) -- (0.215,-0.256) -- (0.237,-0.223) -- (0.258,-0.190) -- (0.279,-0.158) -- (0.300,-0.126) -- (0.320,-0.095) -- (0.340,-0.064) -- (0.360,-0.034) -- (0.380,-0.005) -- (0.400,0.023) -- (0.419,0.050) -- (0.438,0.076) -- (0.458,0.101) -- (0.477,0.124) -- (0.496,0.147) -- (0.515,0.168) -- (0.533,0.188) -- (0.552,0.207) -- (0.570,0.224) -- (0.588,0.240) -- (0.606,0.255) -- (0.624,0.268) -- (0.642,0.281) -- (0.659,0.292) -- (0.677,0.302) -- (0.695,0.310) -- (0.713,0.318) -- (0.732,0.325) -- (0.751,0.331) -- (0.771,0.336) -- (0.792,0.341) -- (0.814,0.344) -- (0.838,0.346) -- (0.864,0.347) -- (0.891,0.347) -- (0.921,0.345) -- (0.953,0.342) -- (0.987,0.337) -- (1.023,0.330) -- (1.062,0.321) -- (1.103,0.310) -- (1.147,0.297) -- (1.193,0.281);
			\end{scope}
			\begin{scope}[shift={(7.94,-5.88)}]
				\shade[inner color=white, outer color=gray!38, even odd rule] (0,0) ellipse (2.31 and 1.33) (-2.198,0.410) .. controls (-0.733,1.092) and (0.733,1.092) .. (2.198,0.410) .. controls (0.733,0.036) and (-0.733,0.036) .. (-2.198,0.410) -- cycle (0,-0.575) ellipse (1.022 and 0.358);
				\draw[surf] (0,0) ellipse (2.31 and 1.33);
				\draw[surf, line join=round] (-2.198,0.410) .. controls (-0.733,1.092) and (0.733,1.092) .. (2.198,0.410) .. controls (0.733,0.036) and (-0.733,0.036) .. (-2.198,0.410) -- cycle;
				\draw[surf] (0,-0.575) ellipse (1.022 and 0.358);
				\fill (-2.198,0.410) circle (3.64pt);
				\fill (2.198,0.410) circle (3.64pt);
			\end{scope}
			\node at (7.94,-8.49) {$N^{-4}$};
		\end{tikzpicture}
		\caption{Nonplanar contractions for the $2$-point correlator of $4$-gluon operators in the fundamental representation: contractions involving a single-pinched vertex with two crossed propagators give rise to a double torus with one pinched cycle (left), whereas contractions of two single-pinched vertices via two crossed propagators give rise to a double torus with two pinched cycles (right).}
		\label{fig:D4nonplanar}
	\end{figure}
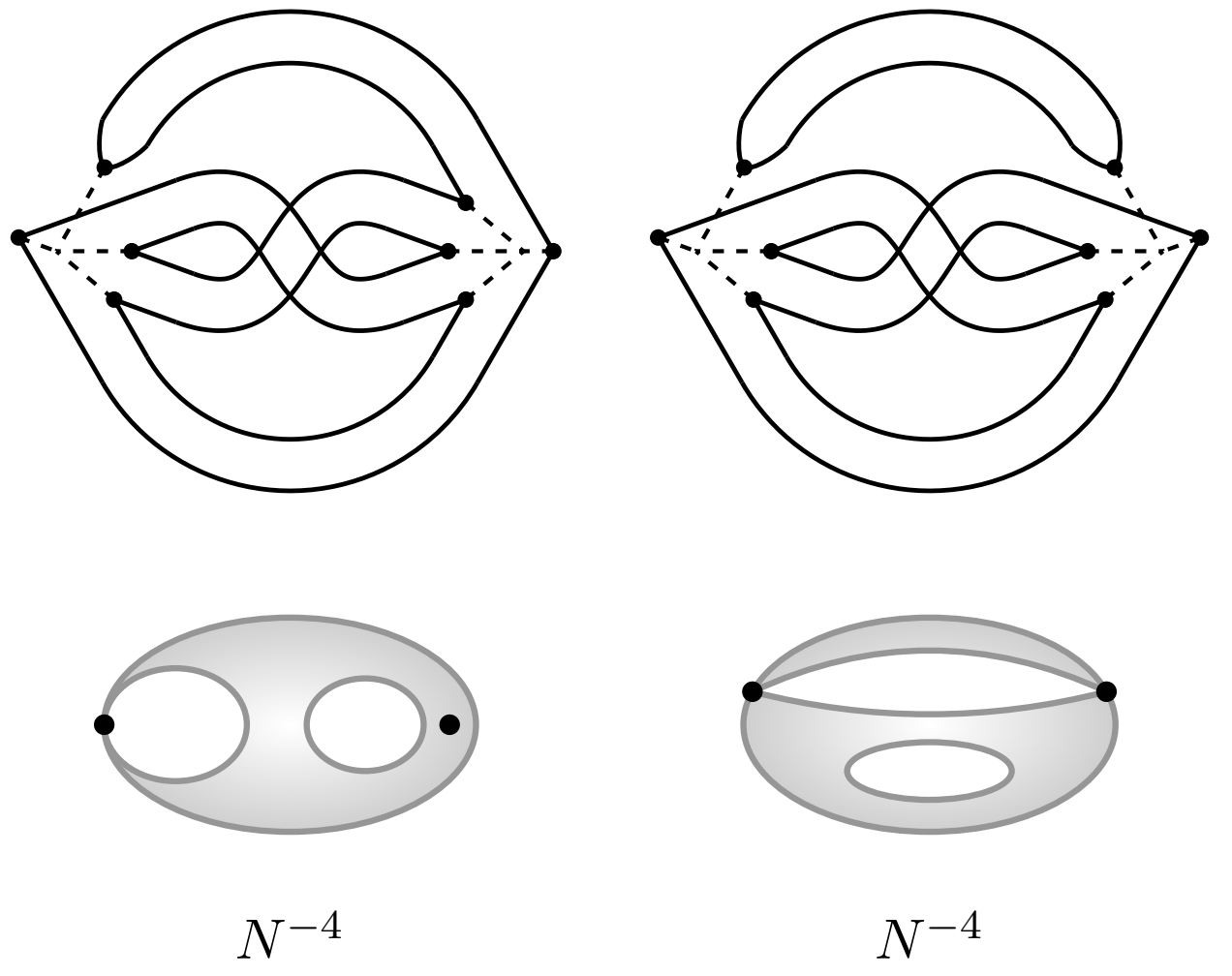

\section{Conclusions}\label{sec:conc}

We have established that the 't Hooft smooth topologies occurring in the large-$N$ expansion of SU($N$) YM-like theories are in fact incomplete, in stark contrast to the bare U($N$) case. In general, new singular topologies corresponding to pinched cycles of Riemann surfaces, possibly colliding with punctures, necessarily arise.\par
Specifically, we have investigated the topologies of Feynman diagrams occurring in the large-$N$ expansion of pure SU($N$) YM theories -- as opposed to the bare U($N$) ones introduced by 't Hooft -- and of the corresponding renormalized U($N$) theories. The difference arises from $\frac{1}{N}$ corrections to the propagators of fields valued in the Lie algebra and to the vertices of local operators.\par
For simplicity, we have discussed only the pure YM case in detail, but the extension to other fields valued in the Lie algebra of SU($N$) holds \emph{verbatim}.\par
For vacuum diagrams in the SU($N$) theory, only the correction to the gluon propagator is relevant; however, its contribution vanishes because it is a U($1$) projector inside U($N$) that drops out when the propagators are attached to the SU($N$) vertices. Hence, vacuum diagrams have the exact same topology as in the 't Hooft expansion of the U($N$) theory.\par
For very special operators whose vertices have a color structure that can be written as nested commutators, the same vanishing mechanism applies, so that the large-$N$ expansions of the SU($N$) and U($N$) theories coincide exactly. However, in all remaining cases, the large-$N$ expansions of the SU($N$) and U($N$) theories differ.\par
The next case in order of difficulty arises in YM theory for $2$-gluon twist-$2$ operators, where extra planar diagrams with the topology of disks occur at subleading $\frac{1}{N}$ order in perturbation theory.\par
For these operators, several choices are possible for the extra $\frac{1}{N}$ corrections, all of which are computationally equivalent in the SU($N$) theory, though not topologically equivalent: we can modify both the propagator and the vertices to ensure the correct SU($N$) projection, only the propagator, or only the vertices. Since the vertices in connected diagrams are always attached to propagators, the final result for the Feynman diagrams is computationally independent of these choices.\par
Yet, the above choices affect the corresponding topologies. Specifically, in the first two cases, new diagrams arise that involve disks disconnected in color space yet connected in space-time. These diagrams have a problematic physical interpretation, and we have therefore discarded the first two choices.
The third choice is the most conservative, because the propagator remains in the standard 't Hooft double-line representation, while only the vertices of local operators are affected by the SU($N$) projection.\par
Hence, we have implemented the third choice and extended our analysis to include many-gluon higher-twist quasi-partonic operators.\par
In this case, although topological disks also occur, they carry marked points associated with the locations of the local operators, so the resulting extra diagrams are always connected in both color space and space-time, giving rise to pinched topologies.\par
Indeed, because of the presence of these marked points, the boundary lines with opposite orientations in these new diagrams can be glued \`a la 't Hooft, giving rise to singular closed Riemann surfaces with pinches possibly colliding with punctures. Moreover, the large-$N$ weights of the extra topologies also match our topological classification, provided that they are viewed as degenerations of smooth surfaces with pinched cycles.\par
Hence, it is clear that the existence of these new topologies has profound implications both for the yet-to-be-found nonperturbative large-$N$ solution in four dimensions and for the existence of a string interpretation of the large-$N$ expansion of gauge theories, a topic partially discussed in \cite{BPSL,BPS42} that we will explore further in a forthcoming paper.
	
		\newpage
	\appendix

	\section{Notation and Wick rotation \label{app:notation}}

We follow the conventions of \cite{Braun:2003rp}. We choose the mostly-minus Minkowskian metric
\begin{equation}
	(g_{\mu\nu}) = \text{diag}(1,-1,-1,-1)
\end{equation}
and define the light-cone coordinates
\begin{equation}
	x^{\pm} = \frac{x^0\pm x^3}{\sqrt{2}}=x_{\mp} \,.
\end{equation}
The corresponding Minkowskian squared distance is
\begin{equation}
	\label{mod2}
	\rvert x \rvert^2 = 2 x^+ x^- -x_{\perp}^2 \,,
\end{equation}
with
\begin{equation}
	x^2_\perp=(x^1)^2+(x^2)^2 \,.
\end{equation}
We denote the derivative with respect to $x^+$ by
\begin{equation}
	\partial_+ = \frac{\partial}{\partial x^+} = \partial_{x^+} =\frac{\partial}{\partial x_-} = \partial_{x_-}
\end{equation}
and define the light-like vectors $n^\mu$ and $\bar{n}^\mu$,
\begin{equation}
	n_\mu n^\mu = \bar{n}_\mu \bar{n}^\mu = 0 \,, \qquad n_\mu \bar{n}^\mu = 1 \,,
\end{equation}
which read $(n^{\mu}) = \frac{1}{\sqrt{2}}(1,0,0,1)$ and $(\bar{n}^{\mu}) = \frac{1}{\sqrt{2}}(1,0,0,-1)$, respectively.
Correspondingly, the Minkowskian metric decomposes into transverse and longitudinal components with respect to these light-like vectors
\begin{equation}
	g_{\mu\nu} =g^\perp_{\mu\nu}+n_\mu \bar{n}_\nu +n_\nu \bar{n}_\mu \,.
\end{equation} 
The Euclidean metric reads
\begin{equation}
	(\delta_{\mu\nu}) = \text{diag}(1,1,1,1) \,,
\end{equation}
with Euclidean squared distance
\begin{equation}
	x^2 = 2x^z x^{\bar{z}}+x_\perp^2 \,,
\end{equation}
where
\begin{equation}
	x^z= \frac{x^4+ix^3}{\sqrt{2}}=\frac{x_4+ix_3}{\sqrt{2}}=x_{\bar z}
\end{equation}
and
\begin{equation}
	x^{\bar z}= \frac{x^4-ix^3}{\sqrt{2}}=\frac{x_4-ix_3}{\sqrt{2}}=x_z \,.
\end{equation}
We define the Wick rotation by
\begin{align}
	\label{wick1}
	&x^0=x_0 \rightarrow - i x^4=-i x_4 
\end{align} 
and
\begin{align}
	\label{wick2}
	&p_0=p^0 \rightarrow i p_4= i p^4 \,.
\end{align} 
Eq.~\eqref{wick1} ensures that $\exp(iS_M)\rightarrow \exp(-S_E)$, where $S_M$ and $S_E$ are the Minkowskian and Euclidean actions, respectively, with $S_E$ being positive definite. Defining $p \cdot x = p_{\mu} x^{\mu}$ and $\langle p x \rangle= p_{\mu} x^{\mu}$ in Minkowskian and Euclidean space-time, respectively, eq.~\eqref{wick2} ensures that under Wick rotation $p \cdot x \rightarrow \langle p x \rangle$, so that the pairings $p \cdot x$ and $\langle p x \rangle$ are independent of the Minkowskian and Euclidean metrics. 
Therefore, by a slight abuse of notation, we also write $p \cdot x$ in Euclidean space-time instead of $\langle p x \rangle$. Moreover, $|x|^2 \rightarrow - x^2$ and $|p|^2 \rightarrow - p^2$. Consequently, the Wick rotation of the scalar propagator of mass $m$ in Minkowskian space-time,
\begin{equation}
	\langle \phi(x) \phi(y)\rangle = \int \frac{d^4p}{(2\pi)^4}\, e^{i p \cdot (x-y)}\, \frac{i}{|p|^2-m^2+i \epsilon}\,,
\end{equation}
reads, in Euclidean space-time,
\begin{equation}
	\langle \phi^E(x) \phi^E(y)\rangle = \int \frac{d^4p}{(2\pi)^4}\, e^{i p \cdot(x-y)}\, \frac{i^2}{- p^2-m^2}= \int \frac{d^4p}{(2\pi)^4}\, e^{i p \cdot(x-y)}\, \frac{1}{p^2+m^2} \,,
\end{equation}
as expected. The Wick rotation of the light-cone coordinates,
\begin{align}
	&x^+=x_-\rightarrow -ix^z= -i x_{\bar z}
\end{align}
and
\begin{align}
	&x^-=x_+\rightarrow -i x^{\bar{z}}= -i x_{z}\,,
\end{align}
implies the Wick rotation of the derivative with respect to $x^+$
\begin{align}
	\label{wickderivative}
	\partial_+\rightarrow i \partial_{z} = i\frac{\partial}{\partial x^z} \,.
\end{align}

\section{Jacobi and Gegenbauer polynomials \label{app:jacobi}}

We recall the definitions and basic properties of the Jacobi and Gegenbauer polynomials. 
For real $x$, the Jacobi polynomials $P^{(\alpha,\beta)}_l(x)$ admit the representation \cite{szego1959orthogonal}
\begin{equation}
	\label{jaco1}
	P^{(\alpha,\beta)}_l(x) = \sum_{k = 0}^{l}{l+\alpha\choose k}{l+\beta\choose k+\beta}\left(\frac{x-1}{2}\right)^{l-k}\left(\frac{x+1}{2}\right)^{k}\,,
\end{equation}
with real $\alpha, \beta$ and $l$ a natural number. Moreover, they satisfy the symmetry property
\begin{align}
	\label{jacox}
	P^{(\alpha,\beta)}_l(-x) = (-1)^{l}P^{(\beta,\alpha)}_l(x) \,.
\end{align} 
The Gegenbauer polynomials $C^{\alpha'}_l(x)$ are a special case of the Jacobi polynomials
\begin{equation} \label{GPol}
	C^{\alpha'}_l(x) = \frac{\Gamma(l+2\alpha')\Gamma(\alpha'+\frac{1}{2})}{\Gamma(2\alpha')\Gamma(l+\alpha'+\frac{1}{2})}P_l^{(\alpha'-\frac{1}{2},\alpha'-\frac{1}{2})}(x) \,,
\end{equation}
with the symmetry property
\begin{align}
	\label{symmgegen}
	C^{\alpha'}_l(-x) = (-1)^l C^{\alpha'}_l(x) \,.
\end{align}
Setting
\begin{equation}
	x = \frac{b-a}{a+b} \,,
\end{equation}
we have
\begin{equation}
	\left(\frac{x-1}{2}\right)^{l-k}\left(\frac{x+1}{2}\right)^{k}= (-1)^{l-k} \frac{a^{l-k} b^k}{(a+b)^l} \,,
\end{equation}
so that eq.~\eqref{jaco1} becomes
\begin{equation}
	\label{jaco2}
	P^{(\alpha,\beta)}_l(x) = \sum_{k = 0}^{l}{l+\alpha\choose k}{l+\beta\choose k+\beta}(-1)^{l-k} \frac{a^{l-k} b^k}{(a+b)^l} \,.
\end{equation}
Moreover, setting $l = J-\alpha'+\frac{1}{2}$ in eq.~\eqref{GPol} and $\alpha=\beta=\alpha'-\frac{1}{2}$ in eq.~\eqref{jaco1}, we obtain
\begin{align}
	C^{\alpha'}_{J-\alpha'+\frac{1}{2}}\left(x\right) =&\frac{\Gamma(J+\frac{1}{2}+\alpha')\Gamma(\alpha'+\frac{1}{2})}{\Gamma(2\alpha')\Gamma(J+1)}\nonumber\\
	&\sum_{k=0}^{J-\alpha'+\frac{1}{2}} {J\choose k}{J\choose k+\alpha'-\frac{1}{2}}(-1)^{J-\alpha'+\frac{1}{2}-k} \frac{a^{J-\alpha'+\frac{1}{2}-k} b^k}{(a+b)^{J-\alpha'+\frac{1}{2}}} \,.
\end{align}
Specializing the above equation to $J = s$ and $\alpha' = \frac{5}{2}$, we get
\begin{equation}
	\label{physicalgegen}
	C^{\frac{5}{2}}_{s-2}\left(x\right) = \frac{\Gamma(s+3)\Gamma(3)}{\Gamma(5)\Gamma(s+1)}\sum_{k=0}^{s-2} {s\choose k}{s\choose k+2}(-1)^{s-k} \frac{a^{s-k-2} b^k}{(a+b)^{s-2}} \,.
\end{equation}
Similarly, for $J = s$ and $\alpha' = \frac{1}{2}$, we obtain
\begin{equation}
	\label{physicalgegen2}
	C^{\frac{1}{2}}_{s}\left(x\right) = \sum_{k=0}^{s} {s\choose k}{s\choose k}(-1)^{s-k} \frac{a^{s-k} b^k}{(a+b)^{s}} \,.
\end{equation}
We now restrict $\alpha, \beta$ to be natural numbers and, correspondingly, $\alpha'$ to be a positive half-integer and $J$ to be a natural number. 
Employing the identity
\begin{align}
	{l+\alpha\choose k}{l+\beta\choose k+\beta}
	=\frac{(l+\beta)!(l+\alpha)!}{l!(l+\alpha+\beta)!}{l\choose k}{l+\beta+\alpha\choose k+\beta} \,,
\end{align}
it follows from eq.~\eqref{jaco2} that
\begin{equation}
	\label{jaco3}
	P^{(\alpha,\beta)}_l(x) = \frac{(l+\beta)!(l+\alpha)!}{l!(l+\alpha+\beta)!}\sum_{k = 0}^{l} {l\choose k}{l+\beta+\alpha\choose k+\beta}(-1)^{l-k} \frac{a^{l-k} b^k}{(a+b)^l} \,.
\end{equation}
Correspondingly, eq.~\eqref{GPol} reads
\begin{align}
	C^{\alpha'}_l\left(x\right) = &\frac{\Gamma(l+2\alpha')\Gamma(\alpha'+\frac{1}{2})}{\Gamma(2\alpha')\Gamma(l+\alpha'+\frac{1}{2})}\frac{(l+\alpha'-\frac{1}{2})!(l+\alpha'-\frac{1}{2})!}{l!(l+2\alpha'-1)!}\nonumber\\
	&\sum_{k = 0}^{l} {l\choose k}{l+2\alpha'-1\choose k+\alpha'-\frac{1}{2}}(-1)^{l-k} \frac{a^{l-k} b^k}{(a+b)^l}\,,
\end{align}
which reduces to
\begin{align} \label{bin}
	C^{\alpha'}_l\left(x\right) = \frac{\Gamma(\alpha'+\frac{1}{2})\Gamma(l+\alpha'+\frac{1}{2})}{\Gamma(2\alpha')\Gamma(l+1)}
	\sum_{k = 0}^{l} {l\choose k}{l+2\alpha'-1\choose k+\alpha'-\frac{1}{2}}(-1)^{l-k} \frac{a^{l-k} b^k}{(a+b)^l} \,.
\end{align}

	\section{U($N$) Yang-Mills theory} \label{app:group}

\subsection{$su(N)$ and $u(N)$ algebras}

The gauge field in SU($N$) YM theory is a Hermitian $N\times N$ matrix \cite{tHooft:1973alw,maltoni2003color}
\begin{equation}
	(A_\mu)^i_{\, j}(x) = (A_\mu^*)^j_{\, i}(x)
\end{equation}
that satisfies the constraint
\begin{equation}
	\Tr A_\mu = 0 \,,
\end{equation}
which holds automatically by setting
\begin{equation}
	\label{eq:adjoint}
	(A_\mu)^i_{\,j}(x) =\sum_{a=1}^{N^2-1} A_\mu^a(x) (\lambda^a)^i_{\,j} \,,
\end{equation}
where $\lambda^a$ are the Gell-Mann matrices of $su(N)$, satisfying
\begin{equation}
	\label{eq:sun}
	\Tr\lambda^a=0,\qquad \Tr \lambda^a\lambda^b=\frac{1}{2}\delta^{ab},\qquad \left[\lambda^a,\lambda^b\right]=i f^{abc}\lambda^c,\qquad (\lambda^a)^\dagger=\lambda^a \,,
\end{equation}
with $f^{abc}$ being the (real) structure constants. Extending the theory to the U($N$) gauge group, we introduce the additional generator $\lambda^0$, which satisfies \cite{maltoni2003color}
\begin{equation}
	\label{eq:u(1)}
	\Tr\lambda^0=\sqrt{\frac{N}{2}},\qquad \Tr \lambda^\alpha\lambda^0 =\frac{1}{2}\delta^{\alpha0},\qquad\left[\lambda^\alpha,\lambda^0\right]=0\,.
\end{equation}
Explicitly,
\begin{equation}
	(\lambda^0)^i_{\,j}=\frac{1}{\sqrt{2N}}\delta^i_{\,j} \,.
\end{equation}
Thus, in the U($N$) theory,
\begin{equation}
	\label{eq:doubled}
	(A_\mu)^i_{\,j}(x) =\sum_{\alpha=0}^{N^2-1} A_\mu^\alpha(x) (\lambda^\alpha)^i_{\,j}=A_\mu^0(x)(\lambda^0)^i_{\,j}+\sum_{a=1}^{N^2-1} A_\mu^a(x) (\lambda^a)^i_{\,j} \,,
\end{equation}
where $\alpha$ runs from $0$ to $N^2-1$, so that
\begin{equation}
	\label{eq:trA0}
	\Tr A_\mu =\sqrt{\frac{N}{2}}A^0_\mu \,.
\end{equation}
Multiplying by $\lambda^{\alpha}$ and taking the trace, we can invert the above relation to obtain
\begin{equation}
	\label{eq:dictonary}
	2\Tr(A_\mu(x)\lambda^{\alpha}) =A_\mu^0(x)\delta^{\alpha 0}+ \sum_{b=1}^{N^2-1}\delta^{\alpha b}A_\mu^b(x) \,.
\end{equation}

\subsection{U($N$) YM action in the light-cone gauge}\label{app:ymlightcone}

We now write down the YM action in the light-cone gauge $A_+ = 0$.
After integrating out $A_-$, it reads \cite{Belitsky:2004sc}
\begin{align}
	\label{eq:compact} 
	S_{YM}(A, \bar A) = &\int\,d^4x\,\biggl[-\bar{A}^0 \square A^0- \bar{A}^a \square A^a -2\frac{g}{\sqrt{N}} f^{abc}(A^a\partial_+\bar{A}^b\bar{\partial}\partial_+^{-1}A^c+\bar{A}^a\partial_+A^b\partial\partial_+^{-1}\bar{A}^c) \nonumber\\
	&-2\frac{g^2}{N}f^{abc}f^{ade}\partial_+^{-1}(A^b\partial_+ \bar{A}^c)\partial_+^{-1}(\bar{A}^d\partial_+A^e)\biggr] \,,
\end{align} 
with $a,b,c,d,e= 1,\ldots, N^2-1$, where the U($1$) part contributes only to the kinetic term, so that the free photon $A^0$ and the interacting gluons $A^a$ decouple. Using eq.~\eqref{eq:doubled}, we write the U($N$) action in the double-line representation
\begin{align}
	S_{YM}(A, \bar A) = &\int d^4x\,\biggl[- \bar{A}^i_{\, j} \square A^j_{\, i} \nonumber\\
	&-2\frac{g}{\sqrt{N}} (\mathcal{V}_3)^{i\,k\,m}_{\,j\,l\,n}\left(A^j_{\, i} \partial_+\bar{A}^l_{\, k} \bar{\partial}\partial_+^{-1}A^n_{\, m} +\bar{A}^j_{\, i} \partial_+A^l_{\, k} \partial\partial_+^{-1}\bar{A}^n_{\, m} \right) \nonumber\\
	&-2\frac{g^2}{N}(\mathcal{V}_4)^{i\,k\,m\,p}_{\,j\,l\,n\,q}\partial_+^{-1}(A^j_{\, i} \partial_+ \bar{A}^l_{\, k}) \partial_+^{-1}(\bar{A}^n_{\, m} \partial_+A^q_{\, p} )\biggr] \,,
\end{align} 
with $\mathcal{V}_3$ and $\mathcal{V}_4$ defined in eq.~\eqref{vertices}. The kinetic term can be separated into pure-trace and traceless contributions, while the vertices automatically project the fields $A$ onto their traceless part, as they are purely nonabelian
\begin{align}
	S_{YM}(A, \bar A) =& -\int d^4x\, \bar{A}^0\square A^0-\int d^4x\bar{A}^j_{\, i} (I-P)^{i\,k}_{\,j\,l}\square A^l_{\, k}\nonumber\\
	&-2\frac{g}{\sqrt{N}} (\mathcal{V}_3)^{i\,k\,m}_{\,j\,l\,n}\left(A^j_{\, i} \partial_+\bar{A}^l_{\, k} \bar{\partial}\partial_+^{-1}A^n_{\, m} +\bar{A}^j_{\, i} \partial_+A^l_{\, k} \partial\partial_+^{-1}\bar{A}^n_{\, m} \right) \nonumber\\
	&-2\frac{g^2}{N}(\mathcal{V}_4)^{i\,k\,m\,p}_{\,j\,l\,n\,q}\partial_+^{-1}(A^j_{\, i} \partial_+ \bar{A}^l_{\, k}) \partial_+^{-1}(\bar{A}^n_{\, m} \partial_+A^q_{\, p} ) \,,
\end{align}
where in double-line notation the color structure of the $su(N)$ kinetic term splits into the difference between the identity $I$ in the $u(N)$ Lie algebra, with components $I^{\,i\,k}_{\,j\,l}=\delta^i_{\, l}\delta^k_{\, j}$, and the $u(1)$ projector $P$, with components $P^{i\,k}_{\,j\,l}=\frac{1}{N}\delta^i_{\, j}\delta^k_{\, l}$. The projectors satisfy
\begin{align}
	\label{eq:projectors}
	\Tr (I-P)=N^2-1, \qquad \qquad P^n = P,\qquad \qquad(I-P)P=0\,.
\end{align}
The vacuum expectation value of a product of local gauge-invariant operators $\mathcal{O}_i(A)$ that do not depend on $A_{-}$ reads
\begin{align}
	&\langle\mathcal{O}_1(x_1)\ldots \mathcal{O}_n(x_n)\rangle=\frac{1}{Z}\int \mathcal{D}A\mathcal{D}\bar{A}\, e^{iS_{YM}(A,\bar A)} \mathcal{O}_1(x_1)\ldots \mathcal{O}_n(x_n)\,.
\end{align}

\subsection{Generating functional for twist-2 operators in U($N$) YM theory}
	
In the U($N$) theory, the connected generating functional of correlators of twist-2 operators splits into the sum of the SU($N$) and U($1$) contributions. For example,
\begin{align}
	\label{eq:WUN}
	&\mathcal{Z}\left[J_{\mathbb{O}}, J_{\mathbb{O}^{U(1)}} \right]=e^{\mathcal{W}\left[J_{\mathbb{O}} ,J_{\mathbb{O}^{U(1)}} \right]}\nonumber\\
	&=\int \prod_{a=1}^{N^2-1}\mathcal{D}A^a\mathcal{D}\bar{A}^a\, e^{-S_{YM}^{SU(N)}(A^a_\mu)}\,
	\exp\Bigg(\frac{1}{2}\int d^4x\sum_{s}\, J_{\mathbb{O}_s} \,\bar{A}^a(x)\delta^{ab}\mathcal{Y}_{s-2}^{\frac{5}{2}}(\overrightarrow{\partial}_+,\overleftarrow{\partial}_+)A^b(x)\Bigg)\nonumber\\
	&\quad\int \mathcal{D}A^0 \mathcal{D}\bar{A}^0\,e^{-S_{YM}^{U(1)}(A_\mu^0)}\,\exp\Bigg(\frac{1}{2}\int d^4x\sum_{s}\, J_{\mathbb{O}^{U(1)}_s} \bar{A}^0(x) \mathcal{Y}_{s-2}^{\frac{5}{2}}(\overrightarrow{\partial}_+,\overleftarrow{\partial}_+)  A^0(x)\Bigg)\nonumber\\
	&=e^{\mathcal{W}^{SU(N)}\left[J_{\mathbb{O}}\right]} \, e^{\mathcal{W}^{U(1)}\left[J_{\mathbb{O}^{U(1)}} \right]} \,.
\end{align}

\section{Collinear quasi-partonic twist-$2$ operators}\label{app:twist2}

At leading order in perturbation theory, suitable linear combinations of the operators below are conserved \cite{makeenko,Belitsky:2007jp} and automatically transform \cite{makeenko,Belitsky:2007jp} as primary operators with respect to the conformal group \cite{Beisert:2004fv}
\begin{align} \label{1}
	&\mathbb{O}^{\mathcal{T}=2}_{s} \quad=\quad \Tr\, F^\mu_{(\rho_1}\overleftarrow{D}_{\rho_2}\ldots \overrightarrow{D}_{\rho_{s-1}}F_{\rho_s)\mu}-\,\text{traces}\qquad\qquad \nonumber\\
	&\tilde{\mathbb{O}}^{\mathcal{T}=2}_{s} \quad=\quad \Tr\, \tilde{F}^\mu_{(\rho_1}\overleftarrow{D}_{\rho_2}\ldots \overrightarrow{D}_{\rho_{s-1}}F_{\rho_s)\mu}-\,\text{traces}\qquad\qquad \nonumber\\
	&\mathbb{S}^{\mathcal{T}=2}_{s} \quad\,\,=\quad \Tr\, (F_{\mu(\nu}+i\tilde{F}_{\mu(\nu})\overleftarrow{D}_{\rho_1}\ldots \overrightarrow{D}_{\rho_{s-2}}(F_{\lambda)\sigma}+i\tilde{F}_{\lambda)\sigma})-\,\text{traces} \,,
\end{align}
where the parentheses denote symmetrization of all enclosed indices and the subtraction of traces ensures that the contraction of any two indices vanishes. 
The restrictions of the conserved balanced operators $\mathbb{O}^{\mathcal{T}=2}_{s}, \tilde{\mathbb{O}}^{\mathcal{T}=2}_{s}$ and the unbalanced operators $\mathbb{S}^{\mathcal{T}=2}_{s}, \bar{\mathbb{S}}^{\mathcal{T}=2}_{s}$ \cite{BPS1} of twist $2$ to the components with maximal-spin projection $s$ along the $p_+$ direction in the standard basis define the gauge-invariant collinear twist-$2$ operators in the light-cone gauge \cite{BPS1}
\begin{align} \label{1000}
	\nonumber
	&\mathbb{O}_{s} = \Tr \partial_{+} \bar{A}(x)(i\overrightarrow{\partial}_++i\overleftarrow{\partial}_+)^{s-2}C^{\frac{5}{2}}_{s-2}\left(\frac{\overrightarrow{\partial}_+-\overleftarrow{\partial}_+}{\overrightarrow{\partial}_++\overleftarrow{\partial}_+}\right)\partial_{+} {A}(x)\\\nonumber
	&\tilde{\mathbb{O}}_{s} = \Tr \partial_{+} \bar{A}(x)(i\overrightarrow{\partial}_++i\overleftarrow{\partial}_+)^{s-2}C^{\frac{5}{2}}_{s-2}\left(\frac{\overrightarrow{\partial}_+-\overleftarrow{\partial}_+}{\overrightarrow{\partial}_++\overleftarrow{\partial}_+}\right)\partial_{+} {A}(x) \\\nonumber
	&\mathbb{S}_{s} =\frac{1}{\sqrt{2}}\Tr \partial_{+} \bar{A}(x)(i\overrightarrow{\partial}_++i\overleftarrow{\partial}_+)^{s-2}C^{\frac{5}{2}}_{s-2}\left(\frac{\overrightarrow{\partial}_+-\overleftarrow{\partial}_+}{\overrightarrow{\partial}_++\overleftarrow{\partial}_+}\right)\partial_{+} \bar{A}(x)\\
	&\bar{\mathbb{S}}_{s} =\frac{1}{\sqrt{2}}\Tr \partial_{+} A(x)(i\overrightarrow{\partial}_++i\overleftarrow{\partial}_+)^{s-2}C^{\frac{5}{2}}_{s-2}\left(\frac{\overrightarrow{\partial}_+-\overleftarrow{\partial}_+}{\overrightarrow{\partial}_++\overleftarrow{\partial}_+}\right)\partial_{+} A(x) \,,
\end{align}
with $C^{\alpha'}_l(x)$ denoting the Gegenbauer polynomials (eq.~\eqref{GPol}). The above construction arises as follows. In a conformal field theory, the propagator of a vector field $V_\mu(x)$ with conformal dimension $1$ is purely longitudinal \cite{makeenko}
\begin{equation}
	\langle V_\mu(x)V_\nu(0)\rangle = \mathcal{C}_2 \left(g_{\mu\nu}-2\frac{x_\mu x_\nu}{\rvert x\rvert^2}\right)\frac{1}{\rvert x\rvert^2}=\frac{\mathcal{C}_2}{2}\partial_\mu\partial_\nu\log \rvert x \rvert ^2 \,.
\end{equation}
Therefore, there is no Lorentz-covariant gauge in which the field $A_{\mu}$ can be primary with respect to the conformal group at lowest order in perturbation theory.\par
However, labeling by $(j_1,j_2)$ the irreducible representations of the Lorentz group, $F_{\mu\nu}$ decomposes \cite{Braun:2008ia} into the $(1,0)\oplus(0,1)$ representations according to
\begin{equation}
	\label{lorentza3}
	F_{a\dot{a}b\dot{b}} = 2(f_{ab}\epsilon_{\dot{a}\dot{b}} - \epsilon_{ab} f_{\dot{a}\dot{b}})
\end{equation}
and
\begin{equation}
	\label{adota3}
	F_{a\dot{a}b\dot{b}} = \sigma_{a\dot{a}}^\mu\sigma_{b\dot{b}}^\nu F_{\mu\nu} \,,
\end{equation}
where\footnote{We write eqs.~\eqref{asd} and \eqref{sd} in the notation of \cite{BPS1}, as opposed to that of \cite{Braun:2008ia}.}
\begin{equation} \label{asd}
	f_{ab} = \frac{i}{2}(\sigma^{\mu\nu})_{ab}F_{\mu\nu}
\end{equation}
and
\begin{equation} \label{sd}
	f_{\dot{a}\dot{b}} = -\frac{i}{2}(\bar{\sigma}^{\mu\nu})_{\dot{a}\dot{b}}F_{\mu\nu} \,,
\end{equation}
with
\begin{equation}
	\label{dagger3}
	\bar{f}_{ab} = f_{\dot{a}\dot{b}} \,.
\end{equation}
It turns out that at leading order in perturbation theory, $f_{ab}$ and $f_{\dot{a}\dot{b}}$ are local primary fields with respect to the conformal group \cite{Beisert:2004fv}.
Moreover, upon projecting onto the maximal-spin component along the $p_+$ direction, the following operators are primaries with respect to the collinear conformal subgroup $SL(2,\mathbb{R})$ \cite{Belitsky:1998gc}
\begin{align} \label{OO}
	\nonumber
	&\mathbb{O}_{s} = \Tr f_{11}(x)(i\overrightarrow{D}_++i\overleftarrow{D}_+)^{s-2}C^{\frac{5}{2}}_{s-2}\left(\frac{\overrightarrow{D}_+-\overleftarrow{D}_+}{\overrightarrow{D}_++\overleftarrow{D}_+}\right) f_{\dot{1}\dot{1}}(x) \qquad s = 2,4,6,\ldots \\\nonumber
	&\tilde{\mathbb{O}}_{s} = \Tr f_{11}(x)(i\overrightarrow{D}_++i\overleftarrow{D}_+)^{s-2}C^{\frac{5}{2}}_{s-2}\left(\frac{\overrightarrow{D}_+-\overleftarrow{D}_+}{\overrightarrow{D}_++\overleftarrow{D}_+}\right) f_{\dot{1}\dot{1}}(x) \qquad s = 3,5,7,\ldots \\\nonumber
	&\mathbb{S}_{s} =\frac{1}{\sqrt{2}}\Tr f_{11}(x)(i\overrightarrow{D}_++i\overleftarrow{D}_+)^{s-2}C^{\frac{5}{2}}_{s-2}\left(\frac{\overrightarrow{D}_+-\overleftarrow{D}_+}{\overrightarrow{D}_++\overleftarrow{D}_+}\right)f_{11}(x) \qquad s = 2,4,6,\ldots\\
	&\bar{\mathbb{S}}_{s} =\frac{1}{\sqrt{2}}\Tr f_{\dot{1}\dot{1}}(x)(i\overrightarrow{D}_++i\overleftarrow{D}_+)^{s-2}C^{\frac{5}{2}}_{s-2}\left(\frac{\overrightarrow{D}_+-\overleftarrow{D}_+}{\overrightarrow{D}_++\overleftarrow{D}_+}\right) f_{\dot{1}\dot{1}}(x)\qquad s = 2,4,6,\ldots \,,
\end{align}
and reduce in the light-cone gauge to the operators in eq.~\eqref{1000} with $f_{11}=- \partial_+ \bar A$. By allowing operator mixing with derivatives of operators of lower spin, they may be extended to primary conformal operators at next-to-leading order in the conformal renormalization scheme \cite{Braun:2003rp}, which differs from the $\overline{MS}$ scheme by a finite renormalization \cite{Braun:2003rp,BPS42}. 
Indeed, the fields $f\equiv f_{11}$ and $\bar f \equiv f_{\dot{1}\dot{1}}$ may be employed in a noncanonical path-integral quantization of YM theory in the light-cone gauge via a change of variables \cite{Belitsky:2004sc}, with action
\begin{align}
	S_{YM}(f,\bar f) = &\int d^4x\,\biggl[\bar{f}^a \partial_+^{-2}\square f^a+2\frac{g}{\sqrt{N}} f^{abc}(\partial_+^{-1}\bar f^a f^b\bar{\partial}\partial_+^{-2}\bar f^c+\partial_+^{-1}f^a \bar f^b\partial\partial_+^{-2}f^c)\nonumber\\
	&-2\frac{g^2}{N}f^{abc}f^{ade}\partial_+^{-1}(\partial_+^{-1}\bar f^b f^c)\partial_+^{-1}(\partial_+^{-1}f^d \bar{f}^e)\biggr] \,.
\end{align}

\section{Collinear quasi-partonic twist-$3$ operators}\label{app:twist3}

Gauge-invariant three-gluon quasi-partonic operators are constructed by decomposing the tensor product $F_{\alpha\beta}\otimes F_{\mu\nu}\otimes F_{\rho\sigma}$ into irreducible representations of the Lorentz group \cite{Braun:2008ia} in the spinorial representation (eqs.~\eqref{lorentza3} and \eqref{adota3}).
Restricting to operators with good transformation properties under the collinear conformal group at lowest order in perturbation theory, it suffices to consider only products of the highest-spin components, $f_{11}$ and $f_{\dot{1}\dot{1}}$
\begin{align}
	&(1,0)\otimes(1,0)\otimes(1,0) = (3,0)\oplus\ldots\\\nonumber
	&(1,0)\otimes(0,1)\otimes(1,0) = (2,1)\oplus\ldots\\\nonumber
	&(0,1)\otimes(1,0)\otimes(0,1) = (1,2)\oplus\ldots\\\nonumber
	&(0,1)\otimes(0,1)\otimes(0,1) = (0,3)\oplus\ldots \,,
\end{align}
where the ellipses indicate terms that do not have maximal collinear spin. Since the trace is cyclic, $\Tr f_{11}(x_1)f_{\dot{1}\dot{1}}(x_2)f_{11}(x_3)=\Tr f_{11}(x_3)f_{11}(x_1)f_{\dot{1}\dot{1}}(x_2)$, the orderings of fields with mixed chirality differ only by the labeling of the points. Hence, there are two independent gauge-invariant operator structures and their Hermitian conjugates
\begin{align}
	\label{naivenbasis}
	&O(x_1,x_2,x_3) =\Tr f_{11}(x_1)f_{11}(x_2)f_{\dot{1}\dot{1}}(x_3)\nonumber\\
	&S(x_1,x_2,x_3)= \Tr f_{11}(x_1)f_{11}(x_2)f_{11}(x_3) \,,
\end{align}
where in the light-cone gauge
\begin{align}
	\label{fdef}
	\nonumber
	& f_{11}=-\partial_{+} \bar{A}\\
	& f_{\dot{1}\dot{1}}=-\partial_{+} A \,.
\end{align}
Following \cite{Ohrndorf:1981qv, Braun:2003rp}, we build good operators for the collinear conformal group out of three fundamental primary fields $\Phi_{j_1}(x_1)$, $\Phi_{j_2}(x_2)$, and $\Phi_{j_3}(x_3)$, with collinear conformal spins $j_1$, $j_2$, and $j_3$ \cite{Braun_1999, Braun_2001}
\begin{align}
	\label{twist3}
	&\mathbb{O}_{\ell,n} =\nonumber\\
	&\Phi_{j_1}(x_1) \mathbb{P}^{(2j_2-1,2j_1-1)}_{n}\left(\frac{D_{x_1^+}-D_{x_2^+}}{D_{x_1^+}+D_{x_2^+}}\right)\Phi_{j_2}(x_2)\mathbb{P}^{(2j_3-1,2j_{12}-1)}_{\ell-n}\left(\frac{D_{x_1^+}+D_{x_2^+}-D_{x_3^+}}{D_{x_1^+}+D_{x_2^+}+D_{x_3^+}}\right)\Phi_{j_3}(x_3)\,,
\end{align}
which depend on two quantum numbers, the degree $\ell$ of the polynomial and $n$, with
\begin{align}
	\label{conf}
	&\ell +j_1+j_2+j_3=J=s+\frac{\tau}{2}\nonumber\\
	&n +j_1+j_2 = j_{12}\,, \qquad n = 0,\ldots,\ell\,,
\end{align}
where $J$, $s$, and $\tau$ are the collinear conformal spin, the collinear spin, and the twist of $\mathbb{O}_{\ell,n}$, respectively, and $j_{12}$ is the collinear conformal spin of the pair of fields at $x_1$ and $x_2$. To simplify the notation, we write
\begin{equation}
	\mathbb{P}^{(\alpha,\beta)}_{n}\left(\frac{X-Y}{X+Y}\right) = i^{n} (X+Y)^{n} P^{(\alpha,\beta)}_{n}\left(\frac{X-Y}{X+Y}\right)\,,
\end{equation}
where $P_n^{(\alpha,\beta)}$ are Jacobi polynomials (appendix~\ref{app:jacobi}).
The collinear spin $s \geq 0$ is the eigenvalue of the spin of the quasi-partonic operator projected onto the light-cone direction $p_+$ \cite{Belitsky:2003sh}. 
By definition, primary fields $\mathcal{O}_{j}(x)$ with collinear conformal spin $j$ have good transformation properties under the collinear conformal group $SL(2,\mathbb{R})$ \cite{Braun:2003rp}, whose generators satisfy 
\begin{align}
	\left[L_0,L_\mp\right]=\mp L_\mp,\qquad\left[L_-,L_+ \right]=-2L_0
\end{align}
and act on primary operators $\mathcal{O}_{j}(x)$ as
\begin{align} 
	\label{confcom}
	&\left[L_+,\mathcal{O}_j(x^+)\right] = -\partial_+\mathcal{O}_j(x^+)\\\nonumber
	&\left[L_-,\mathcal{O}_j(x^+)\right] = (x^{+\,2}\partial_++2jx^+)\mathcal{O}_j(x^+)\\\nonumber
	&\left[L_0,\mathcal{O}_j(x^+)\right] = (x^+ \partial_++j)\mathcal{O}_j(x^+) \,,
\end{align}
with
\begin{equation}
	\partial_+=\frac{\partial}{\partial x^+}=\frac{\partial}{\partial x_-}\,.
\end{equation}
Eq.~\eqref{confcom} must be satisfied by both the fundamental fields $\Phi_{j_i}$, with $j=j_i$, and the operators $\mathbb{O}_{\ell,n}$, with $j=J$, so that
\begin{align}
	&\left[L^2,\mathbb{O}_{\ell,n}(0)\right] = J(J-1)\,\mathbb{O}_{\ell,n}(0)\nonumber\\
	&\left[L_-,\mathbb{O}_{\ell,n}(0)\right] =0\nonumber\\
	&\left[L_0,\mathbb{O}_{\ell,n}(0)\right] = J\,\mathbb{O}_{\ell,n}(0)\,.
\end{align}
These conditions do not uniquely fix the three-field conformal basis. To resolve this degeneracy, following \cite{Braun_1999} and \cite{Braun_2001}, we impose that $\mathbb{O}_{\ell,n}$ diagonalizes the two-particle Casimir operator $L^2_{12}$ of the generators acting on the fields at $x_1$ and $x_2$:
\begin{equation}
	\left[L^2_{12},\mathbb{O}_{\ell,n}(0)\right] = j_{12}(j_{12}-1)\,\mathbb{O}_{\ell,n}(0)\,,
\end{equation}
with $j_{12}$ given in eq.~\eqref{conf}. Then, choosing $\Phi_{\frac{3}{2}} =\partial_+\bar{A}$ or $\Phi_{\frac{3}{2}} =\partial_+A$ in the light-cone gauge, so that $\ell=s-3$ and $j_{12}=n+3$, we obtain
\begin{align}\label{eq:tw3basis}
	&\mathbb{O}_{s,n}(x_1,x_2,x_3) = \nonumber\\
	&\Tr \partial_+\bar{A}(x_1)\mathbb{P}^{(2,2)}_{n}\left(\frac{\partial_{x_1^+}-\partial_{x_2^+}}{\partial_{x_1^+}+\partial_{x_2^+}}\right)\partial_+\bar{A}(x_2)\mathbb{P}^{(2,2n+5)}_{s-n-3}\left(\frac{\partial_{x_1^+}+\partial_{x_2^+}-\partial_{x_3^+}}{\partial_{x_1^+}+\partial_{x_2^+}+\partial_{x_3^+}}\right)\partial_+A(x_3)\nonumber\\
	&\bar{\mathbb{O}}_{s,n}(x_1,x_2,x_3) = \nonumber\\
	&\Tr \partial_+A(x_1)\mathbb{P}^{(2,2)}_{n}\left(\frac{\partial_{x_1^+}-\partial_{x_2^+}}{\partial_{x_1^+}+\partial_{x_2^+}}\right)\partial_+A(x_2)\mathbb{P}^{(2,2n+5)}_{s-n-3}\left(\frac{\partial_{x_1^+}+\partial_{x_2^+}-\partial_{x_3^+}}{\partial_{x_1^+}+\partial_{x_2^+}+\partial_{x_3^+}}\right)\partial_+\bar{A}(x_3)\nonumber\\
	&\mathbb{S}_{s,n}(x_1,x_2,x_3)= \nonumber\\
	&\Tr \partial_+\bar{A}(x_1)\mathbb{P}^{(2,2)}_{n}\left(\frac{\partial_{x_1^+}-\partial_{x_2^+}}{\partial_{x_1^+}+\partial_{x_2^+}}\right)\partial_+\bar{A}(x_2)\mathbb{P}^{(2,2n+5)}_{s-n-3}\left(\frac{\partial_{x_1^+}+\partial_{x_2^+}-\partial_{x_3^+}}{\partial_{x_1^+}+\partial_{x_2^+}+\partial_{x_3^+}}\right)\partial_+\bar{A}(x_3)\,,\nonumber\\
	&\bar{\mathbb{S}}_{s,n}(x_1,x_2,x_3)= \nonumber\\
	&\Tr \partial_+A(x_1)\mathbb{P}^{(2,2)}_{n}\left(\frac{\partial_{x_1^+}-\partial_{x_2^+}}{\partial_{x_1^+}+\partial_{x_2^+}}\right)\partial_+A(x_2)\mathbb{P}^{(2,2n+5)}_{s-n-3}\left(\frac{\partial_{x_1^+}+\partial_{x_2^+}-\partial_{x_3^+}}{\partial_{x_1^+}+\partial_{x_2^+}+\partial_{x_3^+}}\right)\partial_+A(x_3)\,,		
\end{align}
with $n=0,\ldots,s-3$, where the local operators are obtained by setting $x_1=x_2=x_3$. Parity eigenstates are built as the linear combinations
\begin{align}
	&\mathbb{O}^{\pm}_{s,n}(x_1,x_2,x_3) = \mathbb{O}_{s,n}(x_1,x_2,x_3) \pm \bar{\mathbb{O}}_{s,n}(x_1,x_2,x_3) \nonumber\\
	&\mathbb{S}^{\pm}_{s,n}(x_1,x_2,x_3) = \mathbb{S}_{s,n}(x_1,x_2,x_3) \pm \bar{\mathbb{S}}_{s,n}(x_1,x_2,x_3) \,.
\end{align}
In both operators, the fields at $x_1$ and $x_2$ have the same chirality and are therefore identical, while $P^{(2,2)}_n$ has parity $(-1)^n$ under their exchange by eq.~\eqref{jacox}, and the other polynomial is symmetric. Since $\Tr(\lambda^a\lambda^b\lambda^c)=\frac{1}{4}(d^{abc}+if^{abc})$, with $d^{abc}$ symmetric and $f^{abc}$ antisymmetric in $a$ and $b$, the operators with even $n$ contain only $d^{abc}$, and those with odd $n$ contain only $f^{abc}$, i.e., they are odd and even under charge conjugation, respectively.

\section{Euclidean RG-improved asymptotic correlators}  \label{app:C}

\subsection{Operator mixing  \label{app:C1}}

We briefly summarize the construction of RG-improved asymptotic correlators following \cite{BPSpaper2,Bochicchio:2021geometry}.
The renormalized Euclidean correlators
\begin{equation}\label{key}
	\langle [\mathcal{O}_{k_1}](x_1)\ldots [\mathcal{O}_{k_n}](x_n) \rangle = G^{(n)}_{k_1 \ldots k_n}( x_1,\ldots,  x_n; \mu, g(\mu))
\end{equation}
satisfy the Callan-Symanzik equation
\begin{align}\label{CallanSymanzik}
	& \Big(\sum_{\alpha = 1}^n x_\alpha \cdot \frac{\partial}{\partial x_\alpha} + \beta(g)\frac{\partial}{\partial g} + \sum_{\alpha = 1}^n D_{\mathcal{O}_\alpha}\Big)G^{(n)}_{k_1 \ldots k_n} + \sum_a \Big(\gamma_{k_1a}(g) G^{(n)}_{ak_2 \ldots k_n} \nonumber\\
	&\quad+ \gamma_{k_2a}(g) G^{(n)}_{k_1 a k_3 \ldots k_n} + \cdots +\gamma_{k_n a}(g) G^{(n)}_{k_1 \ldots a}\Big) = 0\,,
\end{align}
with solution
\begin{align}\label{csformula}
	&G^{(n)}_{k_1 \ldots k_n}(\lambda x_1,\ldots, \lambda x_n; \mu, g(\mu)) \nonumber \\
	&= \sum_{j_1 \ldots j_n} Z_{k_1 j_1} (\lambda)\ldots Z_{k_n j_n}(\lambda)\hspace{0.1cm} \lambda^{-\sum_{i=1}^nD_{\mathcal{O}_{j_i}}} G^{(n)}_{j_1 \ldots j_n }( x_1, \ldots, x_n; \mu, g(\frac{\mu}{\lambda}))\,,
\end{align}
where $D_{\mathcal{O}_i}$ is the canonical dimension of $\mathcal{O}_i(x)$ and $\gamma(g)=\gamma_0 g^2+ \cdots$ is the matrix of anomalous dimensions, with
\begin{equation}\label{eqZ}
	\Bigg(\frac{\partial}{\partial g} + \frac{\gamma(g)}{\beta(g)}\Bigg)Z(\lambda) = 0
\end{equation}
in matrix notation, and
\begin{equation} \label{ZZ}
	Z(\lambda) = P\exp \Big(\int_{g(\mu)}^{g(\frac{\mu}{\lambda})}\frac{\gamma(g')}{\beta(g')} dg'\Big)\,.
\end{equation}
Eq.~\eqref{csformula} simplifies greatly if a renormalization scheme exists in which $Z(\lambda)$ is diagonalizable to all orders in perturbation theory, and specifically one-loop exact, with eigenvalues $Z_{\mathcal{O}_i}(\lambda)$ \cite{Bochicchio:2021geometry}
\begin{equation} \label{zz}
	Z_{\mathcal{O}_i}(\lambda) = \Bigg(\frac{g(\mu)}{g(\frac{\mu}{\lambda})}\Bigg)^{\frac{\gamma_{0\mathcal{O}_i}}{\beta_0}} \,,
\end{equation}
where $\gamma_{0\mathcal{O}_i}$ are the eigenvalues of $\gamma_0$. In this scheme, eq.~\eqref{csformula} contains only a single term
\begin{align}\label{csformuladiag}
	&G^{(n)}_{j_1 \ldots j_n}(\lambda x_1,\ldots, \lambda x_n; \mu, g(\mu)) \nonumber \\
	&=Z_{\mathcal{O}_{j_1}}(\lambda) \ldots Z_{\mathcal{O}_{j_n}}(\lambda)\,\lambda^{-\sum_{i=1}^nD_{\mathcal{O}_{j_i}}}G^{(n)}_{j_1 \ldots j_n }( x_1, \ldots, x_n; \mu, g(\frac{\mu}{\lambda}))\,.
\end{align}
Then, as $\lambda \rightarrow 0$, in any renormalization scheme, the renormalized correlator on the right-hand side admits the perturbative asymptotic expansion in terms of the renormalized coupling $g(\frac{\mu}{\lambda})$ at the scale $\frac{\mu}{\lambda}$
\begin{align} \label{eq:expansion}
	G^{(n)}_{j_1 \ldots j_n }( x_1, \ldots, x_n; \mu, g(\frac{\mu}{\lambda}))\,\sim\, &{\mathcal G}^{(n,0)}_{j_1 \ldots j_n }( x_1, \ldots, x_n; \mu)+g^2(\frac{\mu}{\lambda}) \, {\mathcal G}^{(n,2)}_{j_1 \ldots j_n }( x_1, \ldots, x_n; \mu)\nonumber\\
	&+ g^4(\frac{\mu}{\lambda}) \, {\mathcal G}^{(n,4)}_{j_1 \ldots j_n }( x_1, \ldots, x_n; \mu)
	+\cdots\,.
\end{align}
Of course, the first term in the above expansion, being independent of the coupling, is the conformal contribution at zero coupling:
\begin{equation}
	{\mathcal G}^{(n,0)}_{j_1 \ldots j_n }( x_1, \ldots, x_n; \mu) = G^{(n)}_{\text{conf} \, j_1 \ldots j_n }( x_1, \ldots, x_n) \, ,
\end{equation}
since the renormalized operators at zero coupling coincide with the conformal ones. 
The higher-order corrections in eq.~\eqref{eq:expansion} arise from the nonconformal interaction due to the nonvanishing beta function, so that the conformal contribution is corrected at higher orders in the renormalized coupling as displayed, with the renormalized operators becoming nonconformal at higher orders in any renormalization scheme.
However, provided that the conformal contribution is nonvanishing, for fixed $x_1, \ldots, x_n$, all higher-order terms in eq.~\eqref{eq:expansion} are suppressed with respect to the conformal one by powers of 
\begin{align} 
	g^2(\frac{\mu}{\lambda}) &\sim \dfrac{1}{\beta_0 \log(\frac{\mu^2}{\lambda^2\Lambda_{YM}^2})}\left(1-\frac{\beta_1}{\beta_0^2}\frac{\log\log(\frac{\mu^2}{\lambda^2\Lambda_{YM}^2})}{\log(\frac{\mu^2}{\lambda^2\Lambda_{YM}^2})}\right)\nonumber\\
	&\sim \dfrac{1}{\beta_0 \log(\frac{1}{\lambda^2})}\left(1-\frac{\beta_1}{\beta_0^2}\frac{\log\log(\frac{1}{\lambda^2})}{\log(\frac{1}{\lambda^2})}\right)
\end{align}
-- i.e., asymptotically by inverse powers of $\log \frac{1}{\lambda}$ -- despite being generally nonconformal.\par
Hence, the corresponding UV asymptotics, for fixed $x_1, \ldots, x_n$, reads\footnote{If $\mathcal{G}^{(n,0)}$ vanishes and $\mathcal{G}^{(n,2)}$ does not, the latter can only be computed on a case-by-case basis, since the YM action vertices would be involved in the computation. Therefore, finding an explicit generalization of eq.~\eqref{eqrg} in this case was outside the scope of \cite{BPSpaper2}.} as $\lambda \rightarrow 0$
\begin{align} \label{eqrg}
	&\langle [\mathcal{O}_{j_1}](\lambda x_1)\ldots[\mathcal{O}_{j_n}](\lambda x_n)\rangle \,\sim \,\frac{Z_{\mathcal{O}_{j_1}}(\lambda)\ldots Z_{\mathcal{O}_{j_n}}(\lambda)}{\lambda^{D_{\mathcal{O}_1}+\cdots+D_{\mathcal{O}_n}}} G^{(n)}_{\text{conf}\,j_1 \ldots j_n }( x_1, \ldots, x_n)\,.
\end{align}
We refer to this scheme as the nonresonant diagonal scheme \cite{Bochicchio:2021geometry}, whose existence follows from the Poincar\'e-Dulac theorem via the following differential-geometric interpretation of operator mixing \cite{Bochicchio:2021geometry}.
We interpret a finite change of basis of the renormalized operators
\begin{equation}\label{linearcomb}
	[\mathcal{O}]'(x) = S(g)[ \mathcal{O}](x)
\end{equation}
in matrix notation as a real-analytic invertible gauge transformation $S(g)$ that depends on $g \equiv g(\mu)$. Then, the matrix
\begin{equation}
	A(g) = -\frac{\gamma(g)}{\beta(g)} = \frac{1}{g} \Big(\frac{\gamma_0}{\beta_0} + \cdots\Big)
\end{equation}
that enters the differential equation for $Z(\lambda)$,
\begin{equation}
	\Big(\frac{\partial}{\partial g} - A(g)\Big) Z(\lambda) = 0\,,
\end{equation}
defines a connection $A(g)$,
\begin{eqnarray} \label{sys2}
	A(g)= \frac{1}{g} \left(A_0 + \sum^{\infty}_ {n=1} A_{2n} g^{2n} \right)\,,
\end{eqnarray}
with a regular singularity at $g = 0$ that transforms as
\begin{equation}
	A'(g) = S(g)A(g)S^{-1}(g) + \frac{\partial S(g)}{\partial g} S^{-1}(g)
\end{equation}
under the gauge transformation $S(g)$, where
\begin{equation}
	\mathcal{D} = \frac{\partial }{\partial g} - A(g)
\end{equation}
is the corresponding covariant derivative.
Consequently, $Z(\lambda)$ is a Wilson line that transforms as
\begin{equation}
	Z'(\lambda) = S(g(\mu))Z(\lambda)S^{-1}(g(\frac{\mu}{\lambda}))\,.
\end{equation}
It follows from the Poincar\'e-Dulac theorem~\cite{Bochicchio:2021geometry} that if any two eigenvalues $\lambda_1, \lambda_2, \ldots$ of the matrix $\frac{\gamma_0}{\beta_0}$, arranged in nonincreasing order $\lambda_1 \geq \lambda_2 \geq \ldots$, do not differ by a positive even integer,
\begin{equation} \label{nr}
	\lambda_i - \lambda_j - 2k \neq 0
\end{equation}
for $i \leq j$ and $k$ a positive integer -- i.e., they are nonresonant -- then there exists a gauge transformation that puts $A(g)$ into the canonical nonresonant form \cite{Bochicchio:2021geometry}
\begin{equation} \label{1loop}
	A'(g) = \frac{\gamma_0}{\beta_0}\frac{1}{g}\,,
\end{equation}
which is one-loop exact to all orders in perturbation theory. Hence, if $\frac{\gamma_0}{\beta_0}$ is also diagonalizable, eq.~\eqref{zz} follows.

\subsection{Nonresonant diagonal renormalization scheme  \label{app:C2}}

To make the present paper self-contained, we provide the order-by-order construction in perturbation theory of the nonresonant diagonal scheme \cite{Bochicchio:2021geometry}.\par
The construction proceeds by induction on $k=1,2, \ldots$ by demonstrating that, once $A_0$ and the first $k-1$ matrix coefficients $A_2,\ldots,A_{2(k-1)}$ in eq.~\eqref{sys2} have been put into the canonical nonresonant form of eq.~\eqref{1loop} -- i.e., $A_0$ diagonal and $A_2,\ldots,A_{2(k-1)}=0$ -- there exists a real-analytic gauge transformation that leaves them invariant and sets the $k$-th coefficient $A_{2k}$ to $0$ as well.\par
The zeroth step of the induction consists of putting $A_0$ into diagonal form -- with its eigenvalues arranged in nonincreasing order -- via a constant gauge transformation.\par
At the $k$-th step, we choose the gauge transformation
\begin{eqnarray}
	S_k(g)=1+ g^{2k} H_{2k}\,,
\end{eqnarray}
where $H_{2k}$ is a matrix to be determined below. Its inverse reads
\begin{eqnarray}
	S^{-1}_k(g)= (1+ g^{2k} H_{2k})^{-1} = 1- g^{2k} H_{2k} + \cdots\,,
\end{eqnarray}
where the dots represent terms of order higher than $g^{2k}$.
The action of $S_k(g)$ on the connection $A(g)$ yields
\begin{align} \label{ind}
	&A'(g) \nonumber\\
	&=  2k g^{2k-1} H_{2k} ( 1+ g^{2k} H_{2k})^{-1}+  (1+ g^{2k} H_{2k}) A(g)( 1+ g^{2k} H_{2k})^{-1} \nonumber \\
	&=  2k g^{2k-1} H_{2k} ( 1+ g^{2k} H_{2k})^{-1} +  (1+ g^{2k} H_{2k})  \frac{1}{g} \left(A_0 + \sum^{\infty}_ {n=1} A_{2n} g^{2n} \right) ( 1+ g^{2k} H_{2k})^{-1} \nonumber \\
	&=  2k g^{2k-1} H_{2k} ( 1- \cdots) +  (1+ g^{2k} H_{2k})  \frac{1}{g} \left(A_0 + \sum^{\infty}_ {n=1} A_{2n} g^{2n} \right)  ( 1- g^{2k} H_{2k}+\cdots) \nonumber \\
	&=  2k g^{2k-1} H_{2k}  +    \frac{1}{g} \left(A_0 + \sum^k_ {n=1} A_{2n} g^{2n} \right) + g^{2k-1} (H_{2k}A_0-A_0H_{2k}) + \cdots\nonumber \\
	&=   g^{2k-1} (2k H_{2k} + H_{2k} A_0 - A_0 H_{2k})  + A_{2(k-1)}(g) + g^{2k-1} A_{2k}+ \cdots\,,
\end{align}
where we have omitted all terms contributing at an order higher than $g^{2k-1}$, with
\begin{align}
	A_{2(k-1)}(g) =  \frac{1}{g} \left(A_0 + \sum^{k-1}_ {n=1} A_{2n} g^{2n} \right)
\end{align}
being the part of $A(g)$ that is unaffected by the gauge transformation $S_k(g)$, which therefore satisfies the induction hypothesis -- i.e., that $A_2, \ldots, A_{2(k-1)}$ vanish.\par
Thus, by eq.~\eqref{ind}, the $k$-th matrix coefficient $A_{2k}$ may be eliminated from the expansion of $A'(g)$ up to order $g^{2k-1}$ provided that there exists a matrix $H_{2k}$ such that
\begin{align}
	A_{2k}+(2k H_{2k} + H_{2k} A_0 - A_0 H_{2k})= A_{2k}+ (2k-\mathrm{ad}_{A_0}) H_{2k}=0\,,
\end{align}
with $\mathrm{ad}_{A_0}Y=[A_0,Y]$.
If the inverse of $\mathrm{ad}_{A_0}-2k$ exists, the unique solution for $H_{2k}$ is
\begin{eqnarray}
	H_{2k}=(\mathrm{ad}_{A_0}-2k)^{-1} A_{2k}\,.
\end{eqnarray}
Hence, to complete the induction, we must demonstrate that if the eigenvalues of $A_0$ are nonresonant, then $\mathrm{ad}_{A_0}-2k$ is invertible, i.e., its kernel is trivial.\par
Now, $\mathrm{ad}_{\Lambda}-2k$, as a linear operator acting on matrices, is diagonal, with eigenvalues $\lambda_{i}-\lambda_{j}-2k$ and eigenvectors given by the matrices $E_{ij}$, whose only nonvanishing entries are $(E_{ij})_{ij}$. The eigenvectors $E_{ij}$, normalized so that $(E_{ij})_{ij}=1$, form an orthonormal basis for the space of matrices. Therefore, $E_{ij}$ belongs to the kernel of $\mathrm{ad}_{\Lambda}-2k$ if and only if $\lambda_{i}-\lambda_{j}-2k=0$. Consequently, since $\lambda_{i}-\lambda_{j}-2k \neq 0$ for every $i,j$ by assumption, the kernel of $\mathrm{ad}_{\Lambda}-2k$ contains only $0$, and the construction is complete.

\subsection{Anomalous dimensions of twist-$2$ operators  \label{app:C3}}

We define the bare collinear twist-$2$ operators with $s \geq 2$ and $k \geq 0$ \cite{Belitsky:1998gc},
\begin{equation}
	\mathcal{O}^{(k)}_{s} = (-i\partial_+)^{k}\mathcal{O}_{s}\,,
\end{equation}
which, at leading order in perturbation theory and for $k>0$, are conformal descendants \cite{Braun:2003rp} of the corresponding primary conformal operator $\mathcal{O}^{(0)}_{s}=\mathcal{O}_{s}$.
As a consequence of operator mixing, we obtain \cite{Belitsky:1998gc,Braun:2003rp} for the renormalized operators
\begin{equation}\label{m}
	[\mathcal{O}^{(k)}_{s}]= \sum^{s}_{i} Z_{si} \mathcal{O}^{(k+s-i)}_{ i} \,,
\end{equation}
where $Z$ is the bare mixing matrix and 
\begin{equation}
	\gamma(g) =- \frac{\partial Z}{\partial \log \mu} Z^{-1}=\sum_{j=0}^{\infty}\gamma_j \, g^{2j+2}
\end{equation}
is the -- generally lower-triangular -- matrix of anomalous dimensions, with $\gamma_0$ being diagonal in the $\overline{MS}$ scheme \cite{Belitsky:1998gc,Braun:2003rp}.
In our notation, the eigenvalue of $\gamma_0$ corresponding to $\mathbb{O}_{s}$, with even $s\geq 2$, reads \cite{Belitsky:1998gc,Belitsky:2004sc}
\begin{align}
	\label{an1}
	\gamma_{0\mathbb{O}_{s}}= \frac{2}{(4 \pi)^2} \Bigg(&4 \psi(s+1) - 4 \psi(1) -\frac{11}{3}- 8 \frac{s^2+s+1}{(s-1)s(s+1)(s+2)}\Bigg) \,,
\end{align}
so that $\gamma_{0\mathbb{O}_{2}}=0$, consistent with the conservation of the stress-energy tensor, whereas the remaining eigenvalues -- $\gamma_{0\mathbb{O}_{s}}$ for $s >2$ -- are positive. 
Moreover, we have
\begin{align}
	\label{an2}
	\gamma_{0s}= \frac{2}{(4 \pi)^2}  \Bigg(&4 \psi(s+1) - 4 \psi(1) -\frac{11}{3} - 8 \frac{s^2+s-2}{(s-1)s(s+1)(s+2)} \Bigg)
\end{align}
for $\tilde{\mathbb{O}}^{(k)}_{s}$ with odd $s\geq 3$. 
The eigenvalues of $\gamma_0$ for $\mathbb{S}^{(k)}_{s}$ and $\bar{\mathbb{S}}^{(k)}_{s}$ are given by \cite{Belitsky:2004sc}
\begin{equation}\label{an3}
	\gamma_{0s}=  \frac{2}{(4 \pi)^2} \left(4\psi(s+1)-4\psi(1)-\frac{11}{3}\right)
\end{equation}
with even $s\geq 2$.\par
From the above equations, it follows that in SU($N$) YM theory, $\gamma_0$ is independent of $N$ and thus has no nonplanar contribution.\par 
We verified numerically \cite{BPSpaper2} that the nonresonant condition in eq.~\eqref{nr} is satisfied up to $s=10^4$ by $\gamma_{0\mathbb{O}_{s}}$ in eq.~\eqref{an1}. Moreover, the nonresonant condition for all collinear twist-$2$ operators in YM theory was subsequently proved \cite{S1,S2}.\par 
Hence, the nonresonant diagonal basis exists for all collinear twist-$2$ operators and reduces to the basis of conformal operators at lowest order in perturbation theory -- where $\gamma_0$ is diagonal -- which may be employed to compute the conformal contribution on the right-hand side of eq.~\eqref{eqrg}.\par

\newpage
\bibliographystyle{JHEP}
\bibliography{mybib} % BibTeX database without .bib extension
\end{document}